\documentclass[aps,preprint,superscriptaddress]{revtex4-1}

\usepackage{graphicx}
\usepackage{amsmath}
\usepackage{amssymb}
\usepackage{slashed}
\usepackage{hyperref}
\hypersetup{hidelinks}

\newcommand{\tr}{\rm tr \,}
\newcommand{\CD}{\hat \partial }

\usepackage{array} 
\newcolumntype{C}{>{$}c<{$}}

\begin{document}


\title{Triangle and box diagrams \\ in coupled-channel systems from the chiral Lagrangian}


\author{Tobias Isken}
\affiliation{GSI Helmholtzzentrum f\"ur Schwerionenforschung GmbH, \\Planckstra\ss e 1, 64291 Darmstadt, Germany}
\affiliation{Helmholtz Forschungsakademie Hessen f\"ur FAIR (HFHF) }
\author{Xiao-Yu Guo}
\affiliation{Beijing University of Technology,\\
  Beijing 100124, China}
\author{Yonggoo Heo}
\affiliation{Bogoliubov Laboratory for Theoretical Physics, Joint Institute for Nuclear Research, RU-141980 Dubna, Moscow region, Russia}
\affiliation{GSI Helmholtzzentrum f\"ur Schwerionenforschung GmbH, \\Planckstra\ss e 1, 64291 Darmstadt, Germany}
\author{Csaba L. Korpa}
\affiliation{Institute of Physics, University of P\'ecs,\\ Ifj\'us\'ag \'utja 6,
7624 P\'ecs, Hungary}
\author{Matthias F.M. Lutz}
\affiliation{GSI Helmholtzzentrum f\"ur Schwerionenforschung GmbH, \\Planckstra\ss e 1, 64291 Darmstadt, Germany}
\date{\today}

\begin{abstract}

We perform an analysis of triangle- and box-loop contributions to the generalized potential in the 
scattering of Goldstone bosons off the $J^P= 0^-$ and $1^-$ charmed mesons. Particular emphasis 
is put on the use of on-shell mass parameters in such contributions in terms of a  
renormalization scheme that ensures the absence of power-counting violating terms. This is achieved with a systematically extended set of Passarino--Veltman basis functions, that leads to manifest power-counting conserving one-loop expressions and avoids the occurrence of superficial kinematical singularities. 
Compact expressions to chiral order three and four are presented that are particularly useful 
in coding such coupled-channel systems. Our formal results are generic and prepare analogous computations for other systems, like meson-baryon scattering from the chiral Lagrangian. 
\end{abstract}


\maketitle
\tableofcontents



\section{Introduction}
\label{sec:1}

The role of left-hand cut contributions in coupled-channel systems receives increasing attention in the hadron physics community as the interplay of modern effective field theory approaches with Lattice QCD simulations requests more and more quantitative and controlled computations. The open-charm sector of QCD not only serves as 
a convenient laboratory since it is largely driven by the symmetries of QCD \cite{Casalbuoni:1996pg,Kolomeitsev:2003ac,Lutz:2015ejy,Chen:2016spr}, but also offers already a sizeable data set from  Lattice QCD simulations \cite{Aoki:2008sm,Mohler:2011ke,Na:2012iu,Kalinowski:2015bwa,Cichy:2016bci,Cheung:2016bym,Moir:2016srx,Cheung:2020mql,Gayer:2021xzv}. We consider such studies as a preparation for the more demanding meson-baryon systems for which the scattering data set from Lattice QCD simulations is significantly more scarce \cite{Fettes:2000xg,Mai:2009ce,Huang:2017bmx,Andersen:2019ktw,Lang:2012db,Detmold:2015qwf,Andersen:2017una,Silvi:2021uya,Bulava:2022vpq,Bulava:2023gfx,Bulava:2023rmn,Alexandrou:2023elk}.

Studies of the quark-mass dependence of the charmed meson masses are the key for the quantitative understanding of the coupled-channel interactions of the latter with the Goldstone bosons of QCD \cite{Guo:2018kno,Guo:2021kdo}. It is useful to acknowledge that 
simultaneous approaches for hadron masses together with their scattering properties are significantly more constrained by QCD as compared to partial studies. Early coupled-channel works in the open-charm sector focused on the s-wave interactions only and ignored the impact from the quark-mass dependence of the charmed meson masses 
\cite{Kolomeitsev:2003ac,Hofmann:2003je,Lutz:2007sk,Liu:2012zya,Altenbuchinger:2013vwa,Cleven:2014oka,Du:2016tgp,Huang:2022cag}. 
Coupled-channel studies of p-wave and d-wave systems are of equal importance, since in Lattice QCD studies or experimental cross section results, a focus on s-wave terms only is not always  possible. For the latter the scattering processes cannot be reliably described by algebraic matrix equations (see e.g. \cite{Kolomeitsev:2003ac,Lutz:2007sk,Guo:2018kno}), that may lead to unitarity but are at odds with the long-range part of the coupled-channel forces as they arise from t- or u-channel exchange processes at the tree- or loop-level \cite{Lutz:2015lca,Lutz:2018kaz,Lutz:2022enz,Korpa:2022voo}. 
We note that a suitable framework for such systems is offered by the generalized potential approach (GPA) as was developed in  
\cite{Gasparyan:2010xz,Danilkin:2010xd,Danilkin:2011fz,Gasparyan:2012km}. It systematically extends the 
applicability domain of the chiral Lagrangian into the resonance region by using an expansion of the 
generalized potential in terms of conformal variables, where the expansion coefficients are well accessible within Chiral Perturbation Theory ($\chi$PT).

In our current formal work we focus on one-loop triangle and box contributions that have not been studied at sufficient rigor from the chiral Lagrangian. While a first estimate of such effects was reported on in \cite{Yao:2015qia,Du:2017ttu} for s-wave scattering in the open-charm system results exist yet for neither s-wave nor p-wave scattering in application of a GPA. It is a challenge to perform such computations in a manner such that on-shell hadron masses can be used in the loop expressions without violating the chiral Ward identities of QCD and the expectations of dimensional counting rules.  Previous works (see e.g. \cite{Fuchs:2003qc,Du:2017ttu}) consider a renormalization of the loop effects using the  extended-on-mass-shell (EOMS) scheme, in which renormalization-scale 
independent results are obtained only if bare hadron masses are used inside the loop expression. We will further develop our scheme and provide explicit expressions for triangle-  and box-loop contributions in the open-charm sector of QCD at chiral order three and four.


\newpage

\section{Scattering from the chiral Lagrangian}
\label{sec:2}

We use the chiral Lagrangian as presented in \cite{Lutz:2022enz} for the two antitriplets of $D$ mesons with $J^P =0^-$ and  $J^P =1^-$ quantum numbers. The $1^-$ states are interpolated in terms of  antisymmetric tensor fields.  The covariant 
derivative involves the chiral connection, and the quark masses enter via the symmetry breaking fields that are linear in the masses, $m_{u,d,s}$, of the up, down and strange quarks. The 
octet of the Goldstone boson fields is encoded into a $3\times3$ matrix. The parameter $f$ is the chiral limit value of the pion-decay constant. Finally the parameters $M$ and $\tilde M$ give the masses of the $D$ and $D^*$ mesons at $m_u=m_d =m_s =0$, where in  the limit of a very large charm-quark mass a common mass arises with $\tilde M/ M \rightarrow 1 $.
The construction of this chiral Lagrangian and implications for the heavy-quark mass limit go back to the early works  \cite{Yan:1992gz,Casalbuoni:1996pg,Kolomeitsev:2003ac,Hofmann:2003je,Lutz:2007sk,Guo:2018kno}. All terms relevant in our current work are recalled in Appendix A.

While the leading order terms introduce the kinetic terms of the mesons with covariant derivatives, the 
first-order interaction terms  provide the 3-point coupling constants of the Goldstone bosons to the charmed mesons parameterized by the low-energy constants (LEC) $g_P$ and $\tilde g_P$. While the decay of the charged $D^*$ meson  implies $ |g_P| = 0.57 \pm 0.07 $ the parameter $\tilde g_P$  cannot be extracted from empirical data directly. The size of $\tilde g_P \simeq g_P$ can be estimated using the heavy-quark spin symmetry of QCD \cite{Yan:1992gz,Casalbuoni:1996pg}. 

Second order terms of the chiral Lagrangian were first studied in \cite{Hofmann:2003je,Lutz:2007sk,Guo:2008gp}.  All parameters $c_i$ and $\tilde c_i$ are expected to scale linearly in the parameter $M$. It holds $\tilde c_i = c_i$  in the heavy-quark mass limit \cite{Lutz:2007sk}. A first estimate of the LEC can be found in \cite{Lutz:2007sk} based on the leading order large-$N_c$ relations. In the combined heavy-quark and large-$N_c$ limit we are left with 4 free parameters only, $c_1, c_3, c_5, c_6$. Additional terms relevant at chiral order three were considered in \cite{Geng:2010vw,Yao:2015qia,Du:2017ttu,Guo:2018kno,Jiang:2019hgs}. A complete list of such terms is given in \cite{Lutz:2022enz}, 
where we note that the LEC with $\tilde g_i \simeq g_i$ do not contribute to the meson masses at the one-loop level. Rather, they are instrumental to achieve a more accurate description of the coupled-channel systems presented here.

\begin{table}[b]
\tabcolsep=4.1mm
\begin{tabular}{|ccccccccccccccccccccccc|}
\hline
\multicolumn{3}{|c|}{$(\frac{1}{2},+2)$} &
\multicolumn{3}{c|}{$(0,+1)$} &
\multicolumn{3}{c|}{$(1,+1)$} &
\multicolumn{3}{c|}{$(\frac12,0)$} &
\multicolumn{3}{c|}{$(\frac32,0)$} &
\multicolumn{3}{c|}{$(0,-1)$} &
\multicolumn{3}{c|}{$(1,-1)$}
\\ \hline
\multicolumn{3}{|c|}{$1\,:\,K\,D_s$} &
\multicolumn{3}{c|}{$\begin{array}{l} 
1\,:\,K\,D \\ 2\,:\,\eta\,D_s 
\end{array}$}&
\multicolumn{3}{c|}{$\begin{array}{c} 1\,:\,\pi\,D_s \\ 
2\,:\,K\,D 
\end{array}$} &
\multicolumn{3}{l|}{$\begin{array}{l} 1\,:\,\pi\,D  \\ 2\,:\, \eta \,D \\ 3\,:\,\bar K\,D_s
\end{array}$} &
\multicolumn{3}{|l|}{$1\,:\, \pi \,D$} &
\multicolumn{3}{l|}{$1\,:\,\bar K\,D$} &
\multicolumn{3}{l|}{$1\,:\,\bar K \,D$}
\\\hline
\end{tabular}
\caption{Coupled-channel states with $(I,S)$ as introduced in \cite{Kolomeitsev:2003ac}. For a channel index $a \leftrightarrow QH$ the meson content $Q \in \{\pi, K, \bar K, \eta \}$ and $ H\in \{D, D_s\}= [0^-]$ is specified. }
\label{tab:states}
\end{table}

We consider the scattering of the Goldstone bosons off the charmed meson states with $J^P = 0^-$ and $J^P=1^-$. 
The corresponding scattering amplitudes are characterized by their isospin (I) and strangeness (S) quantum numbers. For simplicity we recall the scattering processes of the $J^P = 0^-$ states first. 
The tree-level scattering amplitudes at leading and subleading chiral orders take the form
\begin{eqnarray}
&& f^2\,T^{(1)}_{ab}(s,t,u)= \frac{s-u}{4}\,C_{\rm WT} + g_P^2\, \sum_{H \in [1^-]
 } \,C^{(s)}_{H}\,\frac{s\,(\bar q\cdot q) -(\bar q \cdot w)\,(w\cdot q)}{s- M_H^2}
 \nonumber\\
&& \qquad \qquad +\,g_P^2\, \sum_{H \in [1^-]} \,C^{(u)}_{H}\, \frac{u\,(\bar q\cdot q) -(\bar q \cdot (p-\bar q))\,((\bar p- q)\cdot q)}{u - M_H^2}\,,
\nonumber\\
&& f^2\,T^{(2)}_{ab}(s,t,u)= 2\, c_0 \,B_0 \,\Big( 2\,m\,C^{(\pi)}_0 + (m+m_s)\,C^{(K)}_0 \Big) + 
2\, c_1 \,B_0\, \Big( 2\,m\,C^{(\pi)}_1 + (m+m_s)\,C^{(K)}_1 \Big) 
\nonumber\\
&& \qquad \qquad +\,4\,( \bar q\cdot q)\,\Big(c_2\,C_2 + c_3\,C_3 \Big)+
\frac{(s-u)^2}{4\,M^2}\,\Big( c_4\, C_{2} + c_5\,C_3 \Big) \,,
\nonumber\\
&&  f^2\,T^{(3)}_{ab}(s,t,u) = (s-u)\,\sum_Q m_Q^2 \,\frac{g_1}{M}\,C^{(1)}_{Q}\, -4\,(\bar q\cdot q)\,(s-u)\, \frac{g_2}{M}  \,C_{\rm WT} 
\nonumber\\
&& \qquad \qquad + \, 
 (s-u)^3\, \frac{g_3}{2\,M^3}\,C_{\rm WT}\,,
\label{res-tree-Q123}
\end{eqnarray}
with Clebsch coefficients $C_{WT},C_Q^{(1)}, C^{(s)}_{H}, C^{(u)}_{H}, C^{(\pi)}_n, C^{(K)}_n$ and $C_n$ detailed already in \cite{Kolomeitsev:2003ac, 
Hofmann:2003je,Guo:2018poy} and the conventional Mandelstam variables $s, t, u$ of two-body scattering. With $q_\nu $ and $\bar q_\nu$ we denote the initial and final 4-momenta of the Goldstone bosons. The Mandelstam variables are $s= w^2=(p+q)^2$, $u= (p-\bar q)^2$ and $t= (\bar q -q)^2$ in our work. The s- and u-channel exchange processes in (\ref{res-tree-Q123}) involve the $J^P= 1^-$ charmed mesons, i.e. the sums run over $H \in \{D^*, D^*_s \} =[1^-]$. The indices $b$ and $a$ specify the initial and final flavor channels of the chosen process. 

We note that the particularly useful combination of Clebsch coefficients
\begin{eqnarray}
C_{u-ch}  = \sum_{H} C^{(u)}_{H} = C_2-2\,C_3-C_{WT} \,,
\label{def-Cuch}
\end{eqnarray}
was introduced in \cite{Hofmann:2003je} for applications in which the mass difference of the $D^*$ and $D^*_s$, or also the difference of the $D$ and $D_s$ masses, 
can be neglected. Depending on the context such Clebsch coefficients are applied also for processes which involve the scattering of the $J^P = 1^-$ charmed states. This is possible since the corresponding interaction vertices have identical flavor structures in the two sectors. 
 
This can be illustrated at the leading orders for the scattering processes involving the $1^-$ charmed mesons.  The scattering amplitudes are characterized by six invariant amplitudes $G_n$, most economically in the following choice
\begin{eqnarray}
&& T^{(0^- \to 1 ^-)}_{ab}(s,t,u) = G_0(s,t,u)\,\epsilon^{ \alpha \beta \mu \nu }\,\bar q_\alpha \,q_\beta  \,\epsilon^\dagger_\mu( \bar p , \bar \lambda )\,p_\nu\, ,
\nonumber\\
&& T^{(1^- \to 1^-)}_{ab}(s,t,u) =  \epsilon^\dagger_\mu( \bar p , \bar \lambda ) \Big[ G_1(s,t,u)\, g^{\mu \nu}   + G_2(s,t,u)  \,\bar q^\mu\,q^\nu + 
G_3 (s,t,u)\, \bar q^\mu \, (q -\bar q)^\nu  
\nonumber\\
&& \qquad \qquad \;    +\, 
G_4 (s,t,u)\,(\bar q -q)^\mu \,q^\nu + 
G_5(s,t,u)\, (\bar q-q)^\mu\,(q-\bar q)^\nu  \,\Big]\,\epsilon_\nu( p , \lambda)\, ,
\nonumber\\
&& \epsilon_{\alpha \beta}(p, \lambda) =\frac{i}{\sqrt{p^2}}\,\Big( p_\alpha\,\epsilon_\beta(p,\lambda) -  p_\beta\,\epsilon_\alpha(p,\lambda)\Big) \,,
\label{res-WT-spin-1}
\end{eqnarray}
where we use $\bar p_\mu $ and $\bar \lambda $ for the momentum and polarization of the produced $D^*$ meson. The wave function, $\epsilon_{\alpha \beta}(p, \lambda)$ of a vector meson as interpolated by an antisymmetric tensor field, is written in terms of the more conventional wave function $\epsilon_\alpha(p, \lambda)$ of a spin-one particle in the vector-field representation. We identify the leading orders tree-level terms with
\allowdisplaybreaks[1]
\begin{eqnarray} 
&& f^2\,G_0(s,t,u) = \frac{g_P\,\tilde g_P}{2}\, \sum_{H \in [1^-]} \,\Big( C^{(u)}_H\,\frac{M_a +u/M_a}{u-M_H^2}-C^{(s)}_H\,\frac{M_a+s/M_a}{s-M_H^2} \Big) -  \frac{2\,c_6}{M_a}\,C_{WT}
\nonumber\\
&& \qquad \qquad \qquad\!  - \,2\,\frac{s-u}{M\,M_a} \,
\Big(  ( g_4 + g_5) \,C_2 -  2\,g_4\,C_3\Big)
+ {\mathcal O} (Q^2)  \,,
\nonumber\\
&& f^2\,G_2(s,t,u)  = -g_P^2 \sum_{H \in [0^- ]} \,\Big(C^{(s)}_H\,\frac{M_a\,M_b}{s-M_H^2} +C^{(u)}_H\,\frac{M_a\,M_b}{u-M_H^2} \Big) 
\nonumber\\
&& \qquad \qquad \qquad\!+ \,\tilde g_P^2 \sum_{H \in [1^-] } \Big( C^{(s)}_H\,\frac{M_a^2+M_b^2}{s-M_H^2}  +C^{(u)}_H\,\frac{M_a^2+M_b^2}{u-M_H^2}  \Big)\,\frac{M_a^2+M_b^2}{4\,M_a\,M_b}
+ {\mathcal O} (Q^2)\,,
\nonumber\\
&& f^2\,G_{3}(s,t,u)  = C_{WT}\,\frac{M_b}{2\,M_a}+g_P^2 \sum_{H \in [ 0^-] } \,C^{(u)}_H\,\frac{M_a\,M_b}{u-M_H^2}  
\nonumber\\
&& \qquad \qquad \qquad\! - \,\tilde g_P^2 \sum_{H \in [ 1^-] } \Big( C^{(s)}_H\,\frac{M_a^2+M_b^2 + (s-u)/2}{s-M_H^2}  +C^{(u)}_H\,\frac{(s -u)/2}{u-M_H^2}  \Big)\,\frac{M_a^2+M_b^2}{4\,M_a\,M_b}
\nonumber\\
&& \qquad \qquad \qquad  \!
 - \,2\, \frac{s-u}{M_a\,M_b/\tilde M} \,\Big(  (\tilde g_4 + \tilde g_5) \,C_2 - 2\,\tilde g_4\,C_3\Big)- 2\,\tilde c_6\,C_{WT }\,\frac{M_b}{M_a} + {\mathcal O} (Q^2) \,,
\nonumber\\
&&  f^2\,G_{4}(s,t,u)  = C_{WT}\,\frac{M_a}{2\,M_b} +g_P^2 \sum_{H \in [ 0^-] } \,C^{(u)}_H\,\frac{M_a\,M_b}{u-M_H^2}  - 2\,\tilde c_6 \,C_{WT }\, \frac{M_a}{M_b}
\nonumber\\
&& \qquad \qquad \qquad\! - \,\tilde g_P^2 \sum_{H \in [ 1^- ] } \Big( C^{(s)}_H\,\frac{M_a^2+M_b^2 + (s-u)/2}{s-M_H^2}  +C^{(u)}_H\,\frac{(s -u)/2}{u-M_H^2}  \Big)\,\frac{M_a^2+M_b^2}{4\,M_a\,M_b}
\nonumber\\
&& \qquad \qquad \qquad  \!
 - \,2\, \frac{s-u}{M_a\,M_b/\tilde M} \,\Big(  (\tilde g_4 + \tilde g_5) \,C_2 - 2\,\tilde g_4\,C_3\Big)+ {\mathcal O} (Q^2) \,, 
\nonumber\\
&& f^2\,G_{5}(s,t,u)  = -C_{WT}\,\frac{M_{a}^2 +M_b^2}{4\,M_a\,M_b} -g_P^2 \sum_{H \in [ 0^-] } \,C^{(u)}_H\,\frac{M_a\,M_b}{u-M_H^2}   + \tilde c_6\, C_{WT }\,\frac{M_{a}^2+M_b^2+s-u}{M_a\,M_b}  
\nonumber\\
&& \qquad \qquad \qquad\! + \,\tilde g_P^2 \sum_{H \in [1^- ] } \Big( C^{(s)}_H\,\frac{M_a^2+M_b^2 +s-u}{s-M_H^2}  \Big)\,\frac{M_a^2+M_b^2}{4\,M_a\,M_b}
\nonumber\\
&& \qquad \qquad \qquad  
 + \,2\, \frac{s-u}{M_a\,M_b/\tilde M} \,\Big(  (\tilde g_4 + \tilde g_5) \,C_2 - 2\,\tilde g_4\,C_3\Big)+ {\mathcal O} (Q^2) \,,
\label{res-T11}
\end{eqnarray}
where we use the notation $Q^n$ to specify the chiral order $n$ of a given term. While the $\tilde g_4$ and $\tilde g_5$ contribute to the $ 0^-\,1^- \to 0^-\,1 ^-$ processes, the heavy-quark symmetry related $g_4$ and $g_5$ contribute to the production processes $ 0^-\,0^- \to 0^-\,1 ^-$. In the heavy-quark mass limit it holds $\tilde g_n = g_n$, in particular for $n=4,5$.

It remains to specify the invariant amplitude $G_1(s,t,u)$ in 
$T^{(1^- \to 1^-)}_{ab}(s,t,u) $. Owing to the heavy-quark symmetry its form can be inferred from its spin-zero partner reaction $T_{ab}(s,t,u) $ 
at least in the heavy-quark mass limit. Indeed we find for our tree-level the expression
\begin{eqnarray}
&&G_1(s,t,u) =\frac{ \tilde g_P^2}{f^2} \sum_{H \in [1^- ]} \, C^{(s)}_H\,\frac{1}{s-M_H^2} \,\frac{M_a^2+M_b^2}{4\,M_a\,M_b} \,\Big\{ \frac{1}{8}\,(s-u)^2 + \frac{(s-u)^3}{32\,(M_a^2+M_b^2)}+ \frac{s-u}{4}\,t
\nonumber\\
&& \qquad \quad-\frac{M_a^2+M_b^2}{2}\,\Big(m_a^2+m_b^2-t  \Big)
-\Big( \frac{\tilde M^2}{M^2} -1\Big)\,\frac{(s-u)^2}{32}
\Big\}
\nonumber\\
&& \quad \! \!+\,\frac{ \tilde g_P^2 }{f^2}\sum_{H \in [1^-] } \, C^{(u)}_H\,\frac{1}{u-M_H^2} \,\frac{M_a^2+M_b^2}{4\,M_a\,M_b} \,\Big\{ \frac{1}{8}\,(s-u)^2 - \frac{(s-u)^3}{32\,(M_a^2+M_b^2)} - \frac{s-u}{4}\,t
\nonumber\\
&& \qquad \quad -\,\frac{M_a^2+M_b^2}{2}\,\Big(m_a^2+m_b^2- t\Big) 
-\Big(  \frac{\tilde M^2}{M^2} -1\Big)\,\frac{(s-u)^2}{32}
\Big\}
\nonumber\\
&& \quad \!\!- \,
\frac{M_{a}^2+M_b^2}{2\,M_a\,M_b}\,\Big\{
T^{(1)}_{ab}(s,t,u) \Big|_{g_P \to 0} +
T^{(2)}_{ab}(s,t,u)\Big|_{c_n \to \tilde c_n} + T^{(3)}_{ab}(s,t,u)\Big|_{g_n \to \tilde g_n}  \Big\}+ {\mathcal O} (Q^4) \,,
\end{eqnarray}
properly truncated at chiral order four. This is so since contributions from the other amplitudes $G_{2,3,4,5}$ are suppressed by two orders in the chiral expansion. Note the presence of the small 4-momenta $\bar q_\mu$ or $q_\mu$ in (\ref{res-WT-spin-1}).  It is evident that analogous relations hold for the loop expressions, as to be derived in our current work. Therefore from now on we focus on the reactions with spin-zero charmed mesons in the initial and final states.


\clearpage

\section{Scattering with Tadpole and Bubble diagrams}

\begin{figure}[t]
\center{
\includegraphics[keepaspectratio,width=0.97\textwidth]{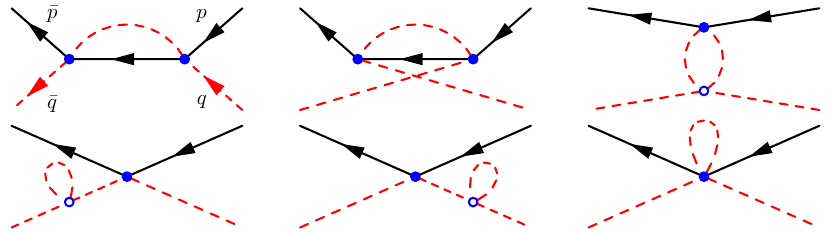} }
\vskip-0.2cm
\caption{\label{fig:1} Dashed lines stand for pion, kaon or eta mesons and solid lines for charmed mesons with $J^P = 0^-$. The vertices are from \cite{Lutz:2022enz}.}
\end{figure}

We discuss one-loop contributions to the two-body scattering amplitudes. At chiral order 3 and 4 there are various types of diagrams to be considered. All one-loop diagrams that contribute at $g_P =0$ have been evaluated in our previous work \cite{Lutz:2022enz}. Such tadpole and bubble loop contributions are recalled in Fig.\ \ref{fig:1}
at order 3 involving leading order vertices only. Corresponding diagrams at order 4 involve subleading order vertices instead. Quite explicit expressions are documented in \cite{Lutz:2022enz}.  

\begin{figure}[t]
\center{
\includegraphics[keepaspectratio,width=0.99\textwidth]{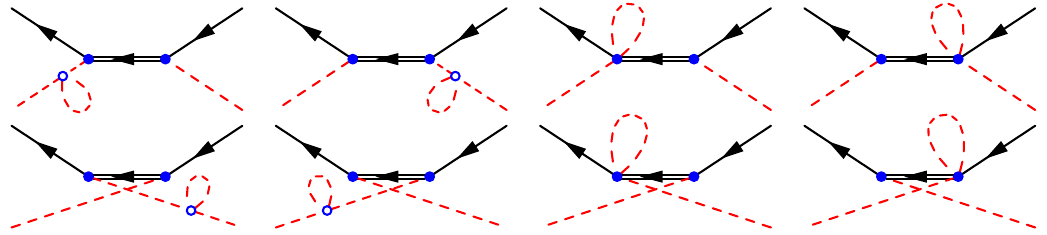} }
\vskip-0.2cm
\caption{\label{fig:2} Dashed lines stand for pion, kaon or eta mesons and solid and double-solid lines for charmed mesons with $J^P = 0^-$ and $J^P = 1^-$ respectively. The vertices are from \cite{Lutz:2022enz}.}
\end{figure}

\begin{figure}[t]
\includegraphics[keepaspectratio,width=0.59\textwidth]{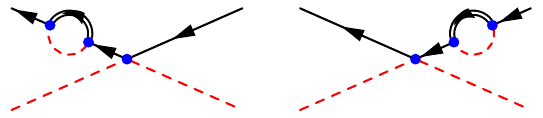}
\vskip-0.2cm
\caption{\label{fig:3} Dashed lines stand for pion, kaon or eta mesons and solid and double-solid lines for charmed mesons with $J^P = 0^-$ and $J^P = 1^-$ respectively. The vertices are from \cite{Lutz:2022enz}.}
\end{figure}

\begin{figure}[t]
\center{
\includegraphics[keepaspectratio,width=0.99\textwidth]{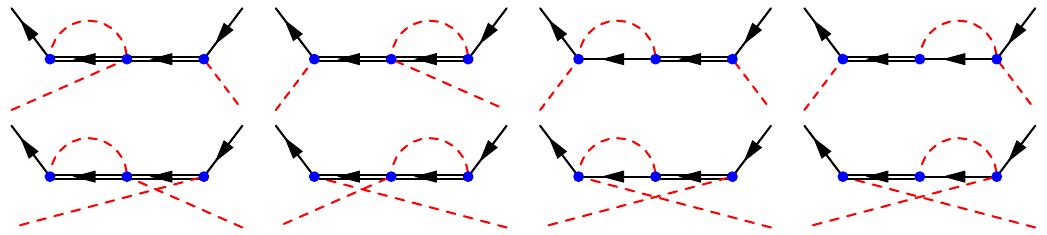}}
\vskip-0.2cm
\caption{\label{fig:4} Dashed lines stand for pion, kaon or eta mesons and solid and double-solid lines for charmed mesons with $J^P = 0^-$ and $J^P = 1^-$ respectively. The vertices are from \cite{Lutz:2022enz}.}
\end{figure}

\begin{figure}[t]
\includegraphics[keepaspectratio,width=0.59\textwidth]{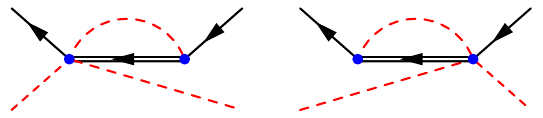}
\vskip-0.2cm
\caption{\label{fig:5} Dashed lines stand for pion, kaon or eta mesons and solid lines and double-solid lines for charmed mesons with $J^P = 0^-$ and $J^P = 1^-$ respectively. The vertices are from \cite{Lutz:2022enz}.}
\end{figure}

An additional set of tadpole, bubble, triangle and box loop diagrams is  proportional to $g_P^2$  has not been documented systematically before.  In  Fig.\ \ref{fig:2}-\ref{fig:7} our target diagrams are shown for the case that initial and final mesons carry $J^P= 0^-$ quantum numbers.  Corresponding diagrams can be drawn for the case in which one or both external lines 
signal a charmed meson with  $J^P= 1^-$. From the form of such diagrams it follows that in the formal limit of a very large mass of the $J^P=1^-$ mesons such contributions may be viewed as a renormalization of tadpole and bubble loop contributions. That was the rationale behind our previous more  phenomenological work, despite the fact that the heavy-quark spin symmetry predicts the mass degeneracy of the $0^-$ and $1^-$ states in the limit of an infinite charm quark mass. Clearly, it is desirable to have a closer look into such diagrams.

We use the conventional Mandelstam variables $s,t$ and $u$ of two-body scattering. The indices $a$ and $b$ specify the final and initial flavor channels of the chosen process.  
The loop functions depend on not only the internal masses, $m_Q, M_H, M_L, M_R$,  but also on external masses
\begin{eqnarray}
 m_a^2= \bar q^2\,,\qquad m_b^2= q^2\,,\qquad M_a^2= \bar p^2\,,\qquad M_b^2=p^2\,,
 \label{def-external-masses}
\end{eqnarray}
where we use small $m$'s for Goldstone boson masses and big $M$'s for the masses for the $0^-$ and $1^-$ charmed mesons. The pairs of initial and final four momenta are $(q_\mu, p_\mu)$ and  $(\bar q_\mu, \bar p_\mu)$ respectively. In turn we may write $ w = q+ p  = \bar q+ \bar p$ with $s = w^2$.

\allowdisplaybreaks[1]
We first consider the tadpole-type diagrams  in Fig.\ \ref{fig:2}
\begin{eqnarray}
&& f^4\,T_{ab}^{\rm tadpole}(s,u) = g_P^2\,\sum_{H \in [1^-]} \,\frac{s\,(\bar q\cdot q) -(\bar q \cdot w)\,(w\cdot q)}{s- M_H^2}\,\Big( \sum_Q C^{(s)}_{QH}\,\bar I_Q 
\nonumber\\
&& \qquad \qquad \qquad \qquad \qquad \qquad  +\,
 C^{(s)}_H\,f^2 \,(Z^{(1)}_a + Z^{(1)}_b-2)/2 \Big)
\nonumber\\
&& \qquad \qquad \qquad \quad   +\,g_P^2\,\sum_{H \in [1^-]} \,\frac{u\,(\bar q\cdot q) -(\bar q \cdot (p-\bar q))\,((\bar p- q)\cdot q)}{u - M_H^2}\,\Big(\sum_Q C^{(u)}_{QH}\,\bar I_Q
\nonumber\\
&& \qquad \qquad \qquad \qquad \qquad \qquad 
+ \,C^{(u)}_H\,f^2\,(Z^{(1)}_a +Z^{(1)}_b -2)/2 \Big)\,,
\nonumber\\
&& f^2\,(Z^{(1)}_\pi - 1 )=    - 8\,L_4 \,(m_\pi^2 +2\,m_K^2  ) - 8\,L_5\,m_\pi^2 + \frac{2}{3}\,\bar I_\pi + \frac{1}{3}\,\bar I_K\,,
\nonumber\\
&& f^2\,(Z^{(1)}_K -1) =  - 12\,L_4 \,(m_\pi^2 +m_\eta^2 ) - 8\,L_5\,m_K^2 + \frac{1}{4}\,\big( \bar I_\pi + \bar I_\eta + 2\,\bar I_K \big)\,,
\nonumber\\
&& f^2\,(Z^{(1)}_\eta \,\, - 1) =   - 24\,L_4 \,(2\,m_K^2 - m_\eta^2 ) - 8\,L_5\,m^2_\eta +\bar I_K \,,
\nonumber\\
&&  \mu^2\, \frac{\text{d}}{\text{d}\mu^2}\, L_4 = - \frac{1}{256\, \pi^2} \,,  \qquad \quad
 \mu^2 \,\frac{\text{d}}{\text{d}\mu^2}\, L_5 = -\frac{3}{256\,\pi^2} \,,
 \qquad \quad   \bar I_Q = \frac{m_Q^2}{(4 \pi )^ 2} \, \log \frac{m_Q^2}{ \mu^2}\,,
\label{ref-wave-function}
\end{eqnarray}
with the wave-function factors of the Goldstone bosons as written in \cite{Lutz:2022enz} by using the LEC $L_4$ and $L_5$ of Gasser and Leutwyler \cite{Gasser:1984gg}. 
While the Clebsch $C^{(s)}_H$ and $C^{(u)}_H$ were given previously, 
the $C^{(s)}_{QH}$ and $C^{(u)}_{QH}$ can easily be expressed in terms of the latter. 
To avoid a proliferation of our notations, $Q$ is used  as a placeholder index for a Goldstone boson  field $\pi, K, \eta$ in (\ref{ref-wave-function}) but also in $Q^n$ together with the chiral order $n$ of a given term (see e.g. (\ref{res-T11})). For the tadpole, $\bar I_Q$,  a conventional $\overline{MS}$ subtraction scheme is used with the renormalization scale $\mu$ of dimensional regularization. 

We aim at a decomposition of the scattering amplitude 
\begin{eqnarray}
&& T_{ab}(s,u) = \sum_{H \in [1^-]} \frac{ (\bar q \cdot q)-(\bar q \cdot w)\, (w \cdot q)/M_H^2 }{s- M_H^2}\,G^{(s)}_{H,ab} 
\nonumber\\
&& \qquad \qquad 
+\, \sum_{H \in [1^-]} \frac{(\bar q\cdot q) -(\bar q \cdot (p-\bar q))\,((\bar p- q)\cdot q)/M_H^2}{u - M_H^2}\,G^{(u)}_{H,ab} + B_{ab} (s,u) \,,
\end{eqnarray}
into s- and u-channel pole terms with on-shell mass $M_H$ and a smooth background term $B_{ab}(s)$. By construction the s-channel pole term  contributes to the $J^P=1^{-}$ partial-wave amplitude only. The u-channel pole is included such that the sum of the two pole terms is compatible with constraints from crossing symmetry. The pole mass, $M_H$ and the residua $G^{(s)}_{H,ab}$ and $G^{(u)}_{H,ab}$ as well as the background term $B_{ab}(s,u)$ receive corrections from loop effects. Given our approximation strategy we will use the physical on-shell mass for $M_H$ and 
\begin{eqnarray}
 G^{(s)}_{H} = g_P^2\,C_H^{(s)}/f^2 \,,\qquad \qquad \qquad \qquad 
 G^{(u)}_{H} = g_P^2\,C_H^{(u)}/f^2 \,,
\end{eqnarray}
from (\ref{res-tree-Q123}). The value $g_P$ may be adjusted as  to recover the empirical decay width of the $D^*\to \pi \,D$ meson. While it would be desirable to 
refine such a scheme, at this stage there is insufficient information available to consider flavor breaking or quark-mass dependence effects in $G^{(s)}_{H,ab}$. It appears impossible to determine corresponding LEC that contribute to $G^{(s)}_{H, ab}$. Therefore we will focus on the 
loop effects in the background term. For the tadpole contribution (\ref{ref-wave-function}) we find 
\begin{eqnarray}
&&f^4\, B^{\rm tadpole}_{ab} (s,u)  =  
g_P^2\,\sum_{H \in [1^-]} \,\Big [ (\bar q\cdot q) - \frac{(s-u)^2}{8\,M^2_{ab}} \Big]\,\Big\{ \sum_Q \big( C^{(s)}_{QH}+ C^{(u)}_{QH}\big)\,\bar I_Q 
\nonumber\\
&& \qquad \qquad \qquad \qquad \;\;+ \,\big(C^{(s)}_H+C^{(u)}_H \big)\,f^2 \,(Z^{(1)}_a + Z^{(1)}_b-2)/2 \Big\} + {\mathcal O} \left( Q^5\right) 
\nonumber\\
&& \qquad= \,
g_P^2\,\sum_{H \in [1^-]} \,\Big [ (\bar q\cdot q) - \frac{(s-u)^2}{8\,M^2_{ab}} \Big]\,
\sum_Q \big( L_4\,C^{(4)}_{QH}+ L_5\,C^{(5)}_{QH}\big)\,m^2_Q   +{\mathcal O} \left( Q^5\right)
\,,
\label{res-B-tadpole}
\end{eqnarray}
where we note that the background term is of chiral order four. This is in contrast to its corresponding contribution to $G^{(s)}_{H}$ and $G^{(u)}_{H}$, which are of chiral order two. In the last line (\ref{res-B-tadpole}) of our rewrite we  observed that the tadpoles in the first line cancel identically with the tadpoles in the wave function terms from (\ref{ref-wave-function}).

We turn to the bubble-type contributions, where we start with the wave-function term 
\begin{eqnarray}
&&  T_{ab}^{\rm bubble}(s,u) = \frac{1}{2}\,\big( T^{(1)}_{ab}(s,u) +  T^{(2)}_{ab}(s,u)\big) \,(Z^{(2)}_a +Z^{(2)}_b- 2)+ {\mathcal O}\left(Q^5 \right)\,,
\label{res-wave-function-2}
\nonumber\\ \nonumber\\
&& Z^{(2)}
 _D = 1  + 4\,B_0\,\Big( (2\,\zeta_0-\zeta_1)\, (m_s+2\,m) +m\,\zeta_1 \Big)
\nonumber\\
&& \quad \!
- \, \frac{g_P^2}{16\,f^2}\,\Big\{ 12\, \Big( \bar I_\pi + 2\,m_\pi^2\,\bar I_{D^*}/M_{D^*}^2-
2\,(m_\pi^2  +r \,M_D^2  )\,\bar I_{\pi D^*}^{(D)}
-\big(4\,m_\pi^2-r^2 \,M_D^2\big) \,Z_{\pi D^*}^{(D)} \Big)
\nonumber\\
&& \qquad \qquad +\, 8\,\Big(\bar I_K + 2\,m_K^2\,\bar I_{D^*_s}/M_{D^*_s}^2 -
2\,(m_K^2  +r \,M_D^2  )\,\bar I_{K D_s^*}^{(D)}
 -\big(4\,m_K^2-r^2 M_D^2\big)\,Z_{\pi D^*}^{(D)}\Big) 
\nonumber\\
&&  \qquad \qquad
+ \, \frac{4}{3}\,\Big(\bar I_\eta + 2\,m_\eta^2\,\bar I_{D^*}/M_{D^*}^2-  2\,(m_\eta ^2  +r \,M_D^2  )\,\bar I_{\eta D^*}^{(D)}
 -\big(4\,m_\eta^2-r^2 M_D^2\big)\,Z_{\pi D^*}^{(D)}\Big)\Big\}
\nonumber\\
&& \quad \!
 + \,{\mathcal O} \left( Q^3\right)\,,
\nonumber\\
&& Z^{(2)}_{D_s} = 1  + 4\,B_0\,\Big( (2\,\zeta_0- \zeta_1)\, (m_s+2\,m) +m_s\,\zeta_1 \Big)
\nonumber\\
&& \quad \!  -\, \frac{g_P^2}{16\,f^2}\,\Big\{   16\,\Big(\bar I_K + 2\,m_K^2\,\bar I_{D^*}/M_{D^*}^2 - 2\,(m_K^2  +r \,M_{D_s}^2  )\,\bar I_{K D^*}^{(D_s)}
 -\big(4\,m_K^2-r^2 M_{D_s}^2\big) \,Z_{K D^*}^{(D_s)}\Big)
\nonumber\\
&& \qquad \qquad   +\, \frac{16}{3}\,\Big(\bar I_\eta + 2\,m_\eta^2\,\bar I_{D^*_s}/M_{D^*_s}^2-  2\,(m_\eta ^2  +r \,M_{D_s}^2  )\,\bar I_{\eta D_s^*}^{(D_s)}
 -\big(4\,m_\eta^2-r^2 M_{D_s}^2\big)\,Z_{\eta D^*}^{(D_s)}\Big) \Big\}
\nonumber\\
&& \quad \!+ \,{\mathcal O} \left( Q^3\right)\,,
\nonumber\\
&& Z^{(D)}_{QH} = M_{D}^2\,\frac{\partial\bar I_{Q H}( M_{D}^2 )}{\partial  M_{D}^2 } \,, \qquad \qquad \quad \;
\bar I^{(D)}_{Q H} = \bar I_{Q H}( M_{D}^2 ) \,,
\nonumber\\
&&   \mu^2\, \frac{\text{d}}{\text{d}\mu^2}\, \zeta_0 = - \frac{13}{384\, \pi^2\,f^2}\,g_P^2 \,,  \qquad \qquad
 \mu^2 \,\frac{\text{d}}{\text{d}\mu^2}\, \zeta_1 =- \frac{5}{128\,\pi^2\,f^2} \,g_P^2\,,
\label{res-zeta-running}
\end{eqnarray}
from Fig.\ \ref{fig:3}. It involves the first and second order tree-level expressions $T^{(1)}_{ab}(s,u) $ and $T^{(2)}_{ab}(s,u) $  as recalled in (\ref{res-tree-Q123}) and the LEC $\zeta_0$ and $\zeta_1$ from the chiral Lagrangian. Our result involves a scalar bubble loop function $\bar I_{QR}(M_H^2)$, with its renormalized form given in (\ref{def-bubble}). 
We find that the wave functions,  $ Z^{(2)}_H$, of the heavy fields do not depend on the renormalization scale $\mu$, if we use the summed expressions 
\begin{eqnarray}
&& B_0 \,(2\,m + m_s) \to \frac{3}{16}\,\big(3\,m_\pi^2+4\,m_K^2+ m_\eta^2 \big)\,, \qquad \quad 
B_0\,m \to \frac{1}{80}\,\big(39\,m_\pi^2+ 4\,m_K^2 -3\,m_\eta^2 \big)\,, 
\nonumber\\
&& B_0 \,m_s \to \frac{1}{80}\,\big( -33\,m_\pi^2+ 52\,m_K^2 +21\,m_\eta^2\big) \,,
\end{eqnarray}
in (\ref{res-zeta-running}). 
This is contrasted by the fact that the wave functions, $Z^{(1)}_Q$, of the light fields do depend on $\mu$. We note that an additional subtraction in (\ref{res-zeta-running}) may be useful as to arrive at the wave function factors $Z^{(2)}_D$ and $Z^{(2)}_{D_s}$ to approach one in the chiral limit. 

The evaluation of the loop functions in Fig.\ \ref{fig:3}-\ref{fig:5} (see also
(\ref{res-wave-function-2}), (\ref{def-J-bubble:01})  and (\ref{def-J-bubble})) is straightforward, even if one insists on the use of on-shell meson masses as is highly advisable 
for coupled-channel systems. In previous works \cite{Lutz:2018cqo,Bavontaweepanya:2018yds,Lutz:2020dfi,Sauerwein:2021jxb}
we developed a novel scheme in application of the Passarino--Veltman decomposition scheme \cite{Passarino:1978jh}. In an initial step the one-loop bubble contributions  can be expressed in terms of scalar loop functions
\begin{eqnarray}
&& I_Q \;=\int \frac{d^d l}{(2\pi)^d}\frac{ i\,\mu^{4-d} }{l^2- m_Q^2}\,, \qquad \qquad \qquad  I_H \;=\int \frac{d^d l}{(2\pi)^d}\frac{ i\,\mu^{4-d} }{l^2- M_H^2}\,,
\nonumber\\
&& I_{QH}(w^2) =  \int \frac{d^d l}{(2\pi)^d}\frac{ -i\,\mu^{4-d} }{l^2- m_Q^2}
 \frac{1}{(l+w)^2-M_H^2 } \,,
 \label{def-scalar-bubble}
\end{eqnarray}
introduced in dimensional regularization. 
While we keep the scalar basis functions un-expanded, as to keep their proper causal structure, the kinematical coefficients are expanded in application of power-counting rules. So far we encounter the renormalized tadpole $\bar I_Q$  and bubble $\bar I_{QH}(s)$ 
functions in  (\ref{res-zeta-running}) only, with
\begin{eqnarray}
&& \bar I_{QH}(s) = I_{QH}(s) - \frac{1 - \gamma}{16\,\pi^2} + \frac{I_H}{M_H^2}\,, \qquad \qquad  
\gamma=- \frac{\tilde M^2-M^2}{M^2}\,\log \frac{\tilde M^2-M^2}{\tilde M^2}\,,
\nonumber\\ \nonumber\\
&& \bar I_{QH}(s) = \frac{1}{16\,\pi^2}
\left\{ \gamma -\frac{1}{2} \left( 1+ \frac{m_Q^2-M_H^2}{s} \right) \,\log \left( \frac{m_Q^2}{M_H^2}\right)
\right.
\nonumber\\
&& \;\quad \;\;+\left.
\frac{p_{Q H}}{\sqrt{s}}\,
\left( \log \left(1-\frac{s-2\,p_{Q H}\,\sqrt{s}}{m_Q^2+M_H^2} \right)
-\log \left(1-\frac{s+2\,p_{Q H}\,\sqrt{s}}{m_Q^2+M_H^2} \right)\right)
\right\} \, ,
\nonumber\\
&& p_{Q H}^2 =
\frac{s}{4}-\frac{m_Q^2+M_H^2}{2}+\frac{(m_Q^2-M_H^2)^2}{4\,s}  \,,
\label{def-bubble}
\end{eqnarray} 
where with $M$ and $\tilde M$ we denote  
the chiral limit values of the charmed meson masses with $J^P=0^-$ and $J^P=1^-$ respectively. Additional contributions from scalar tadpole integrals involving the heavy fields are dropped systematically with $\bar I_L=\bar I_R = \bar I_H \to  0$, at least if they occur in a power-counting violating context. By construction it holds 
$\bar I_{QH}(s) \sim Q$ as expected from dimensional counting rules. In a second step we apply the power-counting scheme \cite{Lutz:2022enz} as introduced in terms of on-shell hadron masses
\begin{eqnarray}
&& s+ t +u = M_{ab}^2+ m_{ab}^2\,, \qquad \qquad \qquad \qquad \!m_{ab}^2=m_{ab}^2 \sim m_a^2\sim m_b^2 \sim Q^2\,,
\nonumber\\
&& M_{ab}^2=M^2_a + M^2_b \sim s + u \sim Q^0  \ ,  \qquad  \qquad \frac{s- u}{2\,(M^2_a+M^2_b)^{1/2}} \sim m_{\rm Goldstone} \sim Q^1\,,
\nonumber\\
&& M^2_L - M^2_{ab}/2 = \delta_L + r\, M^2_{ab}/2 \,, \qquad \qquad \quad \; r= \frac{\tilde M^2}{M^2} - 1\,, 
\nonumber\\
&& M^2_R- M^2_{ab}/2=\delta_R + r\, M^2_{ab}/2 \,,
\nonumber\\
&& m_{\rm Quark} \sim t \sim s+u-M^2_a-M^2_b \sim M_a^2-M_b^2 \sim \delta_L \sim \delta_R\sim Q^2\,,
 \label{def-power-counting}
\end{eqnarray}
where we note that $L,R \in [1^-]$ here. Upon a chiral expansion the chiral power of any of the $\delta_{L,R} \sim m_{u,d,s}\sim Q^2$ is confirmed, where we recall that $M_a$ and $M_b$ give the charm meson masses of the final and initial $J^P = 0^-$ fields.  The merit of such a scheme is that our expansion can be set up in a two-step procedure. 
Initially we do not make any assumption on the size of the ratio $r =( \tilde M^2 - M^2)/M^2$ in (\ref{def-power-counting}). 
An application of the counting rules (\ref{def-power-counting}) generates expressions that probe  rational functions of that $r$, as is illustrated in \cite{Lutz:2020dfi,Sauerwein:2021jxb}. There are at least three relevant possibilities implied by either $r \sim Q^2$ or $r \sim Q$ or $r \sim Q^0$. In the first two cases we make contact with the traditional simultaneous expansion in the small up, down and strange  quark masses and in the small inverse of a large charm quark mass. In the third case we may integrate out the $1^-$ fields in terms of the formal request $\tilde M \gg  M$. Most economic would be the case with $r \sim Q^2$ since it would imply $M_{L,R}^2-M_{a,b}^2 \sim Q^2$. Indeed using previous values from  \cite{Lutz:2022enz} for $\tilde M$ and $M$ we obtain the estimate $r \simeq  0.16 $, which may sufficiently support such an assignment parametrically.

Nevertheless, we argue that it is advantageous to keep the size of the ratio $r$ open at first. This entails us to set up the expansion in a manner that 
permits to integrate out the $1^-$ fields efficiently. In order to connect to the chiral domain with $m_\pi \ll \tilde M - M$, we must assume $r \sim Q^0$ at least. Consistency of our results in that chiral domain will demand a further 
set of subtraction terms, as to eliminate power-counting violating contributions.

We turn to the bubble-type diagrams in Fig.\ \ref{fig:4} with
\begin{eqnarray}
&& f^4\, T_{ab}^{\rm bubble}(s,u) =  
 g_P^2 \,\sum_{L,Q H}\, \Big\{ C^{(s)}_{L,QH}\, J^{(s)}_{L,QH}(s,u)  +\,C^{(u)}_{L,QH}\, J^{(u)}_{L,QH}(s,u)  \Big\}
\nonumber\\
&& \qquad \qquad \qquad \quad  +\, g_P^2 \,\sum_{QH,R}\, \Big\{ C^{(s)}_{QH,R}\, J^{(s)}_{QH,R}(s,u)  +\,C^{(u)}_{QH, R}\, J^{(s)}_{QH,R}(s,u)  \Big\} \,,
\nonumber\\ \nonumber\\ 
&& J^{(u)}_{L,QH}(s,u) = J^{(s)}_{L,QH}(u,s)\,,\qquad \qquad  
J^{(u)}_{QH,R}(s,u) = J^{(s)}_{QH,R}(u,s)\,,\qquad \qquad  
\end{eqnarray}
where  $H \in [1^-]$ and $L,R \in [0]$ or $L,R \in[1^-] $. The loops with $L,R \in [0^-]$ and with $L,R \in [1^-]$ are 
\begin{eqnarray}
&&  J^{(s)}_{L \in[0^- ],QH}(s,u)= \int\frac{d^d l}{(2\pi)^d} \,
\frac{-i\,\mu^{4-d}\,(l-\bar q +2\,w)\cdot (l -\bar q) }{l^2-m_Q^2}\, 
\frac{l_\alpha \,(l+w)_\beta}{(l+w)^2 -M_L^2 }  \,S^{\alpha\beta,\mu\nu}_H( w) \, q_{\mu }\,p_{\nu }\, ,
\nonumber\\
&&  J^{(s)}_{QH,R\in [0^-]}(s,u) =
\int\frac{d^d l}{(2\pi)^d} \,
\frac{-i\,\mu^{4-d}\,(l- q +2\,w)\cdot (l - q) }{l^2-m_Q^2}\, 
\frac{l_\alpha \,(l+w)_\beta}{(l+w)^2 -M_R^2}  \,S^{\alpha\beta,\mu\nu}_H( w) \, \bar q_{\mu }\,\bar p_{\nu }\, ,
\nonumber\\ \nonumber\\
&&  J^{(s)}_{L \in[1^-],QH}(s,u)= \int\frac{d^d l}{(2\pi)^d} \,
\frac{i\,\mu^{4-d} }{l^2-m_Q^2}\, 
l_{ \bar \alpha } \,\bar p_{\bar \beta} \,S^{\bar \alpha \bar \beta, \bar \mu \bar \nu }_L(l+\bar p) \,g_{\bar \nu \nu}\,\Big\{  (\bar p +l)_{\bar \mu} \, (l +\bar q)_\mu 
\nonumber\\
&& \qquad \qquad \qquad \qquad \qquad \qquad +\,(l + \bar q)_{\bar \mu} \, w_\mu\,
\Big\} \,S^{\mu\nu, \alpha\beta}_H( w) \, q_{\alpha}\,p_{\beta }\, ,
\nonumber\\
&&  J^{(s)}_{QH,R\in [1^-]}(s,u) = \int\frac{d^d l}{(2\pi)^d} \,
\frac{i\,\mu^{4-d} }{l^2-m_Q^2}\, 
\bar q_{ \bar \alpha } \,\bar p_{\bar \beta} \,S^{\bar \alpha \bar \beta, \bar \mu \bar \nu }_H(w) \,g_{\bar \nu \nu}\,\Big\{  w_{\bar \mu} \, (l + q)_\mu 
\nonumber\\
&& \qquad \qquad \qquad \qquad \qquad \qquad +\,(l +  q)_{\bar \mu} \, (l+p)_\mu\,
\Big\} \,S^{\mu\nu, \alpha\beta}_R( l + p) \, l_{\alpha}\,p_{\beta }\, ,
\nonumber\\
&& M_H^2\,S_H^{\alpha \beta, \mu \nu}(w)=   g^{\mu \alpha}\,g^{\nu \beta} - \frac{g^{\mu \alpha}\,w^\nu\,w^\beta-
g^{\mu \beta}\,w^\nu\,w^\alpha }{w^2-M_H^2 }- (\mu \leftrightarrow \nu) \,,
\label{def-J-bubble:01}
\end{eqnarray}
where both types show a pole at $s=M_H^2$. In the derivation of the bubble-loop functions of Fig.\ \ref{fig:4}
we need to separate their pole contribution first. The background and pole residuum terms can then be expanded according to the power-counting rules. We write 
\begin{eqnarray}
 J_ {QH}(s,u) = \frac{ (\bar q \cdot q) -(\bar q \cdot w)\, (w \cdot q)/M_H^2}{s- M_H^2}\,G_{QH} +f^4\,B_{QH} (s,u) \,,
\end{eqnarray}
and find the somewhat surprising expressions
\begin{eqnarray}
&&f^4\, B^{(s)}_{L \in[0^-],QH}(s,u)= \frac{s-u}{24}\,( \bar q\cdot q ) \, \Big\{ 2\,\bar I_Q -  \Big( 4\,r\,M^2_{ab} + s-u\Big)\,\bar I_{QL}(M_a^2) \Big\} +  {\mathcal O}\left( Q^6 \right) \,,
\nonumber\\
&&f^4\, B^{(s)}_{QH, R \in[0^-]}(s,u)= \frac{s-u}{24}\,( \bar q\cdot q )\,  \Big\{ 2\,\bar I_Q  -  \Big(4\, r\,M^2_{ab} + s-u\Big)\,\bar I_{QR}(M_b^2) \Big\} +  {\mathcal O}\left( Q^6 \right) \,,
\nonumber\\ \nonumber \\
&& f^4\,B^{(s)}_{L \in[1^-],QH}(s,u)= -\frac{M_b^2 - M_a^2}{4}\,(\bar q\cdot q) \, \Big\{\bar I_Q -  r\,M^2_{ab}\,\bar I_{QL}(M_a^2) \Big\} +  {\mathcal O}\left( Q^7 \right) \,,
\nonumber\\
&& f^4\,B^{(s)}_{QH, R \in[1^-]}(s,u)= - \frac{M_a^2 - M_b^2}{4}\,(\bar q\cdot q)\,  \Big\{\bar I_Q 
-  r\,M^2_{ab} \,\bar I_{QR}(M_b^2) \Big\} +  {\mathcal O}\left( Q^7 \right) \,,
\label{res-BLQH}
\end{eqnarray}
where we assumed $r \sim Q$ for simplicity. We return to such an assumption below in the context of the chiral expansion of triangle and box contributions. While dimensional counting rules suggest a leading contribution to $B_H \sim Q^3$ the specifics of such diagrams 
lead to terms of order $Q^5$ and higher only. 
Since we include terms up to order $Q^4$ only in this work, all such contributions can be dropped here. We note that the corresponding contributions to $G_H \sim Q^3$ are also excluded here.

The bubble-type diagrams in Fig.\ \ref{fig:5} are
\begin{eqnarray}
&& f^4\, T_{ab}^{\rm bubble}(s,u) = g_P^2 \,\sum_{n=1}^2 \,\sum_{Q}\, \Big\{ \sum_{R \in [1 ^-]}\,C^{(s-n)}_{Q, R}\, J^{(s-n)}_{Q,R}(s,u) + \sum_{L \in [1 ^-]}\, C^{(s-n)}_{L, Q}\, J^{(s-n)}_{L, Q}(s,u)
\nonumber\\
&& \qquad \qquad \qquad \qquad  +\,\sum_{R\in [1^-]}\,C^{(u-n)}_{Q, R}\, J^{(u-n)}_{Q,R}(s,u)  +\sum_{L \in [1^-]}\,C^{(u-n)}_{L, Q}\, J^{(u-n)}_{L, Q}(s,u) 
\nonumber\\
&& \qquad \qquad \qquad \qquad +\, \sum_{R  \in [1^-]}\,C^{(t-n)}_{Q, R}\, J^{(t-n)}_{Q,R}(s,u)  +\sum_{L \in [1^-]}\,C^{(t-n)}_{L, Q}\, J^{(t-n)}_{L, Q}(s,u) \Big\} \,,
\nonumber\\ \nonumber\\ 
&& J^{(u-n)}_{L,Q}(s,u) = J^{(t-n)}_{L,Q}(u,s)\,,\qquad \qquad  
 J^{(s-2)}_{L,Q}(s,u)  =  J^{(t-2)}_{L,Q}(s,u) -  J^{(u-2)}_{L,Q}(s,u) \,, 
\nonumber\\
&& J^{(u-n)}_{Q,R}(s,u) = J^{(t-n)}_{Q,R}(u,s) \,,\qquad \qquad  
 J^{(s-2)}_{Q,R}(s,u)  =  J^{(t-2)}_{Q,R}(s,u) -  J^{(u-2)}_{Q,R}(s,u),
\end{eqnarray}
with 
\begin{eqnarray}
&& J^{(s-1)}_{L,Q}(s,u)  = \int\frac{d^d l}{(2\pi)^d}\,l_{\alpha } \,\bar{p}_{\beta } \,S^{\alpha\beta,\mu\nu}_L( \bar p -l) \, l_{\mu }\,(q+ \bar{q})_{\nu }\,
\frac{-i\,\mu^{4-d}}{l^2-m_Q^2}\,,
\nonumber\\
&& J^{(s-1)}_{Q,R}(s,u)  = \int\frac{d^d l}{(2\pi)^d}\frac{-i\,\mu^{4-d}}{l^2-m_Q^2}\,l_{\alpha }\,(q+ \bar q)_{\beta } \,S^{\alpha\beta,\mu\nu}_R(p-l )\,l_{\mu } \,p_{\nu } \,,
 \nonumber\\ \nonumber\\
 && J^{(t-1)}_{L,Q}(s,u)  = \int\frac{d^d l}{(2\pi)^d}\,l_{\alpha } \,\bar{p}_{\beta } \,S^{\alpha\beta,\mu\nu}_L( \bar p -l) \, (l+q)_{\mu }\, \bar{q}_{\nu }\,
\frac{-i\,\mu^{4-d}}{l^2-m_Q^2}\,,
\nonumber\\
&& J^{(t-1)}_{Q,R}(s,u)  = \int\frac{d^d l}{(2\pi)^d}\frac{-i\,\mu^{4-d}}{l^2-m_Q^2}\ (l+\bar q)_{\alpha } \,q_{\beta }\,\,S^{\alpha\beta,\mu\nu}_R(p-l )\,l_{\mu } \,p_{\nu } \,,
 \nonumber\\ \nonumber\\
&& J^{(t-2)}_{L,Q}(s,u)  = \int\frac{d^d l}{(2\pi)^d}\,l_{\alpha } \,\bar{p}_{\beta } \,S^{\alpha\beta,\mu\nu}_L( \bar p -l)  \,(l+q)_{\mu }\,\, p_{\nu }\,
\frac{-i\,\mu^{4-d}}{l^2-m_Q^2}\,,
\nonumber\\
&& J^{(t-2)}_{Q,R}(s,u)  = \int\frac{d^d l}{(2\pi)^d}\frac{-i\,\mu^{4-d}}{l^2-m_Q^2}\ (l+\bar q)_{\alpha }\,\bar p_{\beta }\, \,S^{\alpha\beta,\mu\nu}_R(p-l )\,l_{\mu } \,p_{\nu } \,.
 \label{def-J-bubble}
\end{eqnarray}
And
\begin{eqnarray}
&& \bar J_{L,Q}^{(s-1)}(s, u) =\frac{s-u}{8}\,\Big\{ \Big( r^2\,\frac{M^2_{ab}}{2}-4\,m_Q^2 \Big)\,\bar I_{QL}(M_a^2) - r\,\bar I_Q+ r\,m_Q^2\,\bar I_L/M_L^2 \Big\} + {\mathcal O}\left( Q^5 \right) \,,
\nonumber\\
&& \bar J_{Q,R}^{(s-1)}(s,u) =  \frac{s-u}{8}\,\Big\{ \Big(  r^2\,\frac{M^2_{ab}}{2}-4\,m_Q^2\Big)\,\bar I_{QR}(M_b^2) -r\,\bar I_Q+ r\,m_Q^2\,\bar I_R/M_R^2\Big\} + {\mathcal O}\left( Q^5 \right)
\,,
\nonumber\\
&& \bar J_{L,Q}^{(t-1)}(s, u) = \frac{s-u}{16}\,\Big\{ \Big( r^2\,\frac{M^2_{ab}}{2}- 4\,m_Q^2 \Big)\,\bar I_{QL}(M_a^2) - r\,\bar I_Q + r\,m_Q^2\,\bar I_L/M_L^2\Big\} + {\mathcal O}\left( Q^5 \right) \,,
\nonumber\\
&& \bar J_{Q,R}^{(t-1)}(s,u) = \frac{s-u}{16}\,\Big\{ \Big(  r^2\,\frac{M^2_{ab}}{2}-4\,m_Q^2\Big)\,\bar I_{QR}(M_b^2) -r\,\bar I_Q+ r\,m_Q^2\,\bar I_R/M_R^2\Big\} + {\mathcal O}\left( Q^5 \right)
\,,
\nonumber\\
&& \bar J_{L,Q}^{(t-2)}(s, u) = \frac{M^2_{ab}}{8}\,\Big\{ \Big( r^2\,\frac{M^2_{ab}}{2}-4\,m_Q^2 \Big)\,\bar I_{QL}(M_a^2) -r\,\bar I_Q+ r\,m_Q^2\,\bar I_L/M_L^2\Big\} 
\nonumber\\
&& \; +\,\frac{1}{4}\, \Big( \delta_L - m_Q^2 - (M_a^2- M_b^2)/2\Big)\,\Big\{  r\,M^2_{ab}\,\bar I_{QL}(M_a^2) -\bar I_Q+ m_Q^2\,\bar I_L/M_L^2\Big\} 
+ {\mathcal O}\left( Q^5 \right)
\,,
\nonumber\\
&& \bar J_{Q,R}^{(t-2)}(s,u) = \frac{M^2_{ab}}{8}\,\Big\{ \Big(  r^2\,\frac{M^2_{ab}}{2}-4\,m_Q^2\Big)\,\bar I_{QR}(M_b^2) - r\,\bar I_Q + r\,m_Q^2\,\bar I_R/M_R^2\Big\} 
\nonumber\\
&& \; +\, \,\frac{1}{4}\, \Big( \delta_R - m_Q^2 + (M_a^2- M_b^2)/2\Big)\, \Big\{ r\,M^2_{ab}\,\bar I_{QR}(M_b^2) -\bar I_Q+ m_Q^2\,\bar I_R/M_R^2\Big\} 
+ {\mathcal O}\left( Q^5 \right) \,,
\label{res-bubble}
\end{eqnarray}
where we keep in (\ref{res-bubble}) heavy tadpole terms proportional to $r\,m_Q^2\,\bar I_L/M_L^2$ and $r\,m_Q^2 \,\bar I_R/M_R^2$. Their scale dependence cannot be discriminated from the corresponding terms proportional to $r\,\bar I_Q$.  In our scheme  
neither the LEC $c_0$ and $c_1$ nor $g_1$ receive a finite renormalization from the bubble loop terms in (\ref{res-bubble}).

\clearpage

\section{Scattering with Triangle diagrams}

\begin{figure}[t]
\center{
\includegraphics[keepaspectratio,width=0.71\textwidth]{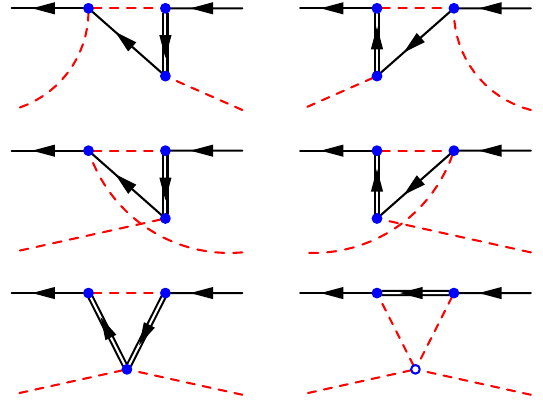} }
\vskip-0.2cm
\caption{\label{fig:6} Dashed lines stand for pion, kaon or eta mesons and solid and double-solid lines for charmed mesons with $J^P = 0^-$ and $J^P = 1^-$ respectively. The vertices are from \cite{Lutz:2022enz}.}
\end{figure}

We turn to the triangle diagrams of Fig.\ \ref{fig:6}  with 
\begin{eqnarray}
&&T^{\rm triangle}_{ab}(s,t,u) =  T^{(s)}_{ab}(s,t,u) + T^{(u)}_{ab}(s,t,u)+ T^{(t)}_{ab}(s,t,u)\,,
\label{def-triangle}
\end{eqnarray}
where we use the conventional Mandelstam variables $s,t$ and $u$ of two-body scattering. The indices $b$ and $a$ specify the initial and final flavor channels of the chosen process.  
The three contributions in (\ref{def-triangle}) correspond to the three rows in Fig.\ \ref{fig:6} in consecutive order. The first term is characterized by its s-channel, the second by its u-channel and the third by its t-channel unitarity cuts. For given isospin (I) and strangeness (S) channel the expressions can conveniently be factorized into universal loop functions and Clebsch coefficients,
\allowdisplaybreaks[1]
\begin{eqnarray}
&& f^4 \,T^{(s)}(s,t,u) = g_P^2\sum_{Q,H\in [0^-]} \Big[ 
 \sum_{L \in [1^-]}  C^{(s)}_{L,QH} \,J^{(s)}_{L,QH}(s,u)  +\sum_{R \in [1^-]}  C^{(s)}_{QH,R} \,J^{(s)}_{QH,R}(s,u)\Big] \,,
\nonumber\\
&& f^4 \,T^{(u)}(s,t,u) = g_P^2\sum_{Q,H\in [0^-]} \Big[ 
 \sum_{L \in [1^-]}  C^{(u)}_{L,QH} \,J^{(u)}_{L,QH}(s,u) +\sum_{R \in [1^-]}  C^{(u)}_{QH,R} \,J^{(u)}_{QH,R}(s,u)  \Big] \,,
\nonumber\\
&& f ^4\,T^{(t)}(s,t,u) =  g_P^2 \sum_{Q,L\in [1^-],\,R\in [1^-]}\,C^{(t)}_{Q,LR} \,J^{(t)}_{Q,LR}(s,u)+
g_P^2 \sum_{H\in [0^-],PQ}\,C^{(s)}_{H,PQ} \,J^{(s)}_{H,PQ}(s,u) 
\nonumber\\
&& \qquad \qquad \qquad \,+\, g_P^2 \sum_{H\in [0^-],PQ}\,C^{(u)}_{H,PQ} \,J^{(u)}_{H,PQ}(s,u) +
g_P^2 \sum_{H\in [0^-],PQ}\,C^{(t)}_{H,PQ} \,J^{(t)}_{H,PQ}( s,u) 
\nonumber\\
&& \qquad \qquad \qquad \,+\,g_P^2 \sum_{H\in [0^-],PQ}\,C^{(\chi)}_{H,PQ} \, J^{(\chi)}_{H,PQ}(s,u) \,,
\nonumber\\
&& \,\quad {\rm with}\qquad  \quad J^{(t)}_{Q,LR}(s,u) = -J^{(t)}_{Q,LR}(u,s)\,,
\qquad  J^{(t)}_{H,PQ}(s,u) = J^{(u)}_{H,PQ}(u,s) \,,
\nonumber\\
&& \qquad  \qquad \qquad 
J^{(s)}_{H,PQ}(s,u) = J^{(s)}_{H,PQ}(u,s) \,,\qquad  \;\;\,
J^{(\chi )}_{H,PQ}(s,u) = J^{(\chi)}_{H,PQ}(u,s) \,,
\label{def-Tsut} 
\end{eqnarray}
where the Clebsch coefficients depend on the isospin and strangeness of the intermediate $(L,Q, H,R)$ and external $(a,b)$ mesons. While $Q, P$ are placeholder indices for a Goldstone boson in (\ref{def-Tsut}), the indices $L,R$ and $H$ refer to the heavy fields with $J^P=1^-$ and $J^P=0^-$ quantum numbers.
The loop functions depend on not only the internal masses, $m_Q, M_H$ and $M_L, M_R$, but also on external masses $m_b, M_b$ and $m_a, M_a$, where we use small $m$'s for Goldstone boson masses and big $M$'s for the masses of  the $0^-$ and $1^-$ charmed mesons. In (\ref{def-Tsut}) it holds  $L,R \in {1^-}$ exclusively. In the s- and u-channel exchange diagrams it holds $H \in [0^-]$, in the t-channel terms $H \in [1^-]$.

The s-channel Clebsch coefficients are readily expressed in terms of the tree-level coefficients $C_{WT}$ and $C^{(u)}_H$ as used in (\ref{res-tree-Q123}) with
\begin{eqnarray}
  \sum_{QH} C^{(s)}_{QH,R} \Big|_{ab} = -\frac{1}{4}\,\sum_{c \leftrightarrow QH} C_{WT} \Big|_{ac}\, C^{(u)}_R  \Big|_{cb} \,,\quad \quad
   \sum_{QH} C^{(s)}_{L,QH} \Big|_{ab} = - \frac{1}{4}\,\sum_{c \leftrightarrow QH}  C^{(u)}_{L}  \Big|_{ac} C_{WT} \Big|_{cb}\, \,,
\end{eqnarray}
where we map the channel index onto its meson content with $c \leftrightarrow QH$ according to Tab.  \ref{tab:states}. The corresponding u-channel Clebsch $C^{(u)}_{L,QH}$ and $C^{(u)}_{QH,R}$  follow from a crossing transformation of $C^{(s)}_{L,QH}$ and $C^{(s)}_{QH,R}$, like the coefficient $C^{(u)}_H$ follows from $C^{(s)}_H$ by such a transformation. 
It is useful to introduce symmetric and antisymmetric combinations with 
\begin{eqnarray}
&&
 \sum_L  C^{(s)}_{L,QH} \,J^{(s)}_{L,QH}(s,u)+  \sum_L  C^{(s)}_{QH,L} \,J^{(s)}_{QH,L}(s,u)  = 
 \sum_L C^{(s,+)}_{L,QH} \,J^{(s,+)}_{L,QH}(s,u)
\nonumber\\
&& \qquad +\, 
 \sum_L  C^{(s,-)}_{L,QH} \,J^{(s,-)}_{L,QH}(s,u) +
 \sum_L C^{(s,+)}_{QH,L} \,J^{(s,+)}_{QH,L}(s,u)+
 \sum_L  C^{(s,-)}_{QH,L} \,J^{(s,-)}_{QH,L}(s,u)
 \,,
\nonumber\\  
&&  C^{(s,\pm)}_{L,QH} = \frac{1}{2}\, \Big( C^{(s)}_{L,QH} \pm  C^{(s)}_{QH,L} \Big)\,, \qquad 
 J^{(s,\pm)}_{L,QH}(s,u) =  \frac{1}{2}\,\Big( J^{(s)}_{L,QH}(s,u) \pm  J^{(s)}_{QH,L}(s,u) \Big) \,,
\nonumber\\  
&&  C^{(s,\pm)}_{QH,L} = C^{(s,\pm)}_{L,QH}\,, \qquad \qquad \qquad \qquad \;
 J^{(s,\pm)}_{QH,L}(s,u) = J^{(s,\pm)}_{L,QH}(s,u) \,,
 \nonumber\\
&& \big[C^{(s,\pm)}_{L,QH}\big]_{ab} = \pm \,\big[ C^{(s,\pm)}_{L,QH} \big]_{ba}\,, \qquad \quad \; \quad \!
\big[J^{(s,\pm)}_{L,QH}(s,u)\big]_{ab} = \pm \,\big[ J^{(s,\pm)}_{L,QH}(s,u) \big]_{ba} \,,
\nonumber\\ 
&& {\rm because } \qquad  \big[J_{QH,L}(s,u) \big]_{ab} =  \big[J_{L,QH}(s,u) \big]_{ba}\,,
\end{eqnarray}
where we used identical summation indices with $R \to L$ in the first line of  (\ref{def-Tsut}). One would expect that it is justified to neglect the mass differences from $M_L$ or $M_R$ in the loop functions summed over $L$ or $R$. This leads to a factorization with the averaged Clebsch structures being implied by $C_{u-ch}$ as recalled already in (\ref{def-Cuch}). In particular we find
\begin{eqnarray}
&& \sum_R  C^{(s)}_{QH,R} +  \sum_L  C^{(s)}_{L,QH} =8 \,C^{(s)}_{QH}-  \frac{1}{4}\,C^{(s-2)}_{QH} + \frac{1}{2}\,\,C^{(s-3)}_{QH}   \,,\qquad 
\nonumber\\ 
&& \sum_R  C^{(u)}_{QH,R} + \sum_L  C^{(u)}_{L,QH}  =8 \,C^{(u)}_{QH} - \frac{1}{4}\, C^{(u-2)}_{QH} + \frac{1}{2}\,\,C^{(u-3)}_{QH}  \,, \qquad 
  \label{res-Cs-Cu}
\end{eqnarray}
with the Clebsch on the r.h.s of (\ref{res-Cs-Cu}) already used in the Appendix of \cite{Lutz:2022enz}.

Similarly, the treatment of the t-channel terms is streamlined upon the introduction of symmetric and antisymmetric Clebsch and loop combinations
\begin{eqnarray}
&& C^{(\cdots  )}_{H,PQ}\,J^{(\cdots  )}_{H,PQ} = 
C^{(\cdots + )}_{H,PQ}\,J^{(\cdots +  )}_{H,PQ} + 
C^{(\cdots - )}_{H,PQ}\,J^{(\cdots - )}_{H,PQ} \,,\qquad \qquad \big[ J^{(\cdots  )}_{H,PQ} \big]_{ab} = \big[ J^{(\cdots  )}_{H,QP} \big]_{ba} \,,
\label{def-pmC-pmJ}
\end{eqnarray}
with
\begin{eqnarray}
&& \big[ C^{(\cdots  \pm )}_{H,PQ} \big]_{ab} = \frac{1}{2}\,\Big( \big[C^{(\cdots  )}_{H,PQ} \big]_{ab} \pm  \big[C^{( \cdots )}_{H,PQ} \big]_{ba}\Big) \,,
\nonumber\\
&& \big[ J\,^{(\cdots  \pm )}_{H,PQ} \big]_{ab} = \frac{1}{2}\,\Big( \big[J\,^{(\cdots  )}_{\!H,PQ} \big]_{ab} \pm  \big[J\,^{( \cdots )}_{\!H,PQ} \big]_{ba}\Big)\,,
\nonumber\\ \nonumber\\
&&  \sum_H  C^{(s \pm )}_{H,PQ} =  \sum_H  \big( C^{(t \pm )}_{H,PQ} + C^{(u \pm)}_{H,PQ} \big) \,,\qquad \qquad 
 C^{(s-)}_{H,PQ} =8\,C^{(-)}_{PQ} \,, 
 \nonumber\\
&& \sum_H  \big( C^{(t +)}_{H,PQ} - C^{(u+)}_{H,PQ}\big) = 16\,C^{(+)}_{PQ} \,, 
 \nonumber\\
&&  \sum_H  C^{(\chi -)}_{H,PQ} = 8\,\Big( 2\,m\,B_0\,C^{(\chi),\pi}_{PQ} +  (m+m_s)\,B_0\,C^{(\chi),K}_{PQ}\Big) \,, 
\end{eqnarray}
in terms of the Clebsch  listed in  Tab. III and Tab. XXI of \cite{Lutz:2022enz}. 

In the following we discuss in depth the computation of the loop functions. It suffices to specify the s- and t-channel loop functions. The u-channel expressions follow from the s-channel loop by the crossing replacement $s \leftrightarrow u$ as is implied by $q_\mu \leftrightarrow -\bar q_\mu$. We find
\begin{eqnarray}
&& J^{(s)}_{L,QH}(s,u)  = \int\frac{d^d l}{(2\pi)^d}\,l_{\alpha } \,\bar{p}_{\beta } \,S^{\alpha\beta,\mu\nu}_L(l+\bar p) \,  \bar{q}_{\mu } \,(l+w)_{\nu }\,
\frac{-i\,\mu^{4-d}}{l^2-m_Q^2}\,\frac{(l+w+p) \cdot (l-q)}{(l+w)^2-M_H^2}\,,
\nonumber\\
&& J^{(s)}_{QH,R}(s,u)  = \int\frac{d^d l}{(2\pi)^d}\frac{-i\,\mu^{4-d}}{l^2-m_Q^2}\,\frac{ (l+w+\bar p) \cdot (l-\bar q)}{(l+w)^2-M_H^2}\,q_{\alpha }\, (l+w)_{\beta } \,S^{\alpha\beta,\mu\nu}_R(l+p)\,l_{\mu } \,p_{\nu } \,,
\nonumber\\
&&  J^{(t)}_{Q,LR}(s,u) =\int\frac{d^d l}{(2\pi)^d}\frac{i\,\mu^{4-d}}{l^2-m_Q^2}\,l_{\alpha } \,\bar{p}_{\beta } \,S^{\alpha\beta, \bar \sigma \bar \tau}_L(l+\bar p)\,
g_{\bar \tau \tau}\,S^{\sigma \tau, \mu\nu}_R(l+p)\,l_{\mu } \,p_{\nu }
\nonumber\\
	&&\qquad\qquad\qquad\quad \times \, \big(\,(l+\bar p )_{\bar \sigma}\, (q+\bar q)_\sigma +
	(q+\bar q)_{\bar \sigma} \,(l +p)_{\sigma}   \big)	\,,
\nonumber\\  \nonumber\\
&&  J^{(s)}_{H,PQ}(s,u) =\int\frac{d^d l}{(2\pi)^d} \frac{-(\bar q\cdot q)- (l+\bar q)\cdot (l +q)}{(l+\bar q)^2-m_P^2}\,\frac{-i\,\mu^{4-d}}{(l +q)^2-m_Q^2}\,
\nonumber\\
	&&\qquad\qquad\qquad\quad \times \, 
(l+\bar q)_{\alpha } \,\bar{p}_{\beta } \, S^{\alpha\beta, \mu \nu}_H(l+ p +q)\, (l+q)_{\mu}\,p_\nu	\,,
\nonumber\\
&&  J^{(t)}_{H,PQ}(s,u) =\int\frac{d^d l}{(2\pi)^d} \frac{( l+q )\cdot q+ \bar q\cdot (l + \bar q)}{(l+\bar q)^2-m_P^2}\,\frac{-i\,\mu^{4-d}}{(l +q)^2-m_Q^2}\,
\nonumber\\
	&&\qquad\qquad\qquad\quad \times \, 
(l+\bar q)_{\alpha } \,\bar{p}_{\beta } \,S^{\alpha\beta, \mu \nu}_H(l+ p+ q)\, (l+q)_{\mu}\,p_\nu	\,,
\nonumber\\
&&  J^{(u )}_{H,PQ}(s,u) =\int\frac{d^d l}{(2\pi)^d} \frac{-( l+ \bar q )\cdot q-  \bar q\cdot (l +q)}{(l+\bar q)^2-m_P^2}\,\frac{-i\,\mu^{4-d}}{(l +q)^2-m_Q^2}\,
\nonumber\\
	&&\qquad\qquad\qquad\quad \times \, 
(l+\bar q)_{\alpha } \,\bar{p}_{\beta } \,S^{\alpha\beta, \mu \nu}_H(l+ p+q)\, (l+q)_{\mu}\,p_\nu	\,,
\nonumber\\
&&  J^{(\chi )}_{H,PQ}(s,u) =\int\frac{d^d l}{(2\pi)^d} \frac{1}{(l+\bar q)^2-m_P^2}\,\frac{-i\,\mu^{4-d}}{(l +q)^2-m_Q^2}\,
\nonumber\\
	&&\qquad\qquad\qquad\quad \times \, 
(l+\bar q)_{\alpha } \,\bar{p}_{\beta } \,S^{\alpha\beta, \mu \nu}_H(l+ p+q) \, (l+q)_{\mu}\,p_\nu	\,,
	\label{def-J-triangle}
\end{eqnarray}
where we use $ w = p+q = \bar p+\bar q $ with $q^2=m_b^2,\, p^2=M_b^2$ and $\bar q^2= m_a^2, \,\bar p^2=M_a^2$.  The renormalization scale of dimensional regularization is  $\mu$. Given the shortage of available letters in any  notation scheme we purposely use $\mu$ in two distinct mathematical contexts.  From the specific form of the t-channel loop functions in (\ref{def-J-triangle})  it is evident that they are all invariant under a simultaneous interchange of $ P \leftrightarrow Q$ and $ (q, p) \leftrightarrow (\bar q, \bar p) $, as was used in (\ref{def-pmC-pmJ}).

The proper evaluation of the triangle-loop functions in (\ref{def-J-triangle}) is not quite so straightforward, in particular if one insists on the use of on-shell meson masses. Following previous works \cite{Lutz:2018cqo,Bavontaweepanya:2018yds,Lutz:2020dfi,Sauerwein:2021jxb},
in an initial step, our one-loop triangle contributions in (\ref{def-J-triangle}) are expressed in terms of three scalar loop functions
\begin{eqnarray}
&& I_a \;=\int \frac{d^d l}{(2\pi)^d}\,\frac{ i\,\mu^{4-d} }{l^2- m_a^2}\,,
\nonumber\\
&& I_{ab}(k^2) =  \int \frac{d^d l}{(2\pi)^d}\,\frac{ -i\,\mu^{4-d} }{l^2- m_a^2}\,
 \frac{1}{(l+k)^2-m_b^2 } \,,
 \nonumber\\
&& I_{abc}(\bar p^2,\bar p \cdot p , p^2) =  \int \frac{d^d l}{(2\pi)^d}\,
 \frac{1}{(l+\bar p)^2-m_a^2 }\,\frac{ i\,\mu^{4-d} }{l^2- m_b^2}\, \frac{1}{(l+p)^2-m_c^2 }  \,,
 \label{def-scalar-integrals}
\end{eqnarray}
as introduced in dimensional regularization. The scalar integrals in (\ref{def-scalar-integrals}) 
arise in the evaluation of (\ref{def-J-triangle}) with the indices $a,b,c$  replaced by any of the 
indices $Q, H,L, R$. For instance, we will encounter the tadpole integrals, $I_L , I_H, I_R$ and $I_Q$, where for $L,H,R$ the mass parameter $M_L, M_H, M_R$ are to-be-used in (\ref{def-scalar-integrals}). In case of the index $Q$ the mass parameter $m_Q$ is encountered.  

In the reduction of the triangle loops the following scalar bubble and triangle-loop expressions occur in addition
\begin{eqnarray}
&& I_{LR} = I_{LR}(t) \,, \,\,
\qquad \quad  I_{LH} = I_{LH}(\bar q^2) \,,\qquad \quad I_{RH} = I_{RH}( q^2) \,,
\nonumber\\
&&  I_{QH} = I_{QH}(s) \,, \qquad \quad I_{QL} = I_{QL}(\bar p^2)  \,, \qquad \quad I_{QR} = I_{QR}(p^2) \,,\qquad
\label{def-ILQHR}\\
&&   I_{L,QH } =  I_{LQH}(\bar p^2,\bar p \cdot w, w^2 ) \,, \quad  I_{QH,R} =  I_{HQR}(w^2,w \cdot p, p^2 ) \,,
 \quad  I_{Q,LR} =  I_{LQR}(\bar p^2,\bar p \cdot p, p^2 ) \,,
\nonumber
\end{eqnarray}
where we specify the kinematical points at which such integrals are needed. 

There are a few well-known technical issues to be considered. A straightforward evaluation of the set of diagrams leads to results that suffer from terms that are at odds with their  expected chiral power. There are terms, not only of too low,  but also of too high orders, both of which need to be eliminated as to arrive at consistent results. 
For instance, according to dimensional counting rules one expects for properly  renormalized scalar loop functions 
\begin{eqnarray}
I_Q \to \bar I_Q \sim Q^2\,,\qquad  I_{QH} \to \bar I_{QH}\sim Q^1 \,,\qquad 
I_{QH,R} \to \bar I_{QH,R}  \sim Q^0 \,,\qquad 
\end{eqnarray}
where we use a bar for renormalized quantities. 
As was already pointed out in \cite{Semke:2005sn} loop functions that are  ultraviolet convergent do not give rise to power-counting violating contributions. Indeed, the expected chiral power of the scalar triangle loop can be confirmed by an explicit computation. 

Yet, there is another technical complication that needs to be resolved. Any application of the original Passarino--Veltman decomposition scheme \cite{Passarino:1978jh} requires the knowledge of specific correlations of the scalar basis functions at particular kinematic conditions \cite{Chetyrkin:1981qh,Tkachov:1981wb,Tarasov:1996br,Duplancic:2003tv,Denner:2005nn,Battistel:2006zq,Guillet:2018cdm,Li:2022cbx}. If such relations are ignored results will suffer from kinematical singularities, a potentially pernicious situation. 
Therefore it is useful to extend the set of scalar basis integrals, such that a decomposition arises void of superficial  singularities. This was advocated already in \cite{Lutz:2020dfi,Sauerwein:2021jxb} in studies of axial form factors of the baryons. 
Two examples relevant for scattering at the one-loop level are discussed in detail below. Consider two candidates  for such extra basis functions with
\begin{eqnarray}
&&  I^{(1)}_{L,QH } = \frac{1}{2}\,\frac{ w^2\,(v^2_{QL}\,I_{L,QH} -I_{LH} +I_{QH} ) -  (\bar p \cdot w) \,(v^2_{QH}\,I_{L,QH} - I_{LH}+ I_{QL} )}{\bar p^2 \, w^2-  ( \bar p \cdot w)^2} \,,
\nonumber\\
&&  I^{(1)}_{QH,R } =  \frac{1}{2}\,\frac{ w^2\,(v^2_{QR}\,I_{QH,R} -I_{HR} +I_{QH} ) -  (w \cdot p)  \,(v^2_{QH}\,I_{QH,R} - I_{HR}+ I_{QR} )}{w^2\,p^2- ( w \cdot p)^2} \,,
\nonumber\\
&& v^2_{QL} = \bar p^2- M_L^2 + m_Q^2\,,\qquad v^2_{QR} = p^2- M_R^2 + m_Q^2\,,\qquad   v^2_{QH} = w^2- M_H^2 + m_Q^2\,,
\label{def-1-triangle}
\end{eqnarray}
where we assure that both functions $ I^{(1)}_{L,QH }$ and $I^{(1)}_{QH, R }$ are regular at the problematic threshold conditions $s = ( w\cdot p)^2/p^2$ and $s = (\bar p \cdot w)^2/\bar p^2$. 
The verification of our claim is tedious and asks for a more powerful viewpoint. We will generalize that extra basis functions with $ I^{(n)}_{L,QH }$ and $I^{(n)}_{QH, R } $,  where with the case $n=0$ we recover the original scalar triangles. 
We introduce the set of basis functions 
\begin{eqnarray}
&& I^{(n)}_{L,QH }  =  \frac{1}{16\, \pi^2}\,\int_0^1 d x \int_0^{1-x} d y \,\frac{x^n}{F_{L,QH}(x,y)}\sim Q^0\,,
\nonumber\\
&& I^{(n)}_{QH, R }  =  \frac{1}{16\, \pi^2}\,\int_0^1 d x \int_0^{1-x} d y \,\frac{x^n}{F_{QH,R}(x,y)}\sim Q^0\,,
\nonumber\\ \nonumber\\
&& F_{L,QH}(x,y) = m_Q^2- x\,  v^2_{QL}  - y\,v^2_{QH}  + x^2\,\bar p^2 + y^2\, w^2 + 2\,x\,y\,( \bar p\cdot w )\,,
\nonumber\\
&& F_{QH,R}(x,y) = m_Q^2- x\,  v^2_{QR}  - y\,v^2_{QH}  + x^2\,p^2 + y^2\, w^2 + 2\,x\,y\,( w \cdot p)\,,
\label{def-n-triangle-QH}
\end{eqnarray}
in terms of a Feynman parameter ansatz. We complement our choice of basis functions with 
\begin{eqnarray}
&& I^{(n )}_{Q, \bar  LR }  =  \frac{1}{16\, \pi^2}\,\int_0^1 d x \int_0^{1-x} d y \,\frac{x^n}{F_{Q,LR}(x,y)}\sim Q^0  \,,
\nonumber\\
&& I^{(n )}_{Q,L\bar R }  =  \frac{1}{16\, \pi^2}\,\int_0^1 d x \int_0^{1-x} d y \,\frac{y^n}{F_{Q,LR}(x,y)}\sim Q^0  \,,
\nonumber\\ \nonumber\\
&& I^{(n)}_{H,\bar PQ }  =  \frac{1}{16\, \pi^2}\,\int_0^1 d x \int_0^{1-x} d y \,\frac{x^n}{F_{H,PQ}(x,y)}\sim Q^{-1}\,,
\nonumber\\
&& I^{(n)}_{H,P \bar Q }  =  \frac{1}{16\, \pi^2}\,\int_0^1 d x \int_0^{1-x} d y \,\frac{y^n}{F_{H,PQ}(x,y)}\sim Q^{-1}\,,
\nonumber\\ \nonumber\\
&& F_{Q,LR}(x,y) \,= m_Q^2 - x\,  v^2_{QL}  - y\,v^2_{QR}\,  + x^2\,\bar p^2 + y^2\, p^2 \,+ 2\,x\,y\,( \bar p \cdot  p)\,,
\nonumber\\
&& F_{H,PQ}(x,y) \,= M_H^2 - x\,  v^2_{HP}  - y\,v^2_{HQ}\,  + x^2\,\bar p^2 + y^2\, p^2 \,+ 2\,x\,y\,(  \bar p \cdot  p)\,,
\nonumber\\
&& v^2_{HP} = \bar p^2- m_P^2 + M_H^2\,,\qquad \qquad v^2_{HQ} = p^2- m_Q^2 + M_H^2\,,\qquad  
\label{def-n-triangle-PQ}
\end{eqnarray}
where our sets of basis functions in (\ref{def-n-triangle-PQ}) transform into each other under exchange of
$\bar p \leftrightarrow p$ if combined with $L\leftrightarrow R$ or $P\leftrightarrow Q$. Corresponding pairs of basis loop functions are instrumental since the loop functions (\ref{def-J-triangle}) and (\ref{res-Jtriangle4}) have specific properties under such transformations. 

By construction such functions (\ref{def-n-triangle-QH}) and (\ref{def-n-triangle-PQ}) are void of kinematical constraints. The definitions (\ref{def-n-triangle-QH}) are compatible with (\ref{def-1-triangle}) for $n=1$ and (\ref{def-scalar-integrals}, \ref{def-ILQHR}) for $n= 0$. 
For example, it holds $ I^{(0)}_{L,QH }= I_{L,QH }$ and $ I^{(0 )}_{Q, \bar L  R } = I^{(0 )}_{Q,L\bar R } = I_{Q,LR } $. We note that also for $ I^{(n)}_{Q,LR }$ and $I^{(n)}_{H,PQ }$ or $n > 1$ expressions analogous to (\ref{def-1-triangle}) can be derived, however, they turn more and more tedious as $n$ increases, involving higher degrees of superficial pole structures. The explicit expression for $n=2$ is given in (\ref{res-I2GLR})  of Appendix B. 

We note that while the integral  representations (\ref{def-n-triangle-QH}) and (\ref{def-n-triangle-PQ}) are numerically stable for space-like 4-momenta only, the hierarchy of functions with $n =0,1,2,\cdots $ has identical analytic branch points and cuts as they arise for time-like 4-momenta. The crucial question arises whether a decomposition of the loop functions into our choice of basis loops can be defined in an unambiguous manner. This is indeed the case, a proof of which is provided in Appendix B.

We now assume that a given triangle loop is decomposed into our extended set of scalar loop functions. 
Such expressions are prohibitively involved, and therefore not shown here. A useful reprentation can be obtained nevertheless upon a chiral expansion thereof. This goes in two steps. First we need to expand the coefficients in front of our basis functions in chiral powers. Here we apply the power-counting scheme \cite{Lutz:2022enz} introduced in terms of on-shell hadron masses, as recalled in (\ref{def-power-counting}).

In order to specify the chiral order of a given contribution we need to assign a chiral power to the basis loop functions also. A subtraction scheme for the basis functions such that power-counting respecting renormalized basis loop functions arise is constructed. Such a procedure is  symmetry conserving \cite{Lutz:1999yr,Semke:2005sn,Lutz:2014oxa,Lutz:2020dfi} as long as 
there is an unambiguous prescription how to represent such one-loop contributions  in terms of the set of basis functions. In this case we do not expect any violation of the chiral Ward identities of QCD. 

Following our previous works \cite{Lutz:2018cqo,Bavontaweepanya:2018yds,Lutz:2020dfi,Sauerwein:2021jxb} we introduce renormalized scalar bubble-loop functions that are independent of the renormalization scale. Here it is instrumental to carefully discriminate the light from the heavy particles.
\begin{eqnarray}
&&\bar I_{PQ}=  I_{PQ}(t) + \frac{I_Q}{2\,m_Q^2} +  \frac{I_P}{2\,m_P^2}\,,  \qquad \,\, \qquad 
\bar I_{LR}=  I_{LR}(t) + \frac{I_L}{2\,M_L^2} +  \frac{I_R}{2\,M_R^2}\,, 
\nonumber\\
&&\bar I_{LH}=  I_{LH}(s) + \frac{I_L}{2\,M_L^2} +  \frac{I_H}{2\,M_H^2}\,,  \qquad \qquad 
\bar I_{HR}=  I_{HR}(s) + \frac{I_H}{2\,M_H^2} +  \frac{I_R}{2\,M_R^2}\,, 
 \nonumber\\
&& \bar I_{QL} = I_{QL} - \frac{1 - \gamma}{16\,\pi^2} + \frac{I_L}{M_L^2}\,, \qquad \qquad \qquad  \bar I_{QR} = I_{QR} - \frac{1- \gamma}{16\,\pi^2} + \frac{I_R}{M_R^2}\,,
\nonumber\\
&&  \bar I_{QH} = I_{QH} - \frac{1 - \gamma^H}{16\,\pi^2} + \frac{I_H}{M_H^2}\,, \qquad \qquad \quad \;\, \,  \gamma=- r\,\log \frac{r}{1+r}\,,\qquad 
r= \frac{\tilde M^2 - M^2}{M^2}\,,
\nonumber\\
&& \bar I_{LR} \to  0 \,,\qquad \quad \! \! \bar I_{LH} \to 0 \,,\qquad  \quad \! \! \bar I_{HR} \to  0 \,,\qquad \gamma^{H\in[ 0^-]} = 0\,, \qquad \gamma^{H\in [1^-] } = \gamma \,,
\label{def-bubble-all}
\end{eqnarray} 
where we use $P,Q$ as placeholders for the light fields (Goldstone bosons) but $H, L, R$ as  placeholders for the heavy fields (charmed mesons). An explicit expression for $\bar I_{QH}=\bar I_{QH}(s)$ was already recalled in  (\ref{def-bubble}).
In turn it is left to renormalize the tadpole contributions with
\begin{eqnarray}
&& \bar I_Q = \frac{m_Q^2}{(4 \pi )^ 2} \, \log \frac{m_Q^2}{ \mu^2}\,,
 \label{def-IQ}
\end{eqnarray}
in terms of the renormalization scale $\mu$ of dimensional regularization as implied by
 $\overline{MS}$ subtraction scheme. For the heavy fields their tadpole contributions  are dropped with  $\bar I_{L,H, R} \to 0$ if associated with power-counting violating structures, but kept otherwise. 

It is noteworthy that the scalar triangle loops  are finite and do not show a renormalization scale dependence, i.e. 
\begin{eqnarray}
&& \bar I^{(n)}_{L,QH } =I^{(n)}_{L,QH } - \frac{\gamma_n}{16\,\pi^2\,M^2}\sim Q^0\,, \qquad  \qquad 
 \bar I^{(n)}_{QH,R } = I^{(n)}_{QH,R }- \frac{\gamma_n}{16\,\pi^2\,M^2}\sim Q^0\,,
\nonumber\\
&& \bar I^{(n)}_{Q,LR } = I^{(n)}_{Q,LR }- \frac{\gamma_n}{16\,\pi^2\,M^2} \sim Q^0\,, \qquad  \qquad \;\,
\nonumber\\
&&  \gamma_0= \log \frac{1+ r}{r}\,, \qquad \gamma_1= \frac{1}{2}  - \frac{r}{2}\,\log \frac{1+r}{r}\,,
\qquad \gamma_2= \frac{1}{6}  - \frac{r}{3} + \frac{r^2}{3}\,\log \frac{1+r}{r}\,,
\nonumber\\
&& \gamma_3= \frac{1}{12}  - \frac{r}{8} +\frac{r^2}{4}- \frac{r^3}{4}\,\log \frac{1+r}{r}\,, \qquad \qquad \qquad \!  {\rm with}\qquad 
r= \frac{\tilde M^2 - M^2}{M^2} \,,
\nonumber\\ 
&&\bar I^{(n)}_{H,PQ } = I^{(n)}_{H,PQ }\,.
\label{def-r}
\end{eqnarray} 
All power-counting violating terms are eliminated by our renormalization conditions in which the unbar basis loop functions are replaced by their bar versions. The expectations of dimensional counting rules come true. In particular it holds $\bar I_{QH} \sim Q^1$. Moreover, owing to the additional subtraction terms $\gamma $ we also 
implemented the expectation of counting rules in the chiral domain with $m_Q \ll \tilde M - M$. Here it holds 
\begin{eqnarray}
&& \bar I_{QH} \sim \frac{m_Q^2}{\tilde M^2 -M^2} \sim Q^2 \,,\qquad \qquad \qquad  \qquad \;\;\;\;\,\,\bar I^{(n)}_{Q,LR } \sim  \frac{m_Q^2}{M^2\,(\tilde M^2 -M^2)
} \sim Q^2 \,,
\nonumber\\
&& \bar I^{(n)}_{L,QH } \sim \bar I^{(n)}_{QH, R } \sim \frac{m_Q}{M\,(\tilde M^2 -M^2)}\sim Q^{1}  \,,
\qquad \quad   \bar I^{(n)}_{H,QP } \sim  \frac{1}{M^2- \tilde M^2} \sim Q^0 \,,
\label{res-chiral-domain}
\end{eqnarray}
where we use $s = M^2$ and $t = 0$ together with $H \in [1^-]$ for simplicity.

We emphasize that here the introduction of the extended basis functions in (\ref{def-n-triangle-QH}) plays a crucial role. We substantiate the findings of \cite{Semke:2005sn} that power-counting violating terms stem from divergent structures. Convergent structures are expected not to cause complications, however, this is so only, if contributions are cast in an unambiguous manner into our extended basis functions.  Here we rely heavily on the results of Appendix B, in which the usefulness of our basis functions is proven. By this we can exclude possible cancellations amongst superficially power-counting violating terms. Indeed all our explicit results confirm the power-counting expectation.

After some algebra we arrive at the amazingly compact $Q^3$ terms in the triangle diagrams of Fig.\ \ref{fig:6}. For the s- and u-channel diagrams it holds 
\begin{eqnarray}
&&   \bar J^{(s)}_{QH,R}(s,u) =
\frac{s-u -r\,M^2_{ab}}{32}\,\Big\{ \Big( (s-u)^2 - 8\,m_{b}^2\,M^2_{ab} \Big)\, \bar I^{(1)}_{QH,R}(s)
\nonumber\\
&& \qquad \quad + \,(s-u)\, \Big[ \Big(s -u +r\,M^2_{ab}\Big)\,\bar I_{QH,R}(s) + 2\,\bar   I_{QR}(M_b^2) -2\, \bar I_{QH}(s) \Big ]\Big\}
+ {\mathcal O} (Q^4) \,,
\nonumber\\
&&  \bar J^{(s)}_{L,QH}(s,u) = 
\frac{s-u- r\,M^2_{ab}}{32}\,\Big\{ \Big( (s-u)^2 - 8\,m_{a}^2\,M^2_{ab} \Big)\, \bar I^{(1)}_{L,QH}(s)
\nonumber\\
&& \qquad \quad + \,(s-u)\, \Big[  \Big(s -u +r\,M^2_{ab}\Big)\,\bar I_{L,QH}(s)
+2\,\bar I_{QL}(M_a^2) -2\,\bar I_{QH}(s) \Big]\Big\} + {\mathcal O} (Q^4) \,,
\nonumber\\
&&   \bar J^{(u)}_{QH,R}(s,u) =\bar J^{(s)}_{QH,R}(u,s) \,,\qquad \qquad \qquad \bar J^{(u)}_{L,QH}(s,u) =\bar J^{(s)}_{L,QH}(u,s) \, ,
\label{res-JQHR}
\end{eqnarray}
where we make the kinematical dependencies explicit again. 
The chiral counting rules (\ref{def-power-counting}) together with $r \sim Q$ are used. Given our renormalization scheme no power-counting violating terms arise. At this order there is  no renormalization scale dependence from any of the tadpole contributions.  As a consequence, we do not encounter a renormalization of the third order LEC $g_n$ from such loop contributions.  The order $Q^4$ terms are a bit more tedious and therefore are delegated to the Appendix. Here tadpole and bubble loop contributions are involved. 

We note that the results (\ref{res-JQHR}) have the expected leading scaling behavior in the chiral domain with $\bar J^{(s,u)}_{QH,R}(s,u)\sim Q^4$ and $\bar J^{(s,u)}_{L,QH}(s,u)\sim Q^4$. This is a consequence of the scaling behavior of our basis functions (\ref{res-chiral-domain}) in that domain around $s \sim M_{ab}^2/2$.

We continue with the expansion of  our t-channel loop functions in chiral powers according to (\ref{def-power-counting}) with $r \sim Q$,  where we drop terms only that are of order $Q^4$ or higher. With this we find the compact expressions
\allowdisplaybreaks[1]
\begin{eqnarray}
&&   \bar J^{(t)}_{Q,LR}(s,u) =  \frac{s-u}{4}\,\Big\{  -\bar I_Q +\frac{1}{2}\,m_Q^2\,\Big[ \bar I_L/M_L^2+\bar I_R/M_R^2\Big] 
\nonumber\\
&& \qquad \quad +\,
 \frac{r}{2}\, M^2_{ab}\, \Big[ \bar I_{QL}(M_a^2)+ \bar I_{QR}(M_b^2
 )\Big] + 2\,\Big( m_Q^2 -\frac{r^2}{8}\, M^2_{ab}  \Big)\,M^2_{ab}\, \bar I_{Q,LR}(t)\Big\} + {\mathcal O} (Q^4) \,, 
\nonumber\\ 
&&   \bar J^{(-)}_{H,PQ}(s,u) =  \frac{M^2_{ab}}{16} \,\Big\{ 
r\, \Big[
\bar I_Q + \bar I_P  - m_{PQ}^2\,\bar I_H / M_H^2  \Big] 
\nonumber\\
&& \qquad \quad +\, 2\,(m_P^2- m_Q^2)\, \Big[\bar I_{PH}(M_a^2) - 
\bar I_{QH}(M_b^2) \Big]
\nonumber\\
&& \qquad \quad +\, \Big( 4\,m_{PQ}^2 +2\,m_{ab}^2-6\,t- \frac{r^2}{2}\,M^2_{ab} \Big)\,\Big[\bar I_{QH}(M_a^2) + \bar I_{PH}(M_b^2) \Big]
\nonumber\\
&& \qquad \quad -\, \Big( m_{PQ}^2+m_{ab}^2 -3\,t  \Big)\,\Big[ r\,\bar I_{PQ}(t)
-r\,\bar I_P/(2\,m_P^2) -r\,\bar I_Q/(2\,m_Q^2) +r\,\bar I_H/M_H^2
\nonumber\\
&& \qquad \qquad \qquad \quad  +\, 2\,\Big( m_{PQ}^2-t  -\frac{r^2}{4}\,M^2_{ab}\Big) \bar I_{H,PQ}(t) \Big]  \Big\}+  {\mathcal O} (Q^4) \,,
\nonumber\\
&&   \bar J^{(+)}_{H,PQ}(s,u) = \frac{s-u}{8}\,\Big\{ \frac{5}{6}\,( \bar I_Q + \bar I_P )  - \frac{1}{6}\,\big(5\,m_{PQ}^2 -t \big)\,\bar I_H/M_H^2
+ \big(3\,m_{PQ}^2-t \big) \,\frac{1}{144\,\pi^2} 
\nonumber\\
&& \qquad  \quad 
-\,\frac{1}{6}\,\Big( 4\,m_{PQ}^2 -5\,t - \frac{3} {2} \,r^2\,M^2_{ab}\Big) \,\Big[\bar  I_{PQ}(t) -\bar I_P/(2\,m_P^2)  -\bar I_Q/(2\,m_Q^2) + \bar I_H/M_H^2\Big]
\nonumber\\
&& \qquad  \quad 
-\,\frac{1}{6}\, \big(m_P^2-m_Q^2\big)^2\,\frac{\bar I_{PQ}(t)-\bar I_{PQ}(0)}{ t}
 -\frac{r}{2}\,M^2_{ab}\,\Big[\bar I_{QH}(M_b^2)  +\bar I_{PH}(M_a^2) \Big]
\nonumber\\
&& \qquad  \quad 
+ \,\frac{r}{2}\,M^2_{ab}\,\Big( m_{PQ}^2-t -\frac{r^2}{4}\,M^2_{ab}\Big)\,\bar I_{H,PQ}(t)
\Big\}
\nonumber\\
&& \qquad \quad  +\,  \frac{m_a^2-m_b^2}{8}\, M^2_{ab}\,\Big\{ 
 \bar I_{PH}(M_a^2)-\bar I_{QH}(M_b^2)  + r\,\big( m_P^2-m_Q^2\big)\,\frac{\bar I_{PQ}(t)-\bar I_{PQ}(0)}{2\,t}
 \nonumber\\
&& \qquad \quad 
- \,\Big( m_{PQ}^2-t -\frac{r^2}{4}\,M^2_{ab}\Big)\,\Big[\bar I^{(1)}_{H,\bar PQ}(t) -\bar I^{(1)}_{H, P\bar Q}(t)  \Big]
\Big\} 
+ {\mathcal O} (Q^4)  \,,
\nonumber\\
&&   \bar J^{(\chi)}_{H,PQ}(s,u) = \frac{M^2_{ab}}{8}\,\Big\{ r\, \Big[
 \bar I_{PQ}(t) -\bar I_P/(2\,m_P^2) - \bar I_Q/(2\,m_Q^2)  +\bar I_H/M_H^2 \Big] 
  -  2\,\bar I_{PH}(M_a^2) 
 \nonumber\\
&& \qquad \quad -\, 2\,\bar I_{QH}(M_b^2)  +\, 2\, \Big( m_{PQ}^2-t - \frac{r^2}{4}\,M^2_{ab} \Big) \,\bar I_{H,PQ}(t)
\Big\}
+ {\mathcal O} (Q^2) \,, 
\nonumber\\ \nonumber\\
&& \frac{1}{2}\,\bar J^{(t)}_{H,PQ}(s,u)+ \frac{1}{2}\,\bar J^{(u)}_{H,PQ}(s,u) = \frac{t}{2}\,  \bar J^{(\chi)}_{H,PQ}(s,u) \,,
\nonumber\\
&&  \bar J^{(+ )}_{H,PQ}(s,u)=  \frac{1}{2}\, \bar J^{(t)}_{H,PQ}(s,u) - \frac{1}{2} \,\bar J^{(u)}_{H,PQ}(s,u) \,,\qquad \quad 
\nonumber\\
&&  \bar J^{(-)}_{H,PQ}(s,u) =  \bar J^{(s )}_{H,PQ}(s,u)+
\frac{1}{2}\,\big(  \bar J^{(t)}_{H,PQ}(s,u) + \bar J^{(u)}_{H,PQ}(s,u) \big)\,,
\label{res-Triangle-J}
\end{eqnarray}
where, again, the order $Q^4$ terms are shown in the Appendix. In (\ref{res-Triangle-J}) we detail the particular combinations $ \bar J^{(\pm )}_{H,PQ}(s,u)$ rather than the original 
$J^{(t)}_{H,PQ}(s,u)$ and $J^{(u)}_{H,PQ}(s,u)$  functions. This is convenient since the former have more transparent properties under a crossing transformation. 
The relevant contributions to the reaction amplitudes follow by a  simple rewrite. 

Given our renormalization scheme no power-counting violating terms arise if we insist on $r \sim Q$ or $r \sim Q^2$. At this order there is a renormalization scale dependence in the t-channel loop $ \bar J^{(+)}_{H,PQ}(s,u)$ only, as is implied here by the $\bar I_P, \bar I_Q$ or $\bar I_H/M_H^2 $ tadpoles. Such terms request a scale dependence of the third order LEC $g_1$ and $g_2$. The scale invariance of the loop functions $ \bar J^{(-)}_{H,PQ}(s,u)$ and $ \bar J^{(\chi)}_{H,PQ}(s,u)$ is a consequence of the condition that the LEC $c_0$ and $c_1$ remain untouched by our renormalized loop effects. The derivation of these results is not straightforward as it depends on the difficult-to-control heavy tadpole terms. 
This can be exemplified by the $J^{(t)}_{Q,LR}(s,u)$ loop, for which its coefficient in front of the $\bar I_L$ and $\bar I_R$ depends critically on terms of formally higher order. Via power-counting violating effects such higher order terms do influence the third order heavy tadpole terms as can be easily verified by explicit computations. Its proper and  unambiguous value can be determined only by the request that the third order amplitudes do not depend on the renormalization scale. Any other choice would be at odds with this requirement. 

\begin{table}[]
  \tabcolsep=3.2mm
  \center
  \begin{tabular}{|C||C|C|C||C|C|C||C|C|C|}
    \hline
    (I,\,S)
    & \multicolumn{3}{C||}{
      {C}^{(0)}_{Q}
    }
     & \multicolumn{3}{C||}{
      {C}^{(1)}_{Q}
    }
    & \multicolumn{3}{C|}{
      {C}^{(2 )}_{Q}
    }
    \\
    &
    {\pi} & {K} & {\eta} &
    {\pi} & {K} & {\eta} &
    {\pi} & {K} & {\eta}
    \\
    \hline
    \hline
    (\frac12,\,+2)   &
    0 & -\frac{1}{2} & \frac{1}{3} & 0 & -4 & 0 & 0 & -6 & 0 \\
    \hline
    \hline
    (0,\,+1)         &
   0 & \frac{1}{2} & -\frac{1}{6} & 0 & 8 & 0 & 6 & 6 & 0 \\
    \hline
    &
   0 & 0 & \frac{1}{2 \sqrt{3}} & 0 & 2 \sqrt{3} & 2 \sqrt{3} & \frac{3}{2 }\sqrt{3} & 3\sqrt{3} & \frac{3}{2 }\sqrt{3} \\
    \hline
    &
   0 & 0 & 0 & 0 & 0 & 0 & 0 & 0 & 0 \\
    \hline
    \hline
    (1,\,+1)         &
  0 & 0 & 0 & 0 & 0 & 0 & 0 & 0 & 0 \\
    \hline
    &
 0 & 0 & \frac{1}{6} & 2 & 2 & 0 & \frac{3}{2} & 3 & \frac{3}{2} \\
    \hline
    &
 -\frac{1}{2} & \frac{1}{2} & 0 & 0 & 0 & 0 & -2 & 2 & 0 \\
    \hline
    \hline
    (\frac12,\,0)     &
 \frac{1}{2} & 0 & -\frac{1}{6} & 8 & 0 & 0 & 8 & 4 & 0 \\
    \hline
    &
 0 & 0 & 0 & 0 & 0 & 0 & 0 & 0 & 0 \\
    \hline
    &
 0 & 0 & \frac{1}{2 \sqrt{6}}  & \sqrt{6} & \sqrt{6} & 0 & \frac{3}{2 }\sqrt{\frac{3}{2}} & 3\sqrt{\frac{3}{2}} & \frac{3}{2 }\sqrt{\frac{3}{2}} \\
    \hline
    &
  0 &  0 & 0 & 0 & 0 & 0 & 0 & 0 & 0 \\
    \hline
    &
 0 & 0 & -\frac{1}{2 \sqrt{6}} & 0 & - \sqrt{6} & - \sqrt{6} & -\frac{3}{2 }\sqrt{\frac{3}{2}} & -3\sqrt{\frac{3}{2}} & -\frac{3}{2 }\sqrt{\frac{3}{2}} \\
    \hline
    &
  0 & \frac{1}{2} & -\frac{1}{3} & 0 & 4 & 0 & 0 & 6 & 0 \\
    \hline
    \hline
    (\frac32,\,0)    &    
 -\frac{1}{4} & 0 & \frac{1}{12} & -4 & 0 & 0 & -4 & -2 & 0 \\
    \hline
    \hline
    (0,\,-1)         &
 \frac{3}{4} & -\frac{1}{2} & -\frac{1}{12} & 0 & 4 & 0 & 6 & 0 & 0 \\
    \hline
    \hline
    (1,\,-1)         &
 \frac{1}{4} & -\frac{1}{2} & \frac{1}{12} & 0 & -4 & 0 & -2 & -4 & 0 \\
    \hline
  \end{tabular}
  \caption{
  The coefficients $C^{(0)}_{Q}$, $C^{(1)}_{Q}$ and $C^{(2)}_{Q}$
    from (\ref{def-ClebschCQ012}).
  }
  \label{tab:ClebschC012Q}
\end{table}

The following Clebsch identities are useful in deriving the renormalization scale invariance of the sum of all third order terms.
\begin{eqnarray}
&&  \sum_{Q, LR} \,C^{(t)}_{Q,LR}\,m_Q^2 = \sum_Q C^{(0)}_Q\,m_Q^2 \,,
\nonumber\\
&&  \sum_{H, PQ} \big(C^{(t)}_{H,PQ}- C^{(u)}_{H,PQ}\big)\,(m_{ab}^2) = \frac{3}{2}\,\sum_Q C^{(1)}_Q\,m_Q^2 \,,
\nonumber\\
&&  \sum_{H, PQ} \big(C^{(t)}_{H,PQ}- C^{(u)}_{H,PQ}\big)\,(m_P^2+m_Q^2) = \sum_Q C^{(2)}_Q\,m_Q^2 \,,
\nonumber\\
&&  \sum_{H, PQ} \big(C^{(t)}_{H,PQ}- C^{(u)}_{H,PQ}\big) = 3\, C_{WT} \,,
\label{def-ClebschCQ012}
\end{eqnarray}
with the Clebsch $C_Q^{(1)}$ and $C_{WT}$ specifying the $g_1$ and $g_2$ terms in the third order tree-level contributions (\ref{res-tree-Q123}). The form of $C^{(0)}_Q$,  $C^{(1)}_Q$ and $C^{(2)}_Q$ in Tab. \ref{tab:ClebschC012Q} confirms the necessity of the particular manner how the heavy tadpole terms $\bar I_L, \bar I_R$ and $\bar I_H$ contribute in $ \bar J^{(t)}_{Q,LR}(t)$ and $\bar J^{(+)}_{H,PQ}(s,u)$. Altogether it holds
\begin{eqnarray}
&&  \frac{f^2\, \mu^2}{M}\,\frac{\text{d}}{\text{d}\mu^2}\, g_1 = -\frac{1 - g_P^2}{5
12\,\pi^2}  \,,\qquad 
\frac{f^2\, \mu^2}{M}\,\frac{\text{d}}{\text{d}\mu^2}\, g_2 = -\frac{1- g_P^2}{512\,\pi^2} \,,
\qquad \frac{\text{d}}{\text{d}\mu^2}\, g_{3,4,5} = 0\,.
\label{res-g12345-running}
\end{eqnarray}

We note that in the chiral domain we expect further suppressed results with $ \bar J^{(t)}_{Q,LR}(s,u) \sim Q^5$ and  $ \bar J^{(s,t,u )}_{H,PQ}(s,u) \sim Q^4$. 
Here we encounter superficial power-counting violating terms, which one may or may not eliminate in part by a subtraction scheme similar to the one developed already in  \cite{Lutz:2020dfi,Sauerwein:2021jxb}. 
Let us discuss $ \bar J^{(t)}_{Q,LR}(s,u)$ in more detail. Here we note that the tadpole term $\bar I_Q$ does not depend on either $M_L$ nor $M_R$. In turn, it is more reasonable to keep such superficially power-counting violating (in the chiral domain only) terms. In particular, since their effect cannot be absorbed into existing counter terms. An analogous  phenomenon occurs in the bubble-type contributions. In fact, the sum of both, tadpole and bubble terms, confirms the expected scaling behavior in that domain. 
The remaining terms proportional to the scalar triangle show their expected scaling unambiguously in the chiral domain, so there is no need for any additional subtraction in any case. However, it is useful to observe that without the subtraction terms in 
$  \bar I_{QL}(M_a^2),  \bar I_{QR}(M_b^2)$ and 
$\bar I_{Q,LR}(t)$, there would be a contribution at small quark masses proportional to $(s-u)\,\log (r/(1+r))\,C_{WT}/f^2$, that acts as an unwanted renormalization of the Tomozawa-Weinberg term in (\ref{res-tree-Q123}).

\clearpage

\section{Scattering with Box diagrams}
 
\begin{figure}[t]
\center{
\includegraphics[keepaspectratio,width=0.71\textwidth]{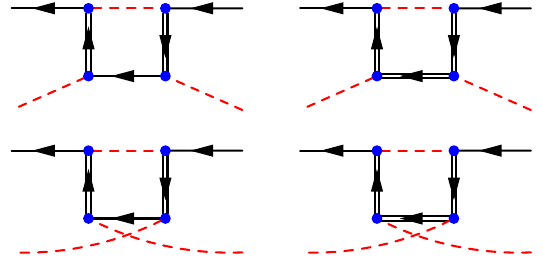} }
\vskip-0.2cm
\caption{\label{fig:7} Dashed lines stand for pion, kaon or eta mesons and solid and double-solid lines for charmed mesons with $J^P = 0^-$ and $J^P = 1^-$ respectively. The vertices are from \cite{Lutz:2022enz}.}
\end{figure}

We consider now the box diagrams of Fig.\ \ref{fig:7}.   
The four contributions in (\ref{def-Tbox}) correspond to the two rows in Fig.\ \ref{fig:7} in consecutive order. The first term is characterized by its s- and t-channel, the second by its u- and t-channel unitarity cuts. The expressions can conveniently be factorized into universal loop functions and Clebsch coefficients,
\begin{eqnarray}
&& f^4 \,T^{\rm box}(s,t,u) = g_P^2\, \sum_{L R\in [1^-]}  \Big[  g_P^2\,\sum_{Q,H \in [0^-]}\,J^{(0)}_{QH,LR}(s,t) + \tilde g_P^2 \,\sum_{Q,H \in [1^-]}\,J^{(1)}_{QH,LR}(s,t) \Big] \,C^{(s)}_{QH,LR}
\nonumber\\
&& \qquad \qquad \qquad \;\, +\,g_P^2\,\sum_{LR \in [1^-]}  \Big[  g_P^2\,\sum_{Q,H \in [0^-]}\,J^{(0)}_{QH,LR}(u,t) + \tilde g_P^2 \,\sum_{Q,H \in [1^-]}\,J^{(1)}_{QH,LR}(u,t)\Big] \,C^{(u)}_{QH,LR} \,,
\nonumber\\
&&  \sum_{QH} C^{(s)}_{QH,LR} \Big|_{ab} = - \sum_{c \leftrightarrow QH}  C^{(u)}_{L}  \Big|_{ac} C^{(u)}_{R} \Big|_{cb}\, ,
\label{def-Tbox} 
\end{eqnarray}
where the Clebsch coefficients depend on the isospin and strangeness of the intermediate $(Q,H,L,R)$ and external $(a,b)$ mesons. The loop functions are expressed in terms of the internal masses, $m_Q, M_H$ and $M_L, M_R$ and external masses $m^2_b=q^2, M^2_b= p^2$ and $m^2_a=\bar q^2, M^ 2_a= \bar p^2$. In (\ref{def-Tbox}) it holds  $L,R \in [1^-]$ always, but $H \in [0^-]$ in $J^{(0)}_{Q H,LR}$ and  $H \in [1^-]$ in $J^{(1)}_{QH,LR}$.

We discuss the computation of the loop functions. It suffices to specify the two s-channel loop functions. The u-channel expressions follow from the s-channel loop by the crossing replacement $s \leftrightarrow u$ as is already implied in (\ref{def-Tbox}). We find
\begin{eqnarray}
&&  J^{(0)}_{QH,LR}(s,t) =\int\frac{d^d l}{(2\pi)^d}\frac{i\,\mu^{4-d}}{l^2-m_Q^2}\, 
\Big( l_{ \bar \alpha}  \, \bar p_{\bar \beta}\,
S^{ \bar \alpha \bar \beta, \bar \mu \bar \nu}_L(l+\bar p)\, (w+ l)_{\bar \mu}\,\bar q_{\bar \nu} 
\nonumber\\
	&&\qquad\qquad\qquad\quad \times  \,
S_H(l+w)\,(w+ l)_{ \mu}\,q_{ \nu} \,
S^{\mu \nu ,\alpha \beta }_R(l+p) \,l_{ \alpha}  \, p_{\beta}
	\Big)	\,,
\nonumber\\
&&  J^{(1)}_{QH,LR}(s,t) =\frac{1}{16}\int\frac{d^d l}{(2\pi)^d}\frac{i\,\mu^{4-d}}{l^2-m_Q^2}\,
\Big( l_{ \bar \alpha}  \, \bar p_{\bar \beta}\,
S^{ \bar \alpha \bar \beta, \bar \mu \bar \nu}_L(l+\bar p)\,\Big[  
 \bar q^{\bar \tau} \,(l+ \bar p)_{\bar \nu}\,
\epsilon_{\bar \mu \bar \tau  \bar \sigma \bar \delta }
\nonumber\\
&& \qquad \qquad \qquad \;\;
-\,\epsilon_{\bar \mu \bar \nu  \bar \sigma \bar \tau }\,\bar q^{\bar \tau} \,(l+ w)_{\bar \delta}\Big] \,
S^{ \bar \sigma \bar \delta, \sigma \delta }_H(l+w)\,\Big[\epsilon_{ \sigma \delta  \mu \tau }\,(l + p)_{ \nu}  \,q^\tau\, 
\nonumber\\
&& \qquad \qquad \qquad \;\;-\, 
(l +w)_{ \delta}  \,q^\tau\,\epsilon_{ \sigma \tau  \mu \nu } \Big]\,S^{\mu \nu ,\alpha \beta  }_R(l+p)  \,
l_{ \alpha}\,p_{ \beta} 	\Big)	\,,
\nonumber\\
&& S_H(l+w) = \frac{1}{(l+w)^2-M_H^2 }\,.
\label{def-J-box}
\end{eqnarray}

Our list of scalar integrals (\ref{def-scalar-integrals}) and (\ref{def-n-triangle-QH}, \ref{def-n-triangle-PQ})
needs an obvious extension with a scalar box integral 
\begin{eqnarray}
&& I_{QH,LR}(s,t) =  \int \frac{d^d l}{(2\pi)^d}\,\frac{- i\,\mu^{4-d} }{l^2- m_Q^2}\,S_H(l+w)\, S_L(l+\bar p) \,S_R(l+p) \,.
 \label{def-scalar-box}
\end{eqnarray}
Like in the case of the diagrams of Fig.\ \ref{fig:6} the proper evaluation of diagrams in Fig.\ \ref{fig:7} 
asks for an extension of the Passarino--Veltman functions. In Appendix D it is proven that the following set 
\begin{eqnarray}
 && I_{QH,LR}^{(n,m)}(s,t) =  \frac{1}{16\,\pi^2} \int_0^1 d x  \int_0^{1-x} d y \,  \int_0^{1-x-y}\,d z\,\frac{x^n\,y^m}{[F_{QH,LR}(x,z,y)]^2  } \,,
 \nonumber\\
 && F_{QH,LR}(x,z,y) = 
  m_Q^2 - x\, (\bar p^2 -M^2_L + m_Q^2) 
- z\,(w^2-M_H^2+ m_Q^2) - y\,(p^2-M_R^2+ m_Q^2)
\nonumber\\ 
&& \qquad \qquad  \;\, + \, x^2\,\bar p^2 + z^2\,w^2 + y^2\, p^2  + \,2\,x\,y\,( \bar p \cdot   p) + 2\,x\,z\,( \bar p \cdot  w)  + 2\,y\,z\,( w \cdot  p) \,,
 \label{def-IQHLR-nm}
 \end{eqnarray}
implies an unambiguous decomposition of the loop diagrams of Fig.\ \ref{fig:7} void of superficial singularities. By construction it holds $I_{QH,LR}^{(0,0)}(s,t) = I_{QH,LR}(s,t)$. Like for the bubble and triangle basis functions we implement a finite subtraction as to be consistent with power-counting expectations in the chiral domain $m_Q\ll \tilde M- M$. Altogether we find
\begin{eqnarray}
&& \bar I^{(m,n)}_{QH,LR } =I^{(m,n)}_{QH,LR } - \frac{\gamma^{H}_{mn}}{16\,\pi^2\,M^4}\sim Q^2\,, 
\nonumber\\
&&  \gamma^{H\in [0^-]}_{00}= \frac{\log (1+ r)}{r^2}\,, \quad  
 \gamma^{H\in [0^-]}_{10}=  \gamma^{H\in [0^-]}_{01}=  -\frac{\log r}{6} +
 \frac{1}{6\,r^3}\,\Big(2-r\, - (2-r^3)\,\log (1+r)\Big) \,,
\nonumber\\
&&  \gamma^{H\in [0^-]}_{11} =   r\,\frac{\log r}{15}
+ \frac{1}{60\,r^4}\,\Big(-6\,r +3\,r^2-2\,r^3+4\,r^4 + (6-r^5)\,\log (1+r)\Big) \,,
\nonumber\\ 
&& \gamma^{H\in [0^-]}_{20} = \gamma^{H\in [0^-]}_{02} = 2\,\gamma^{H\in [0^-]}_{11}  \qquad \qquad \qquad \qquad   \qquad \qquad \qquad \!  {\rm with}\qquad \qquad 
r= \frac{\tilde M^2 - M^2}{M^2} \,,
\nonumber\\ \nonumber\\
&&  \gamma^{H\in [1^-]}_{00}= \frac{1}{2\,r\,(1+r)}\,, \qquad  
 \gamma^{H\in [1^-]}_{10}=  \gamma^{H\in [1^-]}_{01}=  -\frac{1}{6\,(1+r)} + \frac{1}{6}\,\log \frac{1+r}{r} \,,
\nonumber\\
&&  \gamma^{H\in [1^-]}_{11} =   \frac{1}{24}\,\frac{ 1+2\,r}{1+r}
- \frac{r}{12}\, \log \frac{1+r}{r}\,,  \qquad \qquad 
 \gamma^{H\in [1^-]}_{20} = \gamma^{H\in [1^-]}_{02} = 2\,\gamma^{H\in [1^-]}_{11} \,.
\label{def-r-box}
\end{eqnarray}
The two box-loop functions, as properly expressed in 
the particular set of basis functions, are
\begin{eqnarray}
&&\bar J^{(0)}_{QH,LR} (s,t) = \frac{1}{192}\,\Big(( s-u)^2-2\,(t -m_{ab}^2)\, M_{ab}^2 \Big) \Big[ \bar I_{QL}( M_a^2)+\bar I_{QR}(M_b^2)\Big] 
-\frac{(s-u)^2}{64}\,\bar I_{QH}(s)
\nonumber\\
&& \qquad +\,\frac{(s-u)^2}{128}\,
 \Big(s-u +r\,M^2_{ab} \Big)\Big[\bar I_{QH,R}(s)+ \bar I_{L,QH}(s) \Big]
\nonumber\\
&& \qquad + \,\frac{s-u}{128}\,\Big((s-u)^2 - 4\,M_{ ab}^2\,\,m_{ab}^2 \Big)\,\Big[\bar I^{(1)}_{QH,R}(s)+ \bar I^{(1)}_{L,QH}(s) \Big]
\nonumber\\
&& \qquad +\,\frac{1}{96}\,\Big( -(s-u)^3 + r\,M^2_{ab}\,M^2_{ab} \, (t- m_{ab}^2) 
\nonumber\\
&& \qquad \qquad -\, M^2_{ab}\, (s-u)\, (t-m_{ab}^2 + r \,(s-u)/2 )\Big)\,\bar I_{Q,LR}(t)
\nonumber\\
&& \qquad +\,\frac{1}{192}\,\Big( -(s-u)^4 +  M_{ab}^2\,(s-u)^2\,\big(2\,m_Q^2 -(t-m_{ab}^2)  -3\,r\,(s-u)/2 \big)
\nonumber\\
&& \qquad \qquad +\, M^2_{ab}\, M^2_{ab}\, \big( 8\,m_Q^2\,(t-m_{ab}^2) -3\, r^2 \,(s-u)^2/4  \big)\Big)\,\bar I_{QH,LR}(s,t)
\nonumber\\
&& \qquad -\,\frac{s-u}{128}\,  \Big(s-u +r\,M^2_{ab} \Big)\,\Big((s-u)^2 +2\,(t-2\, m_{ab}^2)\,M^2_{ab} \Big)\,\Big[\bar I^{(1,0)}_{QH,LR}(s,t)+ \bar I^{(0,1)}_{QH,LR}(s,t) \Big]
\nonumber\\
&& \qquad + \frac{1}{192}\,\Big( 
12\, \big[ (m_a^2-m_b^2 )\,M_{ab}^2 \big]^2 - \big[(s-u)^2 -4\,m_{ab}^2\, M_{ab}^2  \big] 
\, \big[(s-u)^2 +2\,(t-2\,m_{ab}^2)\, M_{ab}^2  \big]
\nonumber\\
&& \qquad \qquad -\, 4\,t^2\, M_{ab}^2\,M_{ab}^2
\Big) \, \bar I^{(1,1)}_{QH,LR}(s,t) 
\nonumber\\
&& \qquad -\,\frac{(s-u)^2 - 4\,m_{ab}^2\,M^2_{ab} }{384}\,\Big((s-u)^2 + 4 \, (t-m_{ab}^2 )\,\,M_{ab}^2 \Big)\,\Big[\bar I^{(2,0)}_{QH,LR}(s,t)+ \bar I^{(0,2)}_{QH,LR}(s,t) \Big]
\nonumber\\
&& \qquad  +\,\frac{m_a^2-m_b^2}{96}\,M_{ab}^2\,\Big((s-u)^2 + 4\,(t- m_{ab}^2)\,M_{ab}^2 \Big)\,\Big[\bar I^{(2,0)}_{QH,LR}(s,t)- \bar I^{(0,2)}_{QH,LR}(s,t) \Big] 
\nonumber\\
&& \qquad +\,\frac{s-u}{64}\, (m_a^2-m_b^2)\, M^2_{ab}\,\Big\{2\,\bar I^{(1)}_{Q,\bar LR}(t) - 2\,\bar I^{(1)}_{Q, L \bar R}(t) - 2\,\bar I^{(1)}_{L,QH}(s) +2\,
\bar I^{(1)}_{QH,R}(s) 
\nonumber\\
&& \qquad \qquad +\, \big(s-u + r\,M_{ab}^2 \big)\,\Big[\bar I^{(1,0)}_{QH,LR}(s,t)- \bar I^{(0,1)}_{QH,LR}(s,t) \Big]\Big\} +\, {\mathcal O} \left( Q^4\right) \,,
\end{eqnarray}
and 
\begin{eqnarray}
&& \bar J^{(1)}_{QH,LR} (s,t) = 
\frac{m_{ab}^2-t}{16}\,M_{ab}^2\,\bar I_{QH}(s)
\nonumber\\
&& \qquad -\,\frac{1}{192}\,\Big(( s-u)^2-2\,(t-m_{ab}^2)\, M_{ab}^2 \Big) \Big[ \bar I_{QL}( M_a^2)+\bar I_{QR}(M_b^2)\Big] 
\nonumber\\
&& \qquad +\,\frac{1}{128}\,
 \Big(r\,M_{ab}^2 \, (s-u)^2 - (s-u)^3 + 4\,r\, M_{ab}^2\, M_{ab}^2 \,(t-m_{ab}^2 ) \Big)\Big[\bar I_{QH,R}(s)+ \bar I_{L,QH}(s) \Big]
\nonumber\\
&& \qquad - \,\frac{s-u}{128}\,\Big((s-u)^2 +(t- 4\,m_{ab}^2) \,M_{ ab}^2\Big)\,\Big[\bar I^{(1)}_{QH,R}(s)+ \bar I^{(1)}_{L,QH}(s) \Big]
\nonumber\\
&& \qquad +\,\frac{1}{96}\,\Big(
(s-u)^3 +  (s-u) \,\big(t- m_{ab}^2 -  r\,( s-u)/2 \big)\,M_{ab}^2 
\nonumber\\
&& \qquad \qquad -\,  2\,r\, M_{ab}^2\, M_{ab}^2\, (t - m_{ab}^2 )\Big)\,\bar I_{Q,LR}(t)
\nonumber\\
&& \qquad +\,\frac{1}{192}\,\Big( (s-u)^4 +  M_{ab}^2\,(s-u)^2\,\big(4\,m_Q^2 +(t-m_{ab}^2)  - r\,(s-u)/2 \big) 
+4\,r^2\,(\bar q \cdot q)\,\big(M_{ab}^2 \big)^3
\nonumber\\
&& \qquad \qquad +\, M^2_{ab}\, M^2_{ab}\, \big( ( 16\,m_Q^2-2\,r(s-u))\,(t-m_{ab}^2) - r^2 \,(s-u)^2/2  \big)\Big)\,\bar I_{QH,LR}(s,t)
\nonumber\\
&& \qquad +\,\frac{(s-u)^2}{128}\, \Big((s-u)^2 + 2\,(t-2\, m_{ab}^2)\,M^2_{ab} \Big)\,\Big[\bar I^{(1,0)}_{QH,LR}(s,t)+ \bar I^{(0,1)}_{QH,LR}(s,t) \Big]
\nonumber\\
&& \qquad +\, \frac{1}{192}\,\Big( 
\big[(s-u)^2 -4\,m_{ab}^2\, M_{ab}^2  \big] 
\, \big[(s-u)^2 + 2\,(t-2\,m_{ab}^2)\, M_{ab}^2  \big]
\nonumber\\
&& \qquad \qquad + \, 4\,\big[ t^2 -3 \,(m_a^2-m_b^2)^2 \big]\, M_{ab}^2 \,M_{ab}^2
\Big) \, \bar I^{(1,1)}_{QH,LR}(s,t) 
\nonumber\\
&& \qquad +\,\frac{(s-u)^2 - 4\,m_{ab}^2\,M^2_{ab} }{384}\,\Big((s-u)^2 + 4 \, (t -m_{ab}^2)\,\,M_{ab}^2 \Big)\,\Big[\bar I^{(2,0)}_{QH,LR}(s,t)+ \bar I^{(0,2)}_{QH,LR}(s,t) \Big]
\nonumber\\
&& \qquad  -\,\frac{m_a^2-m_b^2}{96}\,M_{ab}^2\,\Big((s-u)^2 + 4\,(t- m_{ab}^2)\,M_{ab}^2 \Big)\,\Big[\bar I^{(2,0)}_{QH,LR}(s,t)- \bar I^{(0,2)}_{QH,LR}(s,t) \Big] 
\nonumber\\
&& \qquad -\,\frac{s-u}{128}\, (m_a^2-m_b^2)\, M^2_{ab}\,\Big\{ 4\,\big(\bar I^{(1)}_{Q,\bar LR}(t) - \bar I^{(1)}_{Q, L \bar R}(t) \big) - 3\,\big( \bar I^{(1)}_{L,QH}(s)-
\bar I^{(1)}_{QH,R}(s) \big)
\nonumber\\
&& \qquad \qquad +\,2\,(s-u )\,\Big[\bar I^{(1,0)}_{QH,LR}(s,t)- \bar I^{(0,1)}_{QH,LR}(s,t) \Big] \Big\} +\, {\mathcal O} \left( Q^4\right) \,,
\end{eqnarray}
where we observe that there is no renormalization scale-dependence generated at this order. 
Corresponding expressions at chiral order four can be found in Appendix E. The merit of our results rests on their compatibility with the expectation of power-counting  
rules, while keeping the on-shell meson masses throughout. Since the scalar basis functions are not further expanded our approximated renormalized expressions enjoy the correct analytic structure as it is requested in local quantum field theories from the micro-causality condition. Since we started with un-renormalized expressions that suffer from large power-counting violating contributions it is absolutely crucial to eliminate the latter in a manner that is sufficiently effective so that a chiral expansion has convincing convergence properties. 

While some readers may be worried about the complexity of our expressions, in particular the fourth order results in Appendix E, we note that a direct decomposition of (\ref{def-J-box}) leads to more than a thousand terms, that cannot be properly expanded into chiral moments. Only with our novel scheme such contributions are cast into useful input for coupled-channel computations, the main target of our developments.

\clearpage

\section{Summary and outlook}

In this work we studied triangle- and box-type contributions to two-body scattering in the context 
of the chiral Lagrangian with a heavy field. The formal developments are detailed at the hand of the open-charm system of QCD, for which we considered third and fourth order contributions formulated in terms of on-shell hadron masses. The challenge has been to explore a novel technique that allows such computations in compliance with chiral-power-counting rules. While such a framework was suggested in application of the Passarino--Veltman reduction scheme, the consideration of triangle and box diagrams leads to additional technical complications that asked for a major extension. 

The problem is well-known in the community: the decomposition of a given loop function into the set of 
scalar basis functions of Passarino and Veltman, avoids superficial singularities only if correlations amongst the basis functions at        
specific kinematical conditions are kept exactly. 
How does this go together with the request to apply a chiral expansion to the loop functions? The simple idea behind such a decomposition of the loop functions, is the possibility to apply a chiral expansion to the coefficient functions, without touching the basis functions themselves. The latter have 
more complicated properties dictated by the micro-causality condition of local quantum field theories, so that it is advantageous to keep their original form.  
The crucial observation pointed out long ago by one of the authors is that power-counting violating terms arise in the relevant basis functions only, that are ultraviolet divergent. Therefore, a suitable subtraction scheme in the Passarino--Veltman functions suffices to restore counting rules upon renormalization.

In the current work we overcome the above-described  challenge by using an extended basis set, constructed such that kinematical constraints are avoided altogether and at the same time consistency with power-counting expectations is observed. We provided a proof that our decomposition is unique and exemplified our novel scheme with explicit expressions at chiral order three and four in the open-charm meson sector of chiral QCD. 

In the next step we will use our results  for an improved description of  s- and p-wave scattering of Goldstone bosons off charmed meson states. This will be important for on-going Lattice QCD computations on CLS ensembles, where owing to their large variety of $\beta$ values a better control of discretization effects is expected. Here a quantitative 
success in the p-wave phase shifts may require the consideration of the left-hand cut contributions in the generalized potential as predicted by the chiral Lagrangian in terms of triangle and box contributions. Moreover, 
with our developments the path for an improved generalized potential approach to meson-baryon scattering based on the chiral Lagrangian is paved. In particular, the left-hand cut contributions can be extracted systematically from expressions as implied by our novel method. 

\clearpage 

\section*{Appendix A: Terms in the chiral Lagraangian  }
The relevant terms in the chiral Lagrangian as used in \cite{Lutz:2022enz}
\begin{eqnarray}
&& {\mathcal L}^{(1)} = 2\,g_P\,\Big\{D_{\mu \nu}\,U^\mu\,(\CD^\nu \bar D)
 - (\CD^\nu D )\,U^\mu\,\bar D_{\mu \nu} \Big\}
\nonumber\\
&& \qquad   -\,\frac{i}{2}\,\tilde g_P\,\epsilon^{\mu \nu \alpha \beta}\,\Big\{
D_{\mu \nu}\,U_\alpha \,
(\CD^\tau \bar D_{\tau \beta} )
+ (\CD^\tau D_{\tau \beta})\,U_\alpha\,\bar D_{\mu \nu}) \Big\} \,,
\nonumber\\
&&\mathcal{L}^{(2)}=-\big( 4\,c_0-2\,c_1\big)\, D \,\bar{D}  \,{\tr} \chi_+ -2\,c_1\,D \,\chi _+\,\bar{D}
\nonumber\\
&& \qquad  - \, \big(8\,c_2+4\,c_3\big)\,D\,\bar{D}\,{\tr }U_{\mu }\,U^{\mu } + 4\,c_3\, D \,U_{\mu }\,U^{\mu  }\,\bar{D} 
\nonumber\\
&& \qquad  -\, \big( 4\,c_4+2\,c_5\big)\, ({\CD_\mu } D)\,({\CD_\nu }\bar{D}) \,{\tr} [ U^{\mu }, \,U^{\nu  }]_+ /M^2
+2\,c_5\,({\CD_\mu } D)\,[ U^{\mu }, \,U^{\nu  }]_+({\CD_\nu }\bar{D})/M^2
\nonumber\\
&& \qquad  -\,i\, c_6\,\epsilon ^{\mu \nu \rho \sigma }\,\big(D\,[U_{\mu },\, U_{\nu }^{ }]_-\bar{D}_{\rho \sigma }-D_{\rho \sigma }\,[U_{\nu }^{ }, U_{\mu }]_-\bar{D}\big)
\nonumber\\
&& \qquad  +\,\big(2\,\tilde{c}_0-\tilde{c}_1\big)\,D^{\mu \nu }\,\bar{D}_{\mu \nu }\,{\tr}\chi _+
 +\tilde{c}_1\,D^{\mu \nu }\,\chi _+\,\bar{D}_{\mu \nu }
\nonumber\\
&&\qquad + \big( 4\,\tilde{c}_2+2\,\tilde{c}_3\big)\,D^{\alpha \beta }\,\bar{D}_{\alpha \beta }\,{\tr }U_{\mu }\,U^{\mu }
-2\,\tilde{c}_3\,D^{\alpha \beta }\,U_{\mu }\,U^{\mu  }\,\bar{D}_{\alpha \beta }
\nonumber\\
&& \qquad  +\,\big(2\,\tilde{c}_4+\tilde{c}_5\big)\, ({\CD_\mu }D^{\alpha \beta })\,({\CD_\nu }\bar{D}_{\alpha \beta } )\,{\tr} [U^{\mu }, \,U^{\nu  }]_+/\tilde M^2
\nonumber\\
&&\qquad - \, \tilde{c}_5 \,({\CD_\mu } D^{\alpha \beta })\,[U^{\mu },\, U^{\nu  }]_+ ({\CD_\nu }\bar{D}_{\alpha \beta })/\tilde M^2
+4\,\tilde{c}_6\,D^{\mu \alpha }\,[U_{\mu },\, U^{\nu }]_-\bar{D}_{\nu \alpha } \,,
\nonumber\\
&& \mathcal{L}^{(3)}=
 \, 4\,g_1\,{D}\,[\chi_-,\,{U}_\nu]_- (\CD^\nu \bar{D})/M
  - 4\,g_2\,{D}\,[{U}_\mu,\,([\CD_\nu,\,{U}^\mu]_- + [\CD^\mu,\,{U}_\nu]_-)]_-( \CD^\nu\bar{D})/M
\nonumber\\ 
&&   \qquad  - \,4\,g_3\,{D}\,[{U}_\mu,\,[\CD_\nu,\,{U}_\rho]_-]_-\,[\CD^\mu,\,[\CD^\nu,\,\CD^\rho]_+]_+\bar{D}/M^3
\nonumber\\
&&\qquad   -\,2\,i\,g_4 \,\epsilon_{\mu \nu \rho \sigma} \,(\CD_\alpha D )\, [U^\mu,\,([\CD^\alpha,\,U^\nu]_- +[\CD^\nu,U^\alpha]_-)]_+\bar D^{\rho\sigma}/M 
\nonumber\\
&&\qquad   -\,2\,i\,g_5 \,\epsilon_{\mu \nu \rho \sigma} \,(\CD_\alpha D ) \, {\tr } [U^\mu,\,([\CD^\alpha,\,U^\nu]_-+[\CD^\nu,U^\alpha]_-)]_+\bar D^{\rho\sigma}/M   
\nonumber\\ 
&&  \qquad  - \,2\,\tilde g_1\,  D_{\mu\nu} \,[\chi_-,\,U_\rho ]_- (\CD^\rho\bar D^{\mu\nu}) /\tilde M    + 2\,\tilde g_2\, D_{\mu\nu} \,[U^\sigma, \,([\CD_\rho,\, U_\sigma ]_- + [\CD_\sigma,\,U_\rho ]_-) ]_- ( \CD^\rho \bar D^{\mu\nu} ) /\tilde M 
\nonumber\\
&&  \qquad
    +\,2\,\tilde g_3\, D_{\alpha\beta}\,[U_\mu,\,[\CD_\rho,\,U_\nu]_-  ]_- [\CD^\mu,\,[\CD^\rho,\,\CD^\nu]_+ ]_+\bar D^{\alpha\beta} /\tilde M^3
\nonumber\\
&&\qquad  -\,2\,\tilde g_4\,\Big( D^{\alpha\beta}\,[U_\alpha,\,([\CD_\beta,\,U_\mu]_-+[\CD_\mu,\,U_\beta]_-)]_+( \CD_\nu\bar D^{\mu\nu} )  
\nonumber\\
&& \quad \qquad \qquad -\,D^{\alpha\beta}\,[U_\nu,\,([\CD_\beta,\,U_\mu]_-+[\CD_\mu,\,U_\beta]_-)]_+(\CD_\alpha\bar D^{\mu\nu} )\Big) /  \tilde M  
\nonumber\\
&&\qquad  -\,2\,\tilde g_5\,\Big( D^{\alpha\beta} \, {\tr }[U_\alpha,\,([\CD_\beta,\,U_\mu]_-+[\CD_\mu,\,U_\beta]_-)]_+(\CD_\nu\bar D^{\mu\nu}  ) 
\nonumber\\
&& \quad \qquad \qquad -\,D^{\alpha\beta}\, {\tr }[U_\nu,\,([\CD_\beta,\,U_\mu]_-+[\CD_\mu,\,U_\beta]_-)]_+(\CD_\alpha\bar D^{\mu\nu} )\Big) /  \tilde M  +{\rm h.c.}
   \,,
\label{chiral Lagrangian}
\end{eqnarray}
where 
\begin{eqnarray}
&& U_\mu = {\textstyle \frac{1}{2}}\,e^{-i\,\frac{\Phi}{2\,f}} \left(
    \partial_\mu \,e^{i\,\frac{\Phi}{f}} \right) e^{-i\,\frac{\Phi}{2\,f}} \,, \qquad \qquad 
    \Gamma_\mu ={\textstyle \frac{1}{2}}\,e^{-i\,\frac{\Phi}{2\,f}} \,\partial_\mu  \,e^{+i\,\frac{\Phi}{2\,f}}
+{\textstyle \frac{1}{2}}\, e^{+i\,\frac{\Phi}{2\,f}} \,\partial_\mu \,e^{-i\,\frac{\Phi}{2\,f}}\,,
\nonumber\\
&& \chi_\pm = {\textstyle \frac{1}{2}} \left(
e^{+i\,\frac{\Phi}{2\,f}} \,\chi_0 \,e^{+i\,\frac{\Phi}{2\,f}}
\pm e^{-i\,\frac{\Phi}{2\,f}} \,\chi_0 \,e^{-i\,\frac{\Phi}{2\,f}}
\right) \,, \qquad \chi_0 =2\,B_0\, {\rm diag} (m_u,m_d,m_s) \,,
\nonumber\\
&& \CD_\mu \bar D = \partial_\mu \, \bar D + \Gamma_\mu\,\bar D \,, \qquad \qquad \qquad \quad \;\; 
\CD_\mu D = \partial_\mu \,D  - D\,\Gamma_\mu \,.
\label{def-chi}
\end{eqnarray}

\section*{Appendix B: Scalar triangle loops}
\label{test}

We begin with an over-complete basis of scalar triangle-loop terms of the generic form 
\begin{eqnarray}
 J^{(a)}_{fki} =\int  \frac{d^d l}{(2\pi)^d}\,\frac{(\bar p \cdot l)^f \,\mu^{4-d}\,(l\cdot Q)^a\,(l^2 )^k \,(l\cdot p)^i }
{((l-\bar p)^2- M^2_L)\,(l^2-m_G^2)\,((l-p)^2-M^2_R)} \,,
\label{def-Cfkia}
\end{eqnarray}
into which each of the introduced diagram expressions (\ref{def-J-triangle}) and (\ref{def-J-triangle-appendix}) can be decomposed upon performing the contraction of the Lorentz indices. Without loss of generality we may assume $f=k=i=0$ in the following. All other cases can be related to the particular choice study, where we assume $Q_\mu=\bar q_\mu +q_\mu $. Such a reduction generates additional bubble- and tadpole-type integrals only, which do not cause any complications related to the introduction of our basis integrals (\ref{def-n-triangle-QH}) and (\ref{def-n-triangle-PQ}). 

The target function is analyzed in terms of a conventional Feynman parameter ansatz
\begin{eqnarray}
&& i\, J^{(a)}_{000} =  \sum_{b=0}^{a/2}\sum_{m +n +2\,b= a}\,(Q\cdot \bar p)^m\,(Q^2)^{b}\,(Q\cdot p)^n\,C^{(b)}_{mn}\,I_b(m,n) \,\quad {\rm with } \quad n \geq 0  \; {\rm and} \; m \geq 0 \,, 
\nonumber\\
&& F(x,y) =  m_G^2 - x\, (\bar p^2 - M^2_L + m_G^2)  - y\,(p^2-M^2_R+ m_G^2)\,  + x^2\,\bar p^2 + y^2\, p^2 \,+ 2\,x\,y\,( \bar p \cdot  p) \,,
\nonumber\\
&&I_b(m,n) = i\,  \int \frac{d^d l}{(2\pi)^d}\,\int_0^1 d x \int_0^{1-x} d y \,\frac{2\,x^m\,(l^2)^{b}\,y^n}{(l^2 -F(x,y) )^3 } \,\mu^{4-d}\,,
\end{eqnarray}
with $F(x,y)=F_{G,LR}(x,y)$ of (\ref{def-n-triangle-PQ}) and some suitable  real-valued coefficients
\begin{eqnarray}
&& C^{(b)}_{mn} = \Bigg(\begin{array}{c} m+n+ 2\,b \\ 2\,b \end{array}\Bigg) \,
 \Bigg(\begin{array}{c} m+n \\ m \end{array}\Bigg) \, x_b \,, \qquad x_0 = 1\,,\qquad x_1= \frac{1}{d}\,, \qquad x_2 = \frac{3}{d\,(d+2)} \,,\quad 
\nonumber\\
 &&  x_3 =\frac{15}{d\,(d+2)\,(d+4)}  \,,\qquad  x_4 =\frac{105}{d\,(d+2)\,(d+4)\,(d+6)}\,,\qquad 
 x_{b+1}= \frac{2\,b+1}{d +2\,n}\,x_b\,,
 \nonumber\\
 && x_b = \frac{(2 \,b -1 ) !!}{2^b\,(b+ 1) ! } \,\Big( 1- \frac{d-4}{2}\,\Big( 1- \sum_{k=1}^{b+1} \,\frac{1}{k} \Big) + {\mathcal O }\big((d-4)^2\big) \Big) \,.
\end{eqnarray}
The summation over the integers $b, m,n$ start at zero. 
We split the integral into a convergent and scale-dependent piece with
\begin{eqnarray}
 &&  i\,J^{(a)}_{000} =  \frac{1}{16\,\pi^2}\int_0^1 d x \int_0^{1-x} d y \, \Bigg( \frac{N_a(x,y)}{F(x,y)  } + 
 L_a(x,y)\,\Big\{ D+ \log \frac{F(x,y)}{\mu^2} \Big\}\Bigg) + {\mathcal O} \left( d- 4 \right) \,,
  \nonumber\\
&& N_a(x,y) =   \sum_{b=0}^{a/2}\sum_{m +n +2\,b= a}\,(Q\cdot \bar p)^m\,(Q\cdot p)^n \,x^m\,y^n\,C^{(b)}_{mn} \,\big[ Q^2\,F(x,y) \big]^{b}\,, 
\nonumber\\
&& L_a(x,y) =  \sum_{b=1}^{a/2}\sum_{m +n +2\,b= a }\,\,(Q\cdot \bar p)^m\,(Q\cdot p)^n\,x^m\,y^n \,C^{(b)}_{mn} \,[Q^2\,F(x,y) ]^{b} \,\frac{b + b^2+ b \,( d-4)}{F(x,y)}\,, 
 \nonumber\\
 && D = \frac{2}{d-4} + \gamma_E -1 -\log (4\,\pi ) \,,
\end{eqnarray}
where we expand around $d=4$. 
A further step
\begin{eqnarray}
 &&  i\,J^{(a)}_{000} =  \frac{1}{16\,\pi^2} \int_0^1 d x  \Bigg( \int_0^{1-x} d y \, \frac{\tilde N_a(x,y)}{F(x,y)  } +
 \tilde L_a(x,1-x)\,\Big\{D+ \log \frac{F(x,1-x)}{\mu^2} \Big\} 
  \nonumber\\
&& \qquad \qquad \qquad - \, \tilde L_a(x,0)\,\Big\{ D+ \log \frac{F(x,0)}{\mu^2} \Big\} \Bigg)  + {\mathcal O} \left( d- 4 \right)\,,
 \nonumber\\
&&  \tilde N_a(x,y) = N_a(x,y) - \tilde L_a(x,y)\, \partial_y F(x,y) \,, \qquad \quad \partial_y \tilde L_a(x,y) = L_a(x,y)\,,
\label{def-tildeN}
 \end{eqnarray}
shows that all scalar-triangle-type contributions take the form  $I_0(m,n)$ with $m+ n\leq a$ always. This is so since in the vicinity of $d \sim 4$ it holds
\begin{eqnarray}
&& I_{0}(m,n) = \frac{1}{16\,\pi^{2}}\,\int_{0}^{1}\,d x\int_{0}^{1-x}\,d y\,\frac{x^m\,y^n}{F(x,y) }+ {\mathcal O} \left( d-4\right)\,.
\label{def-I0mn}
\end{eqnarray}
The remaining terms can be expressed in terms of bubble-type contributions. We assume  the scale independent contributions as implied by $\tilde N (x,y)$ to comply with the expectation of dimensional counting rules, while possible power-counting violating terms stem from the bubble-type contributions. They take the form
\begin{eqnarray}
&&  I_{GL}^{(n)}(\bar p^2) =  -\frac{1}{16\,\pi^2} \int_0^1 d x \,x^n\, \Big\{ 1+ D + \log \frac{F(x,0)}{\mu^2}\Big\} + {\mathcal O} \left( d- 4 \right) \,,
\nonumber\\
&&  I^{(0)}_{GL}(\bar p^2)= \int  \frac{d^d l}{(2\pi)^d}\,\frac{-i\, \,\mu^{4-d}}
{((l-\bar p)^2- M^2_L)\,(l^2-m_G^2)} \,,\qquad 
\label{def-I0n}
\end{eqnarray}
where we celebrate the recursion relation
\begin{eqnarray}
&&2\, I^{(1)}_{GL} (\bar p^2) = I^{(0)}_{GL} (\bar p^2) - \frac{ M^2_L-m_G^2}{\bar p^2}\,\Big\{ I^{(0)}_{GL} (\bar p^2) -  I^{(0)}_{GL} (0) \Big\} \,, 
\nonumber\\
&& \Delta I^{(n)}_{GL}(\bar p^2) =   \frac{I^{(n)}_{GL}(\bar p^2) - I^{(n)}_{GL}(0)}{\bar p^2} \,, \qquad \qquad 
\Delta \Delta I^{(n)}_{GL}(\bar p^2) =   \frac{\Delta I^{(n)}_{GL}(\bar p^2) - \Delta I^{(n)}_{GL}(0)}{\bar p^2} \,,
\nonumber\\
&& (n+1)\,\Delta I^{(n)}_{GL}(\bar p^2) =  
2\,\Delta I^{(1)}_{GL}(\bar p^2) +  m_G^2\,\Big\{n\, \Delta \Delta I^{(n-1)}_{GL}(\bar p^2) - \Delta \Delta I^{(0)}_{GL}(\bar p^2) \Big\} 
\nonumber\\
&& \qquad \qquad \qquad \quad \;\;- \, M^2_L \,\sum_{k=1 }^{n-1} (k+1)\,\Delta \Delta I^{(k)}_{GL}(\bar p^2) \,,
\label{res-recursion}
\end{eqnarray}
that demonstrates our claim on the nature of such contributions. 

The corresponding log terms  involving $F(x,1-x)$ in (\ref{def-tildeN}) follow upon the substitutions $\bar p \to \bar p -p $ and $m_G \to M_R$ in (\ref{def-I0n}) and (\ref{res-recursion}). In particular we find
\begin{eqnarray}
&&  I_{LR}^{(n)}( (\bar p -p)^2) =  -\frac{1}{16\,\pi^2} \int_0^1 d x \,x^n\, \Big\{ 1+ D + \log \frac{F(x,1-x)}{\mu^2}\Big\} + {\mathcal O} \left( d- 4 \right) \,,
\nonumber\\
&&  I^{(0)}_{LR}( (\bar p -p)^2)= \int  \frac{d^d l}{(2\pi)^d}\,\frac{-i\, \,\mu^{4-d}}
{((l-\bar p)^2-M^2_L)\,((l-p)^2-M^2_R)} \,.\qquad 
\end{eqnarray}
It remains to investigate the functions $I_0(m,n)$, for which we claim in (\ref{def-n-triangle-QH}) that it suffices to include a particular subset in our set of extended basis functions. This will be shown in the following by means of recursion equations that relate $I_0(m,n)$ for different choices of $m$ and $n$. 

\newcommand{\vvLQ}{v^{2}_{LQ}}
\newcommand{\vvQR}{v^{2}_{QR}}
\newcommand{\pfpf}{\bar{p}^{2}}
\newcommand{\pipi}{p^{2}}
\newcommand{\pfpi}{\bar{p}\cdot p}
\newcommand{\pfw}{\bar{p}\cdot w}
\newcommand{\piw}{p\cdot w}
\newcommand{\ww}{w^{2}}
\newcommand{\rr}{r^{2}}
We derive by suitable partial-integrations 
\begin{eqnarray}
&& I_{0}(m+1,n) = -\frac{\pfpi}{\pfpf}\,I_{0}(m,n+1)+\frac{\bar p^2- M^2_L+ m_G^2}{2\,\pfpf}\,I_{0}(m,n)
\nonumber\\
&&\qquad-\frac{1}{2\,\pfpf}\, \Sigma(m,n) +\frac{\delta_{m,0}}{2\,\pfpf}\,I_{QR}^{(n)}-\frac{m}{d\,\pfpf}\,I_1(m-1,n)\,,
\nonumber\\
&& I_{0}(m,n+1) = -\frac{\pfpi}{\pipi}\,I_{0}(m+1,n)+\frac{p^2- M^2_R+m_G^2}{2\,\pipi}\,I_{0}(m,n)
\nonumber\\
& &\qquad-\frac{1}{2\,\pipi}\, \Sigma(m,n) +\frac{\delta_{n,0}}{2\,\pipi}\,I_{QL}^{(m)}-\frac{n}{d\,\pipi}\,I_1(m,n-1)\,,
\nonumber\\
&& \Sigma(m,n) = \sum_{k=0}^{n}
\Bigg( \begin{array}{c} n\\ k\end{array}\Bigg) (-1)^{k}\,I^{(m+k)}_{LR}\,,
\label{res-partial-integration-triangle}
\end{eqnarray}
which imply the desired recursions upon the elimination of the structure $I_1(m,n)$. 
The system (\ref{res-partial-integration-triangle}) can be solved by iteration most economically. 
It is useful to consider first the case $m=0$ for arbitrary $n$  in the expressions $I_0(m+1,n)$ for which the $I_1(-1,n)$ contribution vanishes identically. Given $I_0(0,n)$ we obtain all $I_0(1,n)$. Similarly,  from the second equation in  (\ref{res-partial-integration-triangle}) we find $I_0(m,1)$ from the set of all $I_0(m,0)$ unambiguously. In the next step, we consider the second equation at $m\to m-1 $ and $n \to n+1$, so that we can eliminate the common $I_1(m-1,n)$ term from both equations. 
The resulting equation can be used to determine $I_0(m,n+1)$ from $I_0(m,n)$ or alternatively $I_0(m+1,n)$ from $I_0(m,n)$ by iteration. 

Our basis functions in (\ref{def-n-triangle-QH}) are introduced with the particular choice $I^{(n)}_{G,\bar LR} =I_0(n,0)$  and  $I^{(n)}_{G, L \bar R} =I_0(0,n)$  in (\ref{def-Cfkia}). Within such a scheme we derive for a  $(l \cdot Q)$ in the numerator of (\ref{def-Cfkia}) the following result
\begin{eqnarray}
&&i\,J^{(1)}_{000} = i\,J_{G, LR}^{(1)} = \frac{Q \cdot p}{2\,p^2}\, \Big( I^{(0)}_{GL}-I^{(0)}_{LR} + \big( p^2-M_R^2 +m_G ^2 \big)\,I^{(0)}_{G,LR} \Big)
\nonumber\\
&& \qquad \quad\; +\, 
 \Big(\bar p \cdot Q - \frac{(\bar p \cdot p)\,(Q \cdot p)}{p^2} \Big)\,I^{(1)}_{G,\bar LR} 
 \nonumber\\
&& \qquad \quad\; = \frac{\bar p  \cdot Q}{2\,\bar p^2}\, \Big( I^{(0)}_{GR}-I^{(0)}_{LR} + \big( \bar p^2-M_L^2 +m_G ^2 \big)\,I^{(0)}_{G,LR} \Big)
\nonumber\\
&& \qquad \quad\; +\, 
 \Big( p \cdot Q - \frac{(\bar p \cdot p)\,(Q \cdot \bar p)}{\bar p^2} \Big)\,I^{(1)}_{G,L \bar R} 
 \,,
\end{eqnarray}
where we observe that our result is invariant under the simultaneous exchange of $L \leftrightarrow R$ and $\bar p \leftrightarrow p$.
It is emphasized that if and only if our result is expressed in terms of $I_0(0,0)$ and $I_0(1,0)$ (or $I_0(0,1)$) alone, a power-counting respecting expression is obtained with $J_{G, LR}^{(1)} \sim Q^2$ using $m_G \sim Q$. 

The proper evaluation of $J^{(2)}_{000}=J_{G, LR}^{(2)}$ with $(l \cdot Q)^2$ in the numerator of (\ref{def-Cfkia}) is slightly more tedious.  It involves the additional basis functions $I^{(2)}_{G,\bar LR}$ and $I^{(2)}_{G,L \bar R}$ of (\ref{def-n-triangle-PQ}) for which we derive an explicit representation
\begin{eqnarray}
&& \Big( (\bar p \cdot p)^2/p^2- \bar p^2\Big)\,I^{(2)}_{G,\bar LR} = 
  \frac{1}{8}\Big(4\,m_G^2- (p^2-M_R^2+m_G^2)^2/p^2 \Big)\,I^{(0)}_{G,LR}
 \nonumber\\
&& \qquad   -\, \frac{3}{4}\,\Big( \bar p^2 - M_L^2 + m_G^2- (\bar p \cdot p)\,(p^2-M_R^2+m_G^2)/p^2 \Big)\,I^{(1)}_{G,\bar LR}
\nonumber\\
&& \qquad +\,\frac{1}{2}\,(\bar p \cdot p)\,\Big( I^{(1)}_{LG}- I^{(1)}_{LR} \Big)/p^2
- \frac{1}{4\,(d-2)}\,(p^2-M_R^2+ m_G^2)\,\Big( I^{(0)}_{LG}- I^{(0)}_{LR}\Big)/p^2 
\nonumber\\
&& \qquad
-\,\frac{1}{2\,(d-2)}\,I^{(0)}_{LR} + \frac{1}{2}\,I^{(1)}_{LR}
+ {\mathcal O} \left( d-4\right) \,,
\label{res-I2GLR}
\end{eqnarray}
which is an extension of (\ref{def-1-triangle}). Our result (\ref{res-I2GLR}) illustrates the necessity to include $I^{(2)}_{G,\bar LR} $ into our set of basis functions, as it is instrumental to avoid the kinematical singularity at $(\bar p \cdot p)^2/p^2 = \bar p^2$. Note that from (\ref{def-n-triangle-PQ})
it follows that $I^{(2)}_{G,\bar LR} $ is regular at such kinematical conditions. 

A direct application of (\ref{def-tildeN}) leads to a form for $ J_{000}^{(2)} $ that appears power-counting violating. The source of this complication is traced to its 
$I_1(0,0) \sim Q^2$ contribution which should not be derived from (\ref{res-partial-integration-triangle}). Instead, it is well-expressed in application of 
\begin{eqnarray}
&& i\, J^{(0)}_{010} = I_1(0,0)+ \Big[(\bar p\cdot \bar p)\,I_{0}(2 ,0  )
 +2\, (\bar p\cdot  p)\,I_{0}(1 ,1 ) +  ( p\cdot  p)\,I_{0}(0 ,2) \Big]
\nonumber\\
&& \qquad \;\; = m_G^2\,I^{(0)}_{G,LR} - I^{(0)}_{LR} \,,
\end{eqnarray}
where we observe that the particular combination 
\begin{eqnarray}
i\,J_{000}^{(2)} -  \frac{Q^2}{d}\,I_1(0,0)  = ( \bar p \cdot Q)^2 \,I_0(2,0)  +2\, (\bar p \cdot Q)\,(p\cdot Q)\,I_0(1,1) +  ( p \cdot Q)^2 \,I_0(0,2) \,,
\label{}
\end{eqnarray}
does not involve the term $I_1(0,0)$ by construction. As a consequence we find 
\begin{eqnarray}
&& i\, J_{000}^{(2)} =  i\, J_{G,LR}^{(2)}= \Big( (\bar p \cdot Q)^2-  \frac{Q^2 }{d}\,(\bar p\cdot \bar p) \Big)\,I_{0}(2 ,0  ) + \Big( (Q\cdot p )^2-  \frac{Q^2 }{d}\,( p\cdot  p)\Big)\,I_{0}(0 ,2) 
\nonumber\\
&& \qquad \;\; +\,2\,\Big((\bar p \cdot Q)\,(Q\cdot p) -  \frac{Q^2 }{d}\, (\bar p\cdot  p) \Big)\,I_{0}(1 ,1 ) 
+ \frac{Q^2 }{d}\,\Big( m_G^2\,I^{(0)}_{G,LR} - I^{(0)}_{LR} \Big) \,,
\label{res-J2-intermediate}
\end{eqnarray}
an expression that appears at odds with dimensional counting.  From (\ref{res-J2-intermediate}) we would see $J_{000}^{(2)} \sim  Q^2$ rather than the expected $\sim Q^4$. Our final expression $\sim Q ^4$ follows in application of (\ref{res-partial-integration-triangle}), which leads to our result in terms of $I^{(0)}_{G,LR}=I_0(0,0)$ and $I^{(n)}_{G,\bar LR}=I_0(n,0)$ only. 
Using (\ref{res-partial-integration-triangle}) 
we rewrite (\ref{res-J2-intermediate}) into the form
\allowdisplaybreaks[1]
\begin{eqnarray}
&&  i\,J_{G, LR}^{(2)} =  \frac{4\,(Q \cdot p)^2 - Q^2\,p^2}{12\,p^2\,p^2}\, (p^2 -M_R^2+ m_G^2)\, \Big( I^{(0)}_{LG} -I^{(0)}_{LR} + (p^2 -M_R^2+ m_G^2)\,I^{(0)}_{G,LR}  \Big)
\nonumber\\
&& \qquad \quad \; +\, \frac{6\,(\bar p\cdot Q) \,p^2 (Q\cdot p) - (\bar p \cdot p) \,(p^2\,Q^2 + 2\,(p\cdot Q)^2 )}{6\,p^2\,p^2}\, \Big( I^{(1)}_{LG}-I^{(1)}_{LR} \Big)
\nonumber\\
&& \qquad \quad \; -\, \frac{Q^2\,p^2+2\, (Q \cdot p)^2}{6\,p^2}\,I^{(0)}_{LR} -  \frac{Q^2\,p^2- 4\, (Q \cdot p)^2}{6\,p^2}\,I^{(1)}_{LR}
-\,m_G^2\,\frac{ ( Q\cdot p)^2-Q^2\,p^2  }{3\,p^2}\,I^{(0)}_{G,LR} 
\nonumber\\
&& \qquad \quad \;+ \,(Q \cdot p)\,(p^2-M_R^2+m_G^2)\,\frac{ (\bar p \cdot Q)\,p^2 - (\bar p \cdot p)\, (Q \cdot p)}{p^2\,p^2}\,I^{(1)}_{G,\bar LR}
\nonumber\\
&& \qquad \quad\; +\, \Big\{ (\bar p \cdot Q)^2- \frac{1}{3}\, \bar p^2\,\,Q^2 
 +  \frac{ (\bar p \cdot \bar p)\,(Q \cdot p) - 6\,(\bar p \cdot p)\,(\bar p \cdot Q )}{3\,p^2}\,(Q \cdot p)
\nonumber\\
&& \qquad \qquad \qquad   +\,
 (\bar p \cdot  p)^2\,\frac{Q^2\,p^2 +2\,(Q \cdot p)^2}{3\,p^2\,p^2} \Big\}\,I^{(2)}_{G,\bar LR} 
 \nonumber\\
&& \qquad \quad  +\, \frac{d-4}{18}\,\Big( Q^2- (Q \cdot p)^2/p^2 \Big)\,\Big( I^{(0)}_{LR} + I^{(1)}_{LR} \Big)
+ {\mathcal O} \left( d-4\right)\,,
\label{res-JGLR2}
 \end{eqnarray}
which, in its renormalized form with in particular $\bar I_{LR}^{(n)} \to  0$, confirms the expected chiral power $\sim Q^4$.

While in this Appendix we detailed the derivation of triangle loops of the $Q,LR$ type, corresponding results are implied for the $H,PQ$ loops by simple replacements $L\to P$, $R \to Q$ and $G \to H$. The loop functions of the $L, QH$ follow by  replacing  $R \to H$ with $p \to w $. Similarly, the 
$QH, R$ case is implied by $L \to H$ with $ \bar p \to w $. We note that our result can be readily generalized for the case defined by $ (l \cdot Q)^2 \to (l \cdot \bar q) \, (l\cdot q)$. It suffices to use the replacement  $Q_\mu \,Q_\nu \to (\bar q_\mu \, q_\nu + q_\mu \,\bar q_\nu)/2 $ in (\ref{res-JGLR2}). 

Finally, it is advantageous in some cases, to use a symmetrized version of (\ref{res-JGLR2}) that follows in application of the replacements $L \leftrightarrow R$ and $\bar p \leftrightarrow p$, under which $J_{G, LR}^{(2)}$ is unchanged strictly. Such a form involves both $I^{(n)}_{G, L\bar R}$ and   $I^{(n)}_{G,\bar LR} $, making the right-hand side of the updated form of (\ref{res-JGLR2}) invariant manifestly.

\section*{Appendix C: Fourth order triangle-loop expressions }

In this Appendix we specify the additional one-loop diagrams that involve the LEC $c_n$ and $\tilde c_n$.
Such contributions are implied by Fig.\ \ref{fig:6} via a replacement of the leading order Tomozawa-Weinberg two-body vertex by its subleading order $Q^2$ refinement vertex. 
All such diagrams have a minimal chiral order $Q^4$. It holds 
\begin{eqnarray}
&& T^{(4)}_{ab}(s,t,u)=  T^{(4-s)}_{ab}(s,u)+  T^{(4-u)}_{ab}(s,u)+   T^{(4-t)}_{ab}(s,u)\,,
\nonumber\\ 
&& f^4\,T^{(4-s)}_{ab}(s,t,u)= g_P^2 \,\sum_{Q,H}\,\Big\{\sum_L \Big( c_0\,C^{s-L}_{0,QH} + c_1\,C^{s-L}_{1,QH}\Big)\,J^{(s,1)}_{L,QH}(s,u) 
\nonumber\\
&& \quad +\,\sum_R \Big( c_0\,C^{s-R}_{0,QH} + c_1\,C^{s-R}_{1,QH}\Big)\,J^{(s,1)}_{QH,R}(s,u) 
\nonumber\\
&& \quad  +\,\sum_L\Big(c_2\,C^{s-L}_{2,QH} + c_3\,C^{s-L}_{3,QH} \Big)\, J^{(s,3)}_{L,QH}(s,u) 
+ \sum_R\Big(c_2\,C^{s-R}_{2,QH} + c_3\,C^{s-R}_{3,QH} \Big)\, J^{(s,3)}_{QH,R}(s,u)
\nonumber\\ 
&& \quad  +\,\sum_L\Big(c_4\,C^{s-L}_{2,QH} + c_5\,C^{s-L}_{3,QH} \Big)\, J^{(s,5)}_{L,QH}(s,u) 
+\sum_R\Big(c_4\,C^{s-R}_{2,QH} + c_5\,C^{s-R}_{3,QH} \Big)\, J^{(s,5)}_{QH,R}(s,u) \Big\}\,,
\nonumber\\ 
&& f^4\,T^{(4-t)}_{ab}(s,t,u)= g_P^2 \,\sum_{Q, LR}\,\Big\{  \tilde c_6\,C^{t-LR}_{6,Q}\,J^{(t,6)}_{Q,LR}(s,u)  +
\Big( \tilde c_0\,C^{t-LR}_{0,Q} + \tilde c_1\,C^{t-LR}_{1,Q}\Big)\,J^{(t,1)}_{Q,LR}(s,u) 
\nonumber\\
&& \quad  +\,\Big(\tilde c_2\,C^{t-LR}_{2, Q} + \tilde c_3\,C^{t-LR}_{3,Q} \Big)\, J^{(t,3)}_{Q,LR}(s,u) 
 +\Big(\tilde c_4\,C^{t-LR}_{2,Q} + \tilde c_5\,C^{t-LR}_{3,Q} \Big)\, J^{(t,5)}_{Q,LR}(s,u) \Big\}\,,
\end{eqnarray}
where the missing u-channel term $T^{(4-u)}_{ab}(s,u)$ follows from the s-channel expressions by a crossing transformation of the Clebsch coefficients $C^{s-\cdots}_{\cdots}\to C^{u-\cdots}_{\cdots}$ together with corresponding loop functions $J^{(u,n)}_{\cdots} (s,u)$. 
In the s-channel the Clebsch coefficients are easily accessible in terms of the already recalled Clebsch  in (\ref{res-tree-Q123}) via 
\begin{eqnarray}
&&\sum_{QH}\,C_{n,QH}^{s-L}\Big|_{ab} =-2\,B_0\,\sum_{c  \leftrightarrow QH} C^{(u)}_{L} \Big|_{ac}\, \Big[2\,m\,C^{(\pi)}_n + (m+m_s)\,C^{(K)}_n \Big]_{cb}\,,\qquad  
\nonumber\\ 
&& \sum_{QH}\,C_{n,QH}^{s-R}\Big|_{ab} = -2\,B_0\, \sum_{c \leftrightarrow QH} \Big[2\,m\,C^{(\pi)}_n + (m+m_s)\,C^{(K)}_n \Big]_{ac} \,C^{(u)}_{R} \Big|_{cb} \,,\qquad 
\nonumber\\
&& \sum_{QH}\,C^{s-L}_{n,QH} \Big|_{ab}= -\sum_{c  \leftrightarrow QH}\,C_{L}^{(u)}\Big|_{ac}\,C_n \Big|_{cb}\,,  \quad \quad
\sum_{QH}\, C^{s-R}_{n,QH} \Big|_{ab} =- \sum_{c  \leftrightarrow QH}\,C_n \Big|_{ac} \,C_{R}^{(u)} \Big|_{cb}\,,
 \label{}
\end{eqnarray}
with $n=0,1$ in the first two lines and $n=2,3$ elsewhere. The derivation of the t-channel Clebsch is a bit more tedious. The loop functions in the s-, t- and u-channel have the form
\begin{eqnarray}
&& J^{(s,1)}_{L,QH}(s,u)  = \int\frac{d^d l}{(2\pi)^d}\,l_{\alpha } \,\bar{p}_{\beta } \,S^{\alpha\beta,\mu\nu}_L(l+\bar p) \,  \bar{q}_{\mu } \,(l+w)_{\nu }\,
\frac{i\,\mu^{4-d}}{l^2-m_Q^2}\,S_H(l+w) \,,
\nonumber\\
&& J^{(s,1)}_{QH,R}(s,u)  = \int\frac{d^d l}{(2\pi)^d}\frac{i\,\mu^{4-d}}{l^2-m_Q^2}\,S_H(l+w)\,q_{\alpha }\, (l+w)_{\beta } \,S^{\alpha\beta,\mu\nu}_R(l+p)\,l_{\mu } \,p_{\nu } \,,
\nonumber\\ \nonumber\\
&& J^{(s,3)}_{L,QH}(s,u)  = \int\frac{d^d l}{(2\pi)^d}\,l_{\alpha } \,\bar{p}_{\beta } \,S^{\alpha\beta,\mu\nu}_L(l+\bar p) \,  \bar{q}_{\mu } \,(l+w)_{\nu }\,
\frac{-i\,\mu^{4-d}\,4 \,(l\cdot q)}{l^2-m_Q^2}\,S_H(l+w) \,,
\nonumber\\
&& J^{(s,3)}_{QH,R}(s,u)  = \int\frac{d^d l}{(2\pi)^d}\frac{-i\,\mu^{4-d}\,4\,(\bar q\cdot l)}{l^2-m_Q^2}\,S_H(l+w)\,q_{\alpha }\, (l+w)_{\beta } \,S^{\alpha\beta,\mu\nu}_R(l+p)\,l_{\mu } \,p_{\nu } \,,
\nonumber\\ \nonumber\\
&& J^{(s,5)}_{L,QH}(s,u)  = \int\frac{d^d l}{(2\pi)^d}\,l_{\alpha } \,\bar{p}_{\beta } \,S^{\alpha\beta,\mu\nu}_L(l+\bar p) \,  \bar{q}_{\mu } \,(l+w)_{\nu }\,
\frac{-2\,i\,\mu^{4-d}}{l^2-m_Q^2}\,S_H(l+w)
\nonumber\\
	&&\qquad\qquad\qquad\quad \times \,\Big( ( l^2 +l\cdot w )\,(p\cdot q) +(l\cdot q + w\cdot q)\,(l \cdot p) \Big)/M^2  \,,
\nonumber\\
&& J^{(s,5)}_{QH,R}(s,u)  = \int\frac{d^d l}{(2\pi)^d}\frac{-2\,i\,\mu^{4-d}}{l^2-m_Q^2}\,S_H(l+w)\,q_{\alpha }\, (l+w)_{\beta } \,S^{\alpha\beta,\mu\nu}_R(l+p)\,l_{\mu } \,p_{\nu } 
\nonumber\\
	&&\qquad\qquad\qquad\quad \times \, \Big( ( l^2 +l\cdot w )\,(\bar p\cdot \bar q) +(l\cdot \bar q + w\cdot \bar q)\,(l \cdot \bar p) \Big) /M^2 \,,
\nonumber\\ 
\nonumber\\
&& J^{(t,1)}_{Q,LR}(s,u)  = \int\frac{d^d l}{(2\pi)^d}\,
\frac{i\,\mu^{4-d}\,l_{\alpha } \,\bar{p}_{\beta }}{l^2-m_Q^2} \,S^{\alpha\beta,\bar \sigma \bar \tau }_L(l+\bar p) \,g_{\bar \sigma \sigma}\,g_{\bar \tau \tau }\,S^{\sigma \tau,\mu\nu}_R(l+p)\, l_{\mu } \,p_{\nu } \,,
\nonumber\\
&& J^{(t,3)}_{Q,LR}(s,u)  = \int\frac{d^d l}{(2\pi)^d}\,
\frac{i\,\mu^{4-d}\,l_{\alpha } \,\bar{p}_{\beta }}{l^2-m_Q^2} \,S^{\alpha\beta,\bar \sigma \bar \tau}_L(l+\bar p) \,g_{\bar \sigma \sigma}\,4\,(\bar q \cdot q)\,g_{\bar \tau \tau } \,S^{\sigma \tau,\mu\nu}_R(l+p) \, l_{\mu } \,p_{\nu } \,,
\nonumber\\
&& J^{(t,5)}_{Q,LR}(s,u)  = \int\frac{d^d l}{(2\pi)^d}\,
\frac{i\,\mu^{4-d}\,l_{\alpha } \,\bar{p}_{\beta }}{l^2-m_Q^2} \,S^{\alpha\beta,\bar \sigma  \bar \tau}_L(l+\bar p) \,g_{\bar \sigma \sigma}\,g_{\bar \tau \tau } \,S^{\sigma \tau,\mu\nu}_R(l+p)\,  l_{\mu } \,p_{\nu }
\nonumber\\
	&&\qquad\qquad\qquad\quad \times \,2\,\Big( ( l+ \bar p)\cdot  \bar q  )\,( ( l+ p)\cdot q )	 + ( l+ \bar p)\cdot  q  )\,( ( l+ p)\cdot  \bar q )	
	\Big)/\tilde M^2  \,,  
	\nonumber\\
&& J^{(t,6)}_{Q,LR}(s,u)  = \int\frac{d^d l}{(2\pi)^d}\,
\frac{i\,\mu^{4-d}\,l_{\alpha } \,\bar{p}_{\beta }}{l^2-m_Q^2} \,S^{\alpha\beta,\bar \sigma \bar \tau }_L(l+\bar p) \,g_{\bar \sigma \sigma}\,\big(\bar q_{\bar \tau}\, q_\tau -  q_{\bar \tau}\, \bar q_\tau \big) \,S^{\sigma \tau,\mu\nu}_R(l+p)\, l_{\mu } \,p_{\nu } \,,
\nonumber\\	\nonumber\\
&&  J^{(u,n)}_{QH,R}(s,u) =  J^{(s,n)}_{QH,R}(u,s) \,,\qquad \qquad \qquad  J^{(u,n)}_{L,QH}(s,u) =  J^{(s,n)}_{L,QH}(u,s) \,,
\label{def-J-triangle-appendix}
\end{eqnarray}
for which we derive:
\begin{eqnarray}
&& \bar  J^{(s,1)}_{L,QH}(s,u)  = \frac{s-u}{8}\,\Big[  \bar I_{QL}(M_a^2)-\bar I_{QH}(s) \Big]  + \frac{ s-u}{16}\,\Big(s-u +r\, M^2_{ab} \Big)\, \bar I_{L,QH}(s)
\nonumber\\
&& \qquad \qquad \quad \,
+ \,\frac{1}{16}\,\Big( (s-u)^2- 8\, m_a^2\,M^2_{ab} \Big)\,\bar  I^{(1)}_{L,QH}(s) + {\mathcal O } \left( Q^3\right) \,,
\nonumber\\
&& \bar  J^{(s,3)}_{L,QH}(s,u)  = \frac{1}{24}\,\Big(- 8\,(\bar q \cdot q) + ( s-u )^2 /M^2_{ab}  \Big)\,\Big[ \bar I_Q  -\frac{2}{3}\,\frac{m_Q^2}{16\,\pi^2} \Big]
\nonumber\\
&& \qquad \qquad \quad \, - \,\frac{1}{16}\,(s-u)^3\bar I_{QH}(s)/M^2_{ab}  
+\frac{1}{48}\, \Big( 4\,(s-u)^3 + 8\,r\,(M^2_{ab})^2 \,(\bar q\cdot q)
\nonumber\\
&& \qquad \qquad \qquad  \qquad \quad \, - \,M^2_{ab}\,(s-u)\,(r\,(s-u) + 8\,(\bar q\cdot q) ) \Big)\,\bar I_{QL}(M_a^2)/M^2_{ab}
\nonumber\\
&& \qquad \qquad \quad \, +\, \frac{1}{96}\,\Big(4\,(s-u)^4/M^2_{ab} + (s-u)^2\,( 3\,r\,(s-u) -8\,m_Q^2-8\,(\bar q\cdot q) ) 
\nonumber\\
&& \qquad \qquad \qquad \qquad \quad \, +\,64\,(\bar q\cdot q) \,m^2_Q \,M^2_{ab} \Big)\, \bar I_{L,QH}(s)
\nonumber\\
&& \qquad \qquad  \quad \,
+ \,\frac{1}{32}\,(s-u)\,\Big( 2\,(s-u)^3/M^2_{ab} + r\,(s-u)^2 -8\,(\bar q \cdot q)\,r \,M^2_{ab}
\nonumber\\
&& \qquad \qquad \qquad  \qquad \quad \,
+ \, (s-u)\,( 4\,t -12\, m_a^2 -4\,m_b^2 )\Big)\, \bar I^{(1)}_{L,QH}(s) 
\nonumber\\
&& \qquad \qquad \quad \, + \,\frac{1}{48}\,\Big( 8\,m_a^2\,M^2_{ab}-(s-u)^2 \Big)\, \Big( 8\,(\bar q\cdot q) 
\nonumber\\
&& \qquad \qquad \qquad  \qquad \quad \,
- (s-u)^2/M^2_{ab} \Big) \,\bar I^{(2)}_{L,QH}(s)
+ {\mathcal O } \left( Q^5\right) \,,
\nonumber\\
&&M^2\,\bar  J^{(s,5)}_{L,QH}(s,u) = \frac{1}{32}\,r\,M^2_{ab} \,(s-u)^2 \Big[\bar I_{QH}(s) - \bar I_{QL}(M_a^2)\Big] 
\nonumber\\
&& \qquad \qquad \quad \, -\, \frac{1}{64}\,r\,M^2_{ab} \,(s-u)^2\,\Big( r\,M^2_{ab} + (s-u)  \Big)\,\bar  I_{L,QH}(s)
\nonumber\\
&& \qquad \qquad \quad \,
+ \,\frac{1}{64}\,r\,M^2_{ab} \,(s-u)\,\Big(8\, m_a^2\,M^2_{ab} - (s-u)^2\Big)\,\bar  I^{(1)}_{L,QH}(s) + {\mathcal O } \left( Q^5\right) \,,
\nonumber\\ \nonumber\\
&& \bar  J^{(t,1)}_{Q,LR}(s,u)  = -\frac{1}{2}\,\bar I_Q 
+ \frac{1}{4}\,m_Q^2\,\Big(\bar I_L/M_L^2 + \bar I_R/M_R^2 \Big)
+ \frac{r}{4}\,M_{ab}^2\,\Big[\bar I_{QL}(M_a^2)  + \bar I_{QR}(M_b^2) \Big]
\nonumber\\
&& \qquad \qquad \quad \, +\, M^2_{ab}\,\Big( m_Q^2 - \frac{r^2}{8}\,M^2_{ab}\Big)\,\bar  I_{Q,LR} + {\mathcal O } \left( Q^3\right) \,,
\nonumber\\
&& \bar  J^{(t,3)}_{Q,LR}(s,u) =  4\, (\bar q\cdot q)\,\bar J^{(t,1)}_{Q,LR}(s,u) 
- (\bar q\cdot q)\,m_Q^2\,\Big(\bar I_L/M_L^2 + \bar I_R/M_R^2 \Big) +{\mathcal O } \left( Q^5\right)   \,, \qquad 
\nonumber\\
&& \tilde M^2\,\bar  J^{(t,5)}_{Q,LR}(s,u) = \frac{(s-u)^2}{4}\,\bar J^{(t,1)}_{Q,LR}(s,u) - \frac{(s-u)^2}{16}\ m_Q^2\,\Big(\bar I_L/M_L^2 + \bar I_R/M_R^2 \Big)  +  {\mathcal O } \left( Q^5\right)   \,, 
\nonumber\\
&& \bar  J^{(t,6)}_{Q,LR}(s,u) =  {\mathcal O } \left( Q^5\right)   \,. 
\label{res-Jtriangle4}
\end{eqnarray}
We note that the loop functions $J^{(s,n)}_{QH,R}(s,u)$ follow from $J^{(s,n)}_{L,QH}(s,u)$ in (\ref{res-Jtriangle4}) upon the replacements $ \bar q\leftrightarrow q$ and $\bar p \leftrightarrow p$ and $L \leftrightarrow R$. An example for such a replacement is given in (\ref{res-JQHR}). It remains to detail the fourth order terms supplementing our third order expressions in (\ref{res-Triangle-J}). The following form is established
\allowdisplaybreaks[1]
\begin{eqnarray}
&&  \bar J^{(s),4}_{L,QH}(s,u) = \Big( m_a^2-m_b^2+ t\Big)\, \frac{m_Q^2}{288\,\pi^2}
+ \frac{1}{48}\,\Big(8\,(m_a^2-m_b^2) +3\,r\,( s-u )
 \nonumber\\
&& \qquad  \qquad +\,12\,m_{ab}^2 -4\,t -\,9\,(s-u)^2/M^2_{ab} \Big)\, \bar I_Q 
\nonumber\\
&& \qquad  -\,
\frac{1}{96}\,\big( 3\,r\,(s-u)+ 8\,( m_a^2 -m_b^2) +8\,m^2_{ab}  \big)\,\Big(m_Q^2\,\bar I_L/M_L^2 + m_Q^2\,\bar I_R/M_R^2 \Big)
\nonumber\\
&& \qquad  +\,\frac{1}{96}\,\Big( -3\,(s -u)\,\big(4\,\delta_L - 2\,(M_a^2-M_b^2) + r^2\,M^2_{ab} - 6\,m_{ab}^2 \big)
\nonumber\\
&& \qquad \qquad   
+ \, 2\,(m_a^2-m_b^2)\,\big(s -u - r\,M^2_{ab}  \big) + 
       6\,r\,(M_a^4-M_b^4)  -12\,m_{ab}^2\,r\,M^2_{ab}
\nonumber\\
&& \qquad \qquad    
\,+ \, 9\,r\,(s -u)^2 -
  2\,t\,\big( 8\,(s -u)-5\,r\,M^2_{ab} \big) 
-  6\,(s -u)^3/M^2_{ab}
\Big)\,\bar I_{QL}(M_a^2)
\nonumber\\
&& \qquad+\,\frac{1}{32}\,\Big( (s - u)\,\big( 4\,\delta_L + 8\,m_Q^2 + 2\,m_{ab}^2+4\,t + r\,(s - u) - 3\,(s - u)^2/M^2_{ab} \big) - 2\,t\,r\, M^2_{ab} 
  \nonumber\\
&& \qquad \qquad 
- \, 2\,(m_a^2-m_b^2)\,\big( M^2_{ab}\,r - 3\,(s - u) \big)  
 -   2\,(M_a^2-M_b^2)\,\big( r\,M^2_{ab} + (s - u) \big) 
  \Big) \,\bar I_{QH}(s)
\nonumber\\
&& \qquad
+ \,\frac{1}{192}\,\Big( 6\,r\,(M_a^4 - M_b^4)\,\big( 3\,(s - u)  +r\, M^2_{ab}  \big) 
+
 2\,(m_a^2 - m_b^2)\,\big(16\,m_Q^2\,M^2_{ab}+    3\,r^2\,(M^2_{ab})^2
 \nonumber\\
&& \qquad \qquad 
  - \,6\,(s - u)\,r\,M^2_{ab} + (s - u)^2 \big) 
 +  12\,(s - u)\,r\,M^2_{ab}\,\big(\delta_H - 2\,(\delta_L + m_Q^2) - m_{ab}^2 \big)
 \nonumber\\
&& \qquad \qquad     
      + \,
 3\,  (s - u)^2\,\big(-4\,\delta_H + 8\,m_Q^2 -r^2\, M^2_{ab} + 8\,m_{ab}^2 \big)
    +     9\,r\,(s - u)^3 - 6\,(s - u)^4/M^2_{ab} 
 \nonumber\\
&& \qquad \qquad     
    + \,
      2\,t\,\big(16\,m_Q^2\,M^2_{ab} + 3\,r^2\,(M^2_{ab})^2 - 11\,(s - u)^2 \big)\Big) \bar I_{L,QH}(s)  
\nonumber\\
&& \qquad
+ \,\frac{1}{64}\,  \Big(-16\,m_{ab}^2\,(M_a^4 - M_b^4) + 4\,m_{ab}^2\,r^2\,(M^2_{ab})^2 + 
 4\,(s-u)\,r\,(M_a^4 - M_b^4)
 \nonumber\\
&& \qquad \qquad    -\, 8\,(m_{ab}^2)^2\,M^2_{ab}   -\, 12\,(s - u)\,m_{ab}^2\,r\,M^2_{ab} - (s - u)^2 \,r^2\,M^2_{ab}
 +   10\,m_{ab}^2\,(s - u)^2
 \nonumber\\
&& \qquad \qquad    +\, 3\,r\,(s - u)^3 - 2\,(s - u)^4/M^2_{ab}
+  4\,\delta_L\,\big(8\,m_a^2\,M^2_{ab}  - (s - u)^2 \big)
       \nonumber\\
&& \qquad \qquad        + \,
 2\,t\,\big(4\,M^2_{ab}\,m_{ab}^2 - 5\,(s - u)^2 \big)  + 
      4\,(m_a^2 - m_b^2)\,M^2_{ab}\,\big(-4\,(M_a^2 - M_b^2) + r^2\,M^2_{ab}
       \nonumber\\
&& \qquad \qquad  - \,2\,m_{ab}^2 - 3\,r\,(s - u) + 2\,t \big) \Big) \bar I^{(1)}_{L,QH}(s)
\nonumber\\
&& \qquad
+ \,\frac{1}{24}\, \Big( m_a^2-m_b^2+ t\Big)\, \Big( 8\,m_a^2\,M^2_{ab} -(s-u)^2\Big) \bar I^{(2)}_{L,QH}(s) \,,
\nonumber\\ 
&&   \bar J^{(t),4}_{Q,LR}(s,u) =  \frac{s-u}{16}\,\Big\{ -4\,r\, \bar I_Q
+2\,r\,m_Q^2\,\big( \bar I_L/M_L^2 + \bar I_R/M_R^2 \big)
\nonumber\\ 
&& \qquad +\, 2\,
 \Big(\delta_L+\delta_R -4\,m_Q^2 + 3\,r^2\, M^2_{ab}/4\Big)\, \Big[ \bar I_{QL}(M_a^2) + I_{QR}(M_b^2) \Big]
 \nonumber\\ 
&& \qquad \qquad -\, 2\,r\,M^2_{ab}\,\Big(\delta_L +\delta_R-\, 4\, m_Q^2 + r^2\, M^2_{ab}/4\Big)\,\bar I_{Q,LR}(t) \Big\}
\nonumber\\ 
&& \qquad -\,\frac{m_a^2-m_b^2}{8}\,r\,M_{ab}^2\, \Big[ \bar I_{QL}(M_a^2) -\bar I_{QR}(M_b^2)  \Big]
\nonumber\\
&& \qquad+\,\frac{1}{16}\,(m_a^2-m_b^2)\,M^2_{ab}\,\Big(8\,m_Q^2-  r^2\,M^2_{ab}\Big)\,\Big[ \bar I^{(1 )}_{Q,\bar LR}(t) 
-\bar I^{(1 )}_{Q,L \bar R}(t) \Big]\,, 
\nonumber\\ 
&&   \bar J^{(-),4}_{H,PQ}(s,u) = \frac{1}{16}\, \Big(2\,\delta_H -3\,m_P^2- m_Q^2 -m_{ab}^2 +3\,t -4\,(M_a^2-M_b^2)\Big)\,\bar I_P
\nonumber\\
&& \qquad+\,\frac{1}{16}\, \Big(2\,\delta_H - m_P^2- 3\,m_Q^2 -m_{ab}^2 + 3\,t + 4\,(M_a^2-M_b^2)\Big)\,\bar I_Q 
\nonumber\\
&& \qquad -\, \frac{1}{16}\, \Big( 2\,\big( 2\,\delta_H + m_{ab}^2 -2\,(M_a^2 -M_b^2 )  \big)\,m_{PQ}^2 -4\,\delta_H\,m_{ab}^2 -3\, (m_P^4 + m_Q^4) - 2\,m_P^2\,m_Q^2 
\nonumber\\
&& \qquad \qquad - \, ( m_{PQ}^2 -3\,m_{ab}^2)\,( m_{PQ}^2 -2\,m_{ab}^2 )+3\,(M_a^2-M_b^2)\,(m_P^2-m_Q^2)
\Big) \bar I_H/M_H^2
\nonumber\\
&& \qquad +\, \frac{1}{16}\,(M_a^2-M_b^2)\,\Big(m_{PQ}^2+m_{ab}^2  \Big)\,(m_P^2-m_Q^2)\,
\bigg[\frac{\bar I_{PQ}(t)}{t} \bigg]_{\rm reg }
\nonumber\\
&& \qquad+\, \frac{1}{16}\,\Big(  (m_{PQ}^2-3\, t -2 \,\delta_H )\, (m_{PQ}^2+ m_{ab}^2-3\, t) 
\nonumber\\
&& \qquad \qquad - \,3\,(M_a^2-M_b^2)\, 
(m_P^2-m_Q^2) \Big)\,\Big[\bar I_{PQ}(t) -\bar I_P/(2\,m_P^2)- \bar I_Q/(2\,m_Q^2) \Big]
\nonumber\\
&& \qquad-\,\frac{1}{16}\,r\,M^2_{ab}\, \Big(m_P^2- m_Q^2 +M_a^2 -M_b^2 \Big)\,\Big[\bar I_{QH}(M_b^2) - \bar I_{PH}(M_a^2) \Big]
\nonumber\\
&& \qquad+\,\frac{1}{16}\,r\,M^2_{ab}\, \Big( 2\,m_{PQ}^2 -2\,\delta_H +m_{ab}^2 -3\,t\Big)\,\Big[\bar I_{QH}(M_b^2) + \bar I_{PH}(M_a^2) \Big]
\nonumber\\
&& \qquad  -\, \frac{1}{16}\,r\,M^2_{ab}\,\Big(m_{PQ}^2 + m_{ab}^2-3\,t \Big)\,\Big( m_{PQ}^2-2\,t  -2\,\delta_H\Big) \bar I_{H,PQ}(t)    \,,
\nonumber\\
&&   \bar J^{(+),4}_{H,PQ}(s,u) = \frac{s-u}{16}\,\Big\{ r\,( \bar I_Q + \bar I_P )  -r\, (m_P^2+m_Q^2)\,\bar I_H/M_H^2
\nonumber\\
&& \qquad 
+\,
 \Big( 2\,m_{PQ}^2  -t -2\,\delta_H - 3\,r^2\,M^2_{ab}/4\Big)\,\Big[ \bar I_{QH}(M_b^2)+ \bar I_{PH}(M_a^2)\Big]
\nonumber\\
&& \qquad +\, \Big( M_a^2-M_b^2+m_P^2-m_Q^2 \Big)\,\Big[ \bar I_{PH}(M_a^2)- \bar I_{QH}(M_b^2)\Big]
\nonumber\\
&& \qquad -\,r\,\Big(m_{PQ}^2-2\,\delta_H -2\,t  \Big)\,\Big[\bar  I_{PQ}(t) -\bar I_P/(2\,m_P^2)  -\bar I_Q/(2\,m_Q^2) + \bar I_H/M_H^2\Big]
\nonumber\\
&& \qquad 
- \,\Big( \big(m_{PQ}^2-2\,\delta_H \big)\,\big( m_{PQ}^2-t - 3\,r^2\,M^2_{ab}/4 \big) + t\,r^2\,\,M^2_{ab}
\Big)\,\bar I_{H,PQ}(t)
\nonumber\\
&& \qquad 
+ \,\Big( m_{PQ}^2 - t - r^2\,M^2_{ab}/4\Big)\,\Big(t- M_a^2+M^2_b \Big)\,\bar I^{(1)}_{H,\bar PQ}(t)
\nonumber\\
&& \qquad 
+ \,\Big(m_{PQ}^2 - t - r^2\,M^2_{ab}/4\Big)\,\Big(t+ M_a^2-M^2_b \Big)\,\bar I^{(1)}_{H, P \bar Q}(t)
\Big\}
\nonumber\\
&& \qquad -\,\frac{1}{48}\,(m_a^2-m_b^2)\,(M_a^2-M_b^2) \,
\Big[ \bar I_{PQ} (t) 
-\bar I_P/(2\,m_P^2)  -\bar I_Q/(2\,m_Q^2) + \bar I_H/M_H^2\Big]
\nonumber\\
&& \qquad +\,  \frac{m_a^2-m_b^2}{16}\,\Big\{ 
- \frac{M_a^2-M_b^2}{72\,\pi^2}
+\,2\, r\, M^2_{ab}\,\Big[ \bar I_{PH}(M_a^2)-\bar I_{QH}(M_b^2) \Big]
\nonumber\\
&& \qquad +\,3\,\Big( m_P^2- m_Q^2 \Big)  \,\Big[ \bar I_{PQ} (t) 
-\bar I_P/(2\,m_P^2)  -\bar I_Q/(2\,m_Q^2) + \bar I_H/M_H^2\Big]
\nonumber\\
&& \qquad
 - \,\frac{4}{3}\,\Big( M_a^2-M_b^2\Big)\,\Big(m_P^2-m_Q^2
\Big)^2\,\bigg[\frac{\bar I_{PQ}(t)}{t^2} \bigg]_{\rm reg }
\nonumber\\
&& \qquad +\,\Big(\big( m_P^2-m_Q^2\big) \,  \big( 2\,\delta_H -m_{PQ}^2 \big)
+2\,(M_a^2 -M_b^2)\,m_{PQ}^2/3
\Big)\,\bigg[\frac{\bar I_{PQ}(t)}{t} \bigg]_{\rm reg }
\nonumber\\
&& \qquad
- \,r\,M^2_{ab}\,\Big( m_{PQ}^2- 2\,t -2\,\delta_H\Big)\,\Big[\bar I^{(1)}_{H,\bar PQ}(t) - \bar I^{(1)}_{H,P \bar Q}(t)\Big]
\Big\}  \,,
\nonumber\\
&&   \bar J^{(\chi),2}_{H,PQ}(s,u) =  \frac{1}{8}\,\Big[ \bar I_P + \bar I_Q - \big(2\,m_{PQ}^2-2\,\delta_H -3\,m_{ab}^2 \big)\,\bar I_H/M_H^2\Big]
 \nonumber\\
&& \qquad 
-\, \frac{1}{8}\, (M_a^2-M_b^2 )\,(m_P^2-m_Q^2)\,\bigg[\frac{\bar I_{PQ}(t)}{t} \bigg]_{\rm reg }
 \nonumber\\
&& \qquad 
-\, \frac{1}{8}\,\Big( m_{PQ}^2-3\,t -2\,\delta_H\Big)\,\Big[\bar I_{PQ}(t) -\bar I_P/(2\,m_P^2)- \bar I_Q/(2\,m_Q^2) \Big]
 \nonumber\\
&& \qquad+\, \frac{1}{8}\,r\,M^2_{ab}\,\Big\{ \Big( m_{PQ}^2-2\,t -2\,\delta_H  \Big) \,\bar I_{H,PQ}(t) - \bar I_{PH}(M_a^2) -\bar I_{QH}(M_b^2) \Big\} \,,
\end{eqnarray}
where we use the convenient notation
\begin{eqnarray}
\bigg[ \frac{h(t)}{t^n} \bigg]_{\rm reg } =  \frac{1}{t^n}\,\bigg\{ 1 - \sum_{k=0}^{n-1}\,\frac{t^k}{ k !}\,\left( \frac{d }{d t} \right)^k \bigg|_{t=0} \bigg\}\,h(t) \,.
\label{def-regular}
\end{eqnarray}

\clearpage

\section*{Appendix D: Scalar box loops}

We begin with an over-complete basis of scalar-box loop terms of the generic form 
\begin{eqnarray}
&&  B^{(a,b)}_{ikj}(s,t,u) = \!\int \! \frac{d^d l}{(2\pi)^d}\frac{\mu^{4-d}}{l^2-m_G^2}\frac{ (\bar q \cdot l)^a \,(\bar p \cdot l )^i\,(l^2 )^k\,(l \cdot p)^j\,(q\cdot l)^b}
{((l-\bar p)^2-M_L^2)\,((l-w)^2- M_H^2)((l-p)^2-M_R^2)} \,,
\label{def-Babij}
\end{eqnarray}
with $w = p+q = \bar p + \bar q$.
Let us first discuss the case with $i >0$ or $j> 0$. Here the problem can be reduced to 
$i=j=0$   and $k=0$ always. Using the identities
\begin{eqnarray}
&& l^2= [l^2- m_G^2] + m_G^2 \,,\qquad 
\nonumber\\
&& 2\,(l \cdot \bar p) = -[ (l - \bar p )^2  - M_L^2 ] + [ l^2- m_G^2 ] - M_L^2 + \bar p^2 + m_G^2 \,,  
\nonumber\\
&& 2\,(l \cdot p) = -[ (l - p )^2  - M_R^2 ] + [ l^2- m_G^2 ] - M_R^2 +  p^2 + m_G^2 \,,  
\end{eqnarray}
such terms lead to triangle diagrams already reduced systematically before and  box loops with $i=k=j=0$. 
Upon renormalization such reductions conserve the expected dimensional counting rules with $l^2\sim Q^2$ and $(\bar p \cdot l)  \sim (l \cdot p) \sim Q $. 

We turn to the $(\bar q \cdot l) \sim Q^2$ and $(l \cdot q) \sim Q^2$ structures in (\ref{def-Babij}). At first one may expect that a similar reduction is possible via  $(l \cdot w) \sim Q$ with the cancellations of propagators. While this is possible and results can be derived, the chiral order of such contributions is in conflict with the expectation of dimensional counting rules. This is caused by the required rewrite $(l \cdot q) = (l \cdot w)- (l\cdot  p) \sim Q $ or $(l\cdot \bar q) = (l\cdot w) - (l\cdot \bar p) \sim Q$, expressed as the difference of two order-one expressions.

The task is to derive specific correlations of scalar box, triangle and bubble diagrams, such that the counting results are made explicit. This is illustrated by our previous triangle expressions (\ref{def-1-triangle}) and  (\ref{def-n-triangle-QH}) which imply $ I^{(1)}_ {QH, R}\sim Q^0$. From the explicit representation  (\ref{def-1-triangle})  we find  
the nontrivial counting result
\begin{eqnarray}
&& \big(p^2-M_R^2+m_Q^2 \big)\,I_{QH,R} -I_{HR} +I_{QH} \sim Q^2 \,,
\nonumber\\
&& \big(w^2 \!-M_H^2+m_Q^2 \big)\,I_{QH,R} -I_{HR} +I_{QR} \sim Q^2 \,,
\nonumber\\
&&  {\rm with} \qquad p^2\,w^2 - (w\cdot p)^2 = q^2\,p^2 -(p\cdot q)^2 \sim Q^2 \,.
\end{eqnarray}
In the following we will generalize the triangle definitions  (\ref{def-n-triangle-QH}) to the box case (\ref{def-IQHLR-nm}), from which our desired expressions will follow. 

Consider the particular case 
\begin{eqnarray}
&& -i\, B^{(ab)}_{000} = \sum^{c_1+c_2 \,\leq \, a+b}_{[c],[h],[n], [o], [m]} \,(\bar q\cdot \bar p)^{m_1}\, \,(\bar q \cdot w)^{o_1}\,(\bar q\cdot p)^{n_1}\,
\,(\bar p \cdot q)^{m_2}\,(q\cdot w)^{o_2}\,(p \cdot q)^{n_2}
\nonumber\\
&& \qquad \quad \;\;\, \times \, (\bar q \cdot \bar  q)^{h_1}\, (\bar q\cdot q)^{h_2}\,(q\cdot q)^{h_3} \,C^{(ab)}_{[c][h][m][o][n]}\,I_c(m,o, n) \,,
\nonumber\\
&& \qquad m= m_1+m_2\,,\qquad \; \;\;\; \;o = o_1+o_2 \,,\qquad \; \;\; \;\;n = n_1+n_2\,,
\nonumber\\
&& \qquad  a= m_1 + o_1 + n_1 + 2\,h_1+h_2,\qquad b= m_2+o_2+n_2+ 2\,h_3 +h_2\,,
\nonumber\\
&& \qquad c = (a + b - c_1-c_2)/2\,, \quad\;\;\   c_1 = m_1 + o_1 + n_1 \,,\qquad \;\; \; c_2= m_2 + o_2 + n_2\,,
\nonumber\\
&& \quad {\rm with } \quad m_i \geq 0  \;\; {\rm and} \;\; o_i \geq 0 \quad {\rm and} \;\; n_i \geq 0 \quad {\rm and}  \quad h_i \geq 0 \,,
\nonumber\\
&& F(x,z, y) =  m_Q^2 - x\, (\bar p^2 -M_L^2 + m_G^2) 
- z\,(w^2-M_H^2+ m_G^2) - y\,(p^2-M_R^2+ m_G^2)
\nonumber\\ 
&& \qquad \qquad  \;\, + \, x^2\,\bar p^2 + z^2\,w^2 + y^2\, p^2  + \,2\,x\,y\,( \bar p \cdot  p) + 2\,x\,z\,( \bar p \cdot  w)  + 2\,y\,z\,( w \cdot  p) \,,
\nonumber\\ \nonumber\\
&&I_{c}(m,o,n) =- i\,  \int \frac{d^d l}{(2\pi)^d} \int_0^1 d x \int_0^{1-x} d y \int_0^{1-x-y} d z \,\frac{6\,\mu^{4-d}\,(l^2)^c}{(l^2 -F(x,z, y) )^4 }\,x^m\, z^o \,y^n\,,
\end{eqnarray}
with
\begin{eqnarray}
&& C^{(ab)}_{[c][h][m][o][n]} = \Bigg(\begin{array}{c} a \\ c_1 \end{array}\Bigg) \,
 \Bigg(\begin{array}{c} c_1 \\ o_1 \end{array}\Bigg)\,
  \Bigg(\begin{array}{c} m_1 +n_1 \\ m_1 \end{array}\Bigg)
 \Bigg(\begin{array}{c} b \\ c_2 \end{array}\Bigg) \,
 \Bigg(\begin{array}{c} c_2 \\ o_2 \end{array}\Bigg) 
   \Bigg(\begin{array}{c} m_2 +n_2 \\ m_2 \end{array}\Bigg)\, X_c\,Y^{(ab)}_{[c][h]} \,, \qquad 
\nonumber\\ 
&& X_0 = 1\,,\qquad \qquad \qquad 
 X_{c+1}= \frac{1}{d +2\,c}\,X_c\,,
\nonumber\\
&& Y^{(ab)}_{[c][h]} = 
\Bigg(\begin{array}{c} a-c_1 \\ h_2 \end{array}\Bigg)\,
  \Bigg(\begin{array}{c} b -c_2 \\ h_2 \end{array}\Bigg)\,\big(a- c_1-h_2 -1 \big)!!\, \big(b- c_2-h_2 -1 \big)!! \,h_2 ! \,. 
 \end{eqnarray}

Again we split the integral into a convergent and scale-dependent piece with
\begin{eqnarray}
 && - i\,B^{(ab)}_{000} =  \frac{1}{16\,\pi^2}\int_0^1 d x \int_0^{1-x} d y  \int_0^{1-x-y} \!d z \,\Bigg( \frac{N^{(0)}_{ab}(x,z, y)}{[F(x,z,y)]^2  } + \frac{N^{(1)}_{ab}(x,z, y)}{[F(x,z,y)]^1  }
 \nonumber\\
&& \qquad \qquad \qquad \qquad \qquad + \,
 L_{ab}(x,z,y)\,\Big\{ D+ \log \frac{F(x,z,y)}{\mu^2} \Big\}\Bigg) + {\mathcal O} \left( d- 4 \right) \,,
  \nonumber\\
&& N^{(k)}_{ab}(x,z,y) =   \sum^{c_1+c_2 \,\leq \, a+b}_{[c],[n], [o], [m], c \geq k} \,(\bar q\cdot \bar p)^{m_1}\, \,(\bar q \cdot w)^{o_1}\,(\bar q\cdot p)^{n_1}\, \,(\bar p \cdot q)^{m_2}\,(q\cdot w)^{o_2}\,(p \cdot q)^{n_2}
\nonumber\\
&& \qquad \quad \qquad  \times \, (\bar p \cdot p)^{h_1}\, (w^2)^{h_2}\,(\bar q\cdot q)^{h_3}\,C^{(abc)}_{[m][o][n]}\, \big[ F(x,z,y) \big]^{c -k}\Big(1-k \,( 1+ 3\,c) \Big)\,, 
\nonumber\\
&& L_{ab}(x,z,y) =   \sum^{c_1+c_2 \,\leq \, a+b}_{[c],[n], [o], [m], c\geq 2}\!\! \!\!\!\! \,(\bar q\cdot \bar p)^{m_1}\, \,(\bar q \cdot w)^{o_1}\,(\bar q\cdot p)^{n_1}\, \,(\bar p \cdot q)^{m_2}\,(q\cdot w)^{o_2}\,(p \cdot q)^{n_2}
\nonumber\\
&& \qquad \quad \qquad  \times \, (\bar p \cdot p)^{h_1}\, (w^2)^{h_2}\,(\bar q\cdot q)^{h_3}\,C^{(abc)}_{[m][o][n]}\, \big[ F(x,z,y) \big]^{c} \,c\,(c-1)\,
\,\frac{1+c +  \frac{3}{2}\,( d-4)}{[F(x,z,y)]^2} \,,
 \nonumber\\
 && D = \frac{2}{d-4} + \gamma_E -1 -\log (4\,\pi ) \,,
\end{eqnarray}
where we expand around $d=4$. 
A further step
\begin{eqnarray}
 &&  -i\,B^{(ab)}_{000} =  \frac{1}{16\,\pi^2} \int_0^1 d x  \int_0^{1-x} d y \,  \Bigg\{ \int_0^{1-x-y} d z\,\frac{\tilde N^{(0)}_a(x,z,y)}{[F(x,z,y)]^2  }
 +\frac{ \hat N_{ab}(x,1-x -y, y)}{F(x,1-x-y, y)} 
\nonumber\\
&& \qquad \qquad  -\, 
\frac{ \hat N_{ab}(x,0, y)}{F(x,0, y) }
+
\hat  L_{ab}(x,1-x -y, y)\,\Big\{D+ \log \frac{F(x,1-x-y, y)}{\mu^2} \Big\} 
  \nonumber\\
&& \qquad \qquad  - \, \hat L_{ab}(x,0,y)\,\Big\{ D+ \log \frac{F(x,0,y)}{\mu^2} \Big\} \Bigg\}  + {\mathcal O} \left( d- 4 \right)\,,
 \nonumber\\
&&  \tilde N^{(1)}_{ab}(x,z,y) = N^{(1)}_{ab}(x,z,y) - \hat L_{ab}(x,z,y)\, \partial_z F(x,z,y) \,, \quad \quad 
 \nonumber\\
&&  \tilde N^{(0)}_{ab}(x,z,y) = N^{(0)}_{ab}(x,z,y) - \hat N_{ab}(x,z,y) \, \partial_z F(x,z,y) \,,
\nonumber\\
&& \partial_z \hat L_{ab}(x,z,y) = L_{ab}(x,z,y)\,, \qquad \qquad \qquad 
 \partial_z \hat N_{ab}(x,z,y) = \tilde N^{(1)}_{ab}(x,z,y) \,,
\label{def-tildeN-box}
 \end{eqnarray}
shows that all scalar-box-type contributions take the form  $I_0(m,o,n)$ with $m+o+ n\leq a+b$ always.  This is so since in the vicinity of $d \sim 4$ it holds
\begin{eqnarray}
&& I_{0}(m,o,n) = \frac{1}{16\,\pi^{2}}\,\int_{0}^{1} d x\int_{0}^{1-x} d y\,\int_0^{1-x-y} d z\,\frac{x^m\,z^o\,y^n}{[F(x,z,y)]^2 }+ {\mathcal O} \left( d-4 \right)\,.
\label{def-I0mn}
\end{eqnarray}
All remaining terms have the form of scalar triangle and bubble diagrams as studied in Appendix B already. This follows by a partial integration in $x$ and $y$ of the log-type terms with either $z=0$ or $z=1-x-y$ fixed. We note that the terms with $z=1-x-y$ correspond to triangle loops that involve three heavy particles, and therefore can be dropped altogether in our renormalization scheme.  In contrast, the 
terms with $z=0$ correspond to triangle loops with two heavy and one light meson, contributions analogous to previously considered triangles. 

\newcommand{\vvQH}{v^{2}_{QH}}
We generalize our triangle relations (\ref{res-partial-integration-triangle}) to the box case. Suitable partial-integration lead to 
\begin{eqnarray}
&& I_{0}(m+1,o,n) = -\frac{\pfw}{\pfpf}\,I_{0}(m,o+1,n)-\frac{\pfpi}{\pfpf}\,I_{0}(m,o,n+1)
-\frac{1}{2\,\pfpf}\, \Sigma(m,o,n)
\nonumber\\
&& \qquad \quad 
+\,\frac{\bar p^2-M_L^2+m_G^2}{2\,\pfpf}\,I_{0}(m,o,n)
+\frac{\delta_{m,0}}{2\,\pfpf}\,I_{QH,R}^{(n,o)}-\frac{m}{d\,\pfpf}\,I_{1}(m-1,o,n)\,,
\nonumber\\
&&	I_{0}(m,o,n+1) = -\frac{\piw}{\pipi}\,I_{0}(m,o+1,n)
-\frac{\pfpi}{\pipi}\,I_{0}(m+1,o,n)-\frac{1}{2\,\pipi}\, \Sigma(m,o,n) 
\nonumber\\
&& \qquad \quad +\,\frac{p^2-M_R^2+ m_G^2}{2\,\pipi}\,I_{0}(m,o,n)
 +\,\frac{\delta_{n,0}}{2\,\pipi}\,I_{L,QH}^{(m,o)}-\frac{n}{d\,\pipi}\,I_{1}(m,o,n-1)\,,
\nonumber\\
&&I_{0}(m,o+1,n) = -\frac{\pfw}{\ww}\,I_{0}(m+1,o,n)-\frac{\piw}{\ww}\,I_{0}(m,o,n+1) -\frac{1}{2\,\ww}\, \Sigma(m,o,n) 
\nonumber\\
&& \qquad \quad
+\,\frac{w^2-M_H^2+m_G^2}{2\,\ww}\,I_{0}(m,o,n)
+\frac{\delta_{o,0}}{2\,\ww}\,I_{Q,LR}^{(m,n)}-\frac{o}{d\,\ww}\,I_{1}(m,o-1,n)\,,
\nonumber\\ \nonumber\\
&& \Sigma(m,o,n) = \sum_{k}^{o}(-1)^{k}
 \Bigg(\begin{array}{c} o \\ k \end{array}\Bigg)\,
\sum_{l}^{k}  \Bigg(\begin{array}{c}k\\ l\end{array}\Bigg) \,I_{H,LR}^{(m+k-l,n+l)}\,,
\label{res-partial-integration-box}
\end{eqnarray}
where we encounter the triangle functions $ I_{QH,R}^{(n,o)}$ and $I_{L,QH}^{(m,o)}$ from (\ref{def-n-triangle-QH}) and (\ref{def-I0mn}) with $x^n \to x^n \,y^o$ and $x^m \to x^m \,y^o$ respectively.
Similar  for $ I_{Q,LR}^{(m,n)}$ for  $ I_{H,LR}^{(m,n)}$. Altogether we have
\begin{eqnarray}
&& I_{L,QH}^{(m,n)} =  \frac{1}{16\,\pi^2} \int_0^1 d x  \int_0^{1-x} d y \,\frac{x^m\,y^n}{F(x,y,0)} +{\mathcal O}\left(d-4 \right)\,,
\nonumber\\
&& I_{QH,R}^{(m,n)} =  \frac{1}{16\,\pi^2} \int_0^1 d x  \int_0^{1-x} d y \,\frac{x^m\,y^n}{F(0,y,x)} +{\mathcal O}\left(d-4 \right)\,,
\nonumber\\
&& I_{Q,LR}^{(m,n)} =  \frac{1}{16\,\pi^2} \int_0^1 d x  \int_0^{1-x} d y \,\frac{x^m\,y^n}{F(x,0,y)} +{\mathcal O}\left(d-4 \right)\,,
\nonumber\\
&& I_{H,LR}^{(m,n)} =  \frac{1}{16\,\pi^2} \int_0^1 d x  \int_0^{1-x} d y \,\frac{x^m\,y^n}{F(x,1-x-y,y)}+{\mathcal O}\left(d-4 \right)\,.
\end{eqnarray}
The system (\ref{res-partial-integration-box}) can be used to express box contributions in terms of our extended basis functions $I_{QH,LR}^{(m,n)} =I_0(m,0,n)$. This is readily achieved in application of the third line equation in (\ref{res-partial-integration-box}). While it would be possible to further reduce the size of our extended basis set, this is possible only, at the prize of encountering kinematical singularities. By a suitable combination of the first two lines in  (\ref{res-partial-integration-box}), a relation among $I_{QH,LR}^{(m,n)}$ of different pairs of $m,n$ can be derived. An application of such relations would lead to kinematical singularities at $w \cdot p =0$ or $\bar p \cdot w=0.$

We provide two examples illustrating our advocated rewrite. First we consider the numerator $(l\cdot Q)$ in (\ref{def-Babij}), for which we find
\begin{eqnarray}
&& -i\,\big(B^{(1,0)}_{000} + B^{(0,1)}_{000} \big) = (\bar p \cdot Q)\,I_0(1,0,0 )  +(Q \cdot w)\,I_0(0,1,0) +
(Q \cdot p)\,I_0(0,0,1 )
\nonumber\\
&& \qquad =\,\frac{w \cdot Q}{2\,w^2}\,\big[w^2-M_H^2+ m_G^2 \big]\,I_{GH,LR}(s,t)+ \big[ (\bar p \cdot Q) - (\bar p\cdot w)\,(Q \cdot w)/w^2 \big]\,I^{(1,0)}_{GH,  L R}(s,t) 
\nonumber\\
&& \qquad +\, \big[ ( p \cdot Q) - ( p\cdot w)\,(Q \cdot w)/w^2 \big]\,I^{(0,1)}_{GH, L R}(s,t) + 
\frac{w \cdot Q}{2\,w^2}\,\Big( I_{G,LR}(t) -  I_{H,LR}(t)\Big)
\,, 
\label{red-lQ-box}
\end{eqnarray}
where we used the third identity in (\ref{res-partial-integration-box}). 
While the first line in (\ref{red-lQ-box}) appears to cause a conflict with the expectation of power-counting, a suitable rewrite as implied by  
(\ref{res-partial-integration-box}) leads to expressions that are consistent with this expectation after renormalization. Here we assume that the renormalized triangle loop functions $\bar I_{H,LR} \to 0$ vanish. 

We turn  to a $(l \cdot Q)^2$ numerator in  (\ref{def-Babij}). Following our strategy already used successfully for the corresponding triangle 
case in (\ref{res-JGLR2}), we manipulate that expression in two steps, with the first step involving the intermediate object $B^{(0,0)}_{010}$ and 
\begin{eqnarray}
&& -i\,\Big(  B^{(2,0)}_{000} +2\,B^{(1,1)}_{000} + B^{(0,2)}_{000}
\Big) = \Big[ (\bar p \cdot Q)^2 -\frac{Q^2}{d}\,(\bar p \cdot \bar p)\Big]\,I_0(2,0,0 )  
\nonumber\\
&& \qquad \qquad \qquad  
+ \,\Big[(Q \cdot p)^2- \frac{Q^2}{d}\,(p \cdot p)\Big]\,I_0(0,0,2 ) +
\Big[(Q \cdot w)^2- \frac{Q^2}{d}\,(w \cdot w)\Big]\,I_0(0,2,0) 
\nonumber\\
&& \qquad \qquad \qquad  +\,  2\,\Big[(\bar p \cdot Q)\,( Q \cdot p)
- \frac{Q^2}{d}\,( \bar p \cdot p)\Big] \,I_0(1,0,1 )
\nonumber\\
&& \qquad \qquad \qquad +\,2\,\Big[(\bar p \cdot Q)\,( Q \cdot w)
- \frac{Q^2}{d}\,(\bar p \cdot w)\Big]\,I_0(1,1,0 )
\nonumber\\ 
&& \qquad \qquad \qquad
+\,2\,\Big[( Q \cdot w)\,(Q \cdot p) - \frac{Q^2}{d}\,( w  \cdot p)\Big]\,I_0(0,1,1 ) 
\nonumber\\
&& \qquad \qquad \qquad+\,\frac{Q^2}{d}\,\Big[ m_G^2\,I_{GH,LR}(s,t) -I_{H,LR}(t) \Big]\,.
\label{res-lQlQ-box-1st}
\end{eqnarray}
Like in the triangle case, after the first step, the apparent chiral power of (\ref{res-lQlQ-box-1st}) is at odds with the expectation of dimensional counting rules. 
In the second step we apply the third line of (\ref{res-partial-integration-box}) as to eliminate $I_0(0,2,0)$ in favor of $I_0(1,1,0)$ and $I_0(0,1,1)$ and some triangle contributions. In the final step, upon a further application of the 
 third line of (\ref{res-partial-integration-box}), we are left with $I_0(2,0,0)$, $I_0 (0,0,2)$ and  $I_0(1,0,0)$, $I_0(0,0,1)$ and $I_0(1,0,1)$, all members of our extended basis set (\ref{def-IQHLR-nm}). The coefficients in front of such terms start at chiral order 4 in all cases. Similarly, after an application of (\ref{res-partial-integration-triangle}), the associated triangle basis functions contribute according to their expected chiral order. 
Altogether we find
\begin{eqnarray}
&&  -i\,\big(  B^{(2,0)}_{000} +2\,B^{(1,1)}_{000} + B^{(0,2)}_{000} \Big)_{ \rm ren.} = \frac{1}{12\,s}\,\Big\{ (s -M_H^2 + m_Q^2 ) \,\big[4\,Q_w^2/s -Q^2 \big]
\Big\}\,\bar I_{Q,LR}
\nonumber\\
&& \qquad +\,\frac{1}{6\,s}\,\Big\{ 6\,Q_{\bar p}\,Q_{w}-w_{\bar p}\,\big[2\,Q^2_w/s + Q^2 \big] \Big\}\,\bar I^{(1)}_{Q,\bar LR}+\frac{1}{6\,s}\,\Big\{ 6\,Q_{w}\,Q_{ p}-\big[2\,Q^2_w/s + Q^2 \big]\,w_{ p}   \Big\}  \,\bar I^{(1)}_{Q, L\bar R}
\nonumber\\
&& \qquad +\,\frac{1}{12\,s^2}\,\Big\{ 4\,Q_w^2\,(-M_H^2 + m_Q^2 + s)^2 + 
      s\,\big(-4\,m_Q^2\,Q_w^2 + 4\,m_Q^2\,Q^2\,s
      \nonumber\\
&& \qquad  \qquad \qquad -\, Q^2\,(-M_H^2 + m_Q^2 + s)^2 \big)\Big\}\,\bar I_{QH, LR}
      \nonumber\\
&& \qquad +\,\frac{Q_w}{s^2}\,\Big(s-M_H^2+m_Q^2 \Big)\,\Big\{
\big(Q_{\bar p}\,s-w_{\bar p}\,Q_w \big)\,
  \bar I_{QH,LR}^{(1,  0)} + \big(Q_p\,s -w_p\,Q_w \big)\,
  \bar I_{QH,LR}^{(0,  1)}   \Big\}   
\nonumber\\
&& \qquad +\, \frac{1}{3\,s}\, \Big\{-6\,Q_{\bar p}\,w_{\bar p}\,Q_w + w_{\bar p}^2\,(2\,Q_w^2/s + Q^2) + 
      3\,Q_{\bar p}^2\,s + \bar p^2\,(Q_w^2 - Q^2\,s) \Big\}\,\bar  I_{QH,LR}^{(2,  0)}
\nonumber\\
&& \qquad +\, \frac{1}{3\,s}\,\Big\{ -6\,Q_p\,w_p\,Q_w + w_p^2\,(2\,Q_w^2/s + Q^2) + 
      3\,Q_p^2\,s + p^2\,(Q_w^2 - Q^2\,s) \Big\} \,\bar  I_{QH,LR}^{(0,  2)}
\nonumber\\
&& \qquad +\,\frac{2}{3\,s}\,\Big\{-3\,\big[ Q_{\bar p}\,w_p + Q_p\,w_{\bar p}\big]\,Q_w + (\bar p \cdot p)\,(Q_w^2 - Q^2 \,s) + 3\,Q_{\bar p}\,Q_p\,s
\nonumber\\
&& \qquad \qquad  \qquad +\,
           w_{\bar p}\,w_p\,(2\,Q_w^2/s + Q^2 ) \Big\}\,\bar  I_{QH,LR}^{(1, 1)} + {\mathcal O} \left( d-4\right) \,,
\label{res-JGHLR2}
\end{eqnarray}
where we use the short-hand notation $a_b=a \cdot b$.
 We note that our result can be readily generalized for the case defined by $ (l \cdot Q)^2 \to (l \cdot \bar q) \, (l\cdot q)$. Like in the corresponding triangle case  it suffices to use the replacement  $Q_\mu \,Q_\nu \to (\bar q_\mu \, q_\nu + q_\mu \,\bar q_\nu)/2 $ in (\ref{res-JGHLR2}).

\section*{Appendix E: Fourth order box-loop expressions }

\begin{eqnarray}
&&M_{ab}^2\,\bar J^{(0), 4}_{QH,LR} (s,t) =
-\frac{m_Q^2}{144\,(4\, \pi)^2}\,\Big((s-u)^2+ 4\,M_{ab}^2\,(t- m_{ab}^2 )  \Big)
\nonumber\\
&& \qquad + \,
\frac{1}{96}\,\Big( (s-u)^2 + 4\,M_{ab}^2\,(t-m_{ab}^2)\Big)\,\bar I_Q 
-\,\frac{1}{64}\,\Big((s-u)^3 -2\,M_{ab}^2\,(s-u)\,t \Big)\,\bar I_{QH}(s)
\nonumber\\
&& \qquad 
+\, \frac{m_a^2 - m_b^2}{64}\,(s-u)\,M_{ab}^2\,\Big[ \bar I_{QL}(M_a^2)- \bar I_{QR}(M_b^2)\Big]
\nonumber\\
&& \qquad +\,\frac{M_{ab}^2}{384}\,\Big((s-u)^3/M_{ab}^2 +2\,(s-u)\,\big( m_{ab}^2 -2\,t-r\,(s-u) \big)
\nonumber\\
&& \qquad \quad
-\, 8\,r\,M_{ab}^2\,(t-m_{ab}^2)\Big)\, \Big[ \bar I_{QL}(M_a^2)+ \bar I_{QR}(M_b^2)\Big]
\nonumber\\
&& \qquad +\,\frac{M_{ab}^2}{384}\,
 \Big( (s-u)^2\,\big( -6\,\delta_H +3\,( \delta_L+ \delta_R ) +8\,m_Q^2
  + 7\,m_{ab}^2 + 3\, r \,(s-u ) - 7 \,t \big) 
 \nonumber\\
&& \qquad \quad - \,(s-u)^4/M_{ab}^2 
-2\, M_{ab}^2 \,\big( 8\, m_Q^2 \,(t - m_{ab}^2 ) + 3 \,r \,(s-u)\, t \big)
\Big)\,\Big[ \bar I_{L,QH}(s) + \bar I_{QH,R}(s) \Big]
\nonumber\\
&& \qquad - \,\frac{M_{ab}^2}{128}\,\Big(4 \,(m_a^2-m_b^2)^2 \,M_{ab}^2 + 4 \,(m_a^2-m_b^2)\, (M_a^2-M_b^2) \,M_{ab}^2 
+ (s - u)^2 (t -r (s - u) )
\nonumber\\
&& \qquad \quad + \, M_{ab}^2\, \big(4\, r \,m_{ab}^2\, (s - u) - 4 \,m_{ab}^2\, t - 2 \,r \,(s - u)\, t \big) \Big)\,\Big[ \bar I^{(1)}_{L,QH}(s) +\bar I^{(1)}_{QH,R}(s) \Big]
 \nonumber\\
&& \qquad +\,\frac{1}{384}\,\Big((s-u)^2 - 4 \,M_{ab}^2 \,m_{ab}^2  \Big)\,\Big( (s - u)^2 + 4 \,M_{ab}^2\, ( t- m_{ab}^2 )
\Big)\,\Big[ \bar I^{(2)}_{L,QH}(s) + \bar I^{(2)}_{QH,R}(s) \Big]
 \nonumber\\
&& \qquad +\,\frac{M_{ab}^2}{128}\,(s-u)^2\,\Big( \delta_L -\delta_R +2\,(m_a^2-m_b^2) -(M_a^2-M_b^2)\Big)
\Big[\bar I_{L,QH}(s)- \bar I_{QH,R}(s) \Big]
\nonumber\\
&& \qquad - \,\frac{M_{ab}^2}{128}\,\Big( -(m_a^2-m_b^2) \,(s - u)^2 + (M_a^2-M_b^2) \,\big( 4 \,M_{ab}^2 \,m_{ab}^2 + (s - u)^2 \big) 
 \nonumber\\
&& \qquad \quad +\, 
  2\, (m_a^2-m_b^2)\, M_{ab}^2\, \big(2 \,m_{ab}^2 + r\, (s - u) - 2\, t \big)\Big)\,\Big[ \bar I^{(1)}_{L,QH}(s) -\bar I^{(1)}_{QH,R}(s)\Big]
 \nonumber\\
&& \qquad -\,\frac{M_{ab}^2}{96}\,  (m_a^2-m_b^2)\, \big( (s - u)^2 + 4 \,M_{ab}^2\, (t -m_{ab}^2 ) \big)\,\Big[ \bar I^{(2)}_{L,QH}(s) -  I^{(2)}_{QH,R}(s) \Big]
\nonumber\\
&& \qquad +\,\frac{M_{ab}^2}{384}\,\Big( -2\,( \delta_L +\delta_R)\,\big( (s-u)^2 -2\, M_{ab}^2 \,(t -m_{ab}^2 ) \big)  
+8\,\delta_H\,\big( (s-u)^2 + M_{ab}^2 \,(t -m_{ab}^2 ) \big)  
 \nonumber\\
&& \qquad \qquad +\,8\,(s-u)^4/M_{ab}^2 - r\,(s- u)^3 + (s-u)^2\,\big(16\,t- 24\,m_{ab}^2 - 28\,m_Q^2 +r^2\,M_{ab}^2 \big) 
 \nonumber\\
&& \qquad \qquad 
 +\, 4\,M_{ab}^2\,( t -m_{ab}^2 )\,\big(t -m_{ab}^2 -4\,m_Q^2 + r^2\,M_{ab} \big)
 \nonumber\\
&& \qquad \qquad +\,2\, (s-u)\,r\,M_{ab}^2\,\big(2\,t - m_{ab}^2 \big)
\Big)\,\bar I_{Q,LR}(t)
\nonumber\\
&& \qquad -\,\frac{M_{ab}^2}{96}\,\Big( (s-u)^2 + M_{ab}^2 \,(t - m_{ab}^2)  \Big)\,t\,\Big[\bar I^{(1)}_{Q,\bar LR}(t)+ \bar I^{(1)}_{Q,L\bar R}(t)\Big]
\nonumber\\
&& \qquad +\,\frac{M_{ab}^2}{192}\,\Big( -7\, (m_a^2 -m_b^2) \,(s - u)^2 + 2\, (m_a^2 -m_b^2)\, M_{ab}^2\, (5\, m_{ab}^2 + 3\, r\, (s - u) - 5 \,t)
\nonumber\\
&& \qquad \qquad + \,
 3\, (M_a^2-M_b^2)\, \big((s - u)^2 + 2\, M_{ab}^2\, t\big)  \Big)\,\Big[\bar I^{(1)}_{Q,\bar LR}(t)- \bar I^{(1)}_{Q,L\bar R}(t)\Big]
\nonumber\\
&& \qquad +\,\frac{1}{192}\,\Big( 2\, (s - u)^5 +  (M_{ab}^2)^3\, r\, \big(3 \,r \,(s - u)\, t + 16 \,m_Q^2 (t -m_{ab}^2) \big) /2
\nonumber\\
&& \qquad \qquad +\,
  M_{ab}^2\, (s - u)^3\, \big(8\, \delta_H - 3 \,(\delta_L + \delta_R ) - 18\, m_Q^2 - 14\, m_{ab}^2 
 +     r \,(s - u) + 10\, t \big)/2
 \nonumber\\
&& \qquad \qquad- \,
M_{ab}^2\,M_{ab}^2 \,(s - u) \,\big(-8\, (m_{ab}^2)^2 + 6\, (\delta_L + \delta_R)\, r\, (s - u) 
 \nonumber\\
&& \qquad \qquad  + \,
    14 \,r\, m_{ab}^2\, (s - u) + 3\, r^2\, (s - u)^2 + 4 \,\delta_H \,(4 \,m_{ab}^2 - 3\, r\, (s - u) - 4\, t) 
    + 16\, m_{ab}^2 \,t 
\nonumber\\
&& \qquad \qquad
   -\, 14\, r\, (s - u)\, t - 8 \,t^2 + 16\, m_Q^2\, \big(-3\, m_{ab}^2 + r\, (s - u) + 2 \,t \big) \big)/4\Big)\,\bar I_{QH,LR}(s,t)
\nonumber\\
&& \qquad +\,\frac{1}{256}\,  \Big( 4 \,(m_a^2 -m_b^2)\, M_{ab}^2\,(m_a^2 -m_b^2)\,M_{ab}^2\, \big(r\,M_{ab}^2 - (s - u) \big)
\nonumber\\
&& \qquad \qquad+ \,
 4 \,(m_a^2 -m_b^2)\, M_{ab}^2\,M_{ab}^2\, \big( (M_a^2-M_b^2)\,(2\,(s-u)+ r\,M_{ab}^2) - (\delta_L-\delta_R)\, (s - u) \big)
  \nonumber\\
&& \qquad \qquad
  +\, (4 \,M_{ab}^2\, m_{ab}^2 - (s - u)^2 - 
    2\, M_{ab}^2\, t) \,\big( 2\, (\delta_L +\delta_R -2\,\delta_H)\,M_{ab}^2\, (s - u) 
  \nonumber\\
&& \qquad \qquad    + \, 8\, M_{ab}^2\, m_Q^2\, (s - u)
+ M_{ab}^2\,M_{ab}^2\, r^2 \,(s - u)+ 6 \,M_{ab}^2\, m_{ab}^2\, (s - u)
\nonumber\\
&& \qquad \qquad - \,
     4\, (s - u)^3 - 2\, r\,M_{ab}^2\,M_{ab}^2\, t - 4\, M_{ab}^2 \,(s - u) \,t \big) \Big)\,\Big[\bar I^{(1,0)}_{QH,LR}(s,t)+ \bar I^{(0,1)}_{QH,LR}(s,t) \Big]
\nonumber\\
&& \qquad +\,\frac{M_{ab}^2}{128}\,  \Big( 2\, \big( - (\delta_L-\delta_R)\,t  + \big( \delta_L +\delta_R- 2\,\delta_H\big)\, (m_a^2 -m_b^2) \big)\, M_{ab}^2\, (s - u)
\nonumber\\
&& \qquad \qquad +\, (m_a^2 -m_b^2)\, M_{ab}^2 \,\big( 8\,m_Q^2 -6\,( t-m_{ab}^2)\big)\, (s - u) + 
 (m_a^2 -m_b^2)\,r^2\, M_{ab}^2\,M_{ab}^2\,  (s - u) 
 \nonumber\\
&& \qquad \qquad + \,  (m_a^2 -m_b^2)\, M_{ab}^2\, r \,(s - u)^2 + 
 (M_a^2-M_b^2)\,r\, M_{ab}^2 \,(s - u)^2 - 3\, (m_a^2 -m_b^2) \,(s - u)^3
 \nonumber\\
&& \qquad \qquad + \,(M_a^2-M_b^2) \,(s - u)^3 + 2 \,(M_a^2-M_b^2)\, r\,M_{ab}^2\,M_{ab}^2  \,t 
\nonumber\\
&& \qquad \qquad
+\, 4 \,(M_a^2-M_b^2)  \,M_{ab}^2 \,(s - u) \,t \Big)\,\Big[\bar I^{(1,0)}_{QH,LR}(s,t)- \bar I^{(0,1)}_{QH,LR}(s,t) \Big]   
\nonumber\\
&& \qquad + \,\frac{M_{ab}^2}{192}\,\Big( 12\, (m_a^2 -m_b^2)^2\, \big(r\, M_{ab}^2  - (s - u) \big) + 12\, (m_a^2 -m_b^2)\, (M_a^2-M_b^2)\,(s - u) 
\nonumber\\
&& \qquad \qquad- \,  \big( r\,M_{ab}^2  - 2\,(s - u) \big)\, \big(4 \,m_{ab}^2 - (s - u)^2/M_{ab}^2 \big)^2 
+  4 \, \big(-r\,M_{ab}^2 + (s - u) \big)\, t^2
 \nonumber\\
&& \qquad \qquad + \,
 2\, \big( r\,M_{ab}^2 - \,3\, (s - u) \big)\, \big(4\, m_{ab}^2 - (s - u)^2 /M_{ab}^2 \big)\, t 
\Big) \, \bar I^{(1,1)}_{QH,LR}(s,t) 
\nonumber\\
&& \qquad +\,\frac{M_{ab}^2 }{384}\,\Big( 
 4\, M_{ab}^2\, (s - u)\, \big( 4\, m_{ab}^2 +\, r\, (s - u) - 2\, t\big)\,(2\, m_{ab}^2 - t)
 \nonumber\\
&& \qquad \qquad 
-  (s - u)^3 \,\big(16 \,m_{ab}^2 
+ r\, (s - u) - 8\, t)+ 
 16\, r\,(M_{ab}^2)^2\, m_{ab}^2 \,(t- m_{ab}^2 \big) 
 \nonumber\\
&& \qquad \qquad +\, 2\, (s - u)^5/M_{ab}^2 
  \Big)\,\Big[\bar I^{(2,0)}_{QH,LR}(s,t)+ \bar I^{(0,2)}_{QH,LR}(s,t) \Big]
\nonumber\\
&& \qquad  -\,\frac{M_{ab}^2}{192}\,\Big( - (M_a^2-M_b^2) (s - u) \,\big( (s - u)^2 + 4\, M_{ab}^2\, (t-m_{ab}^2) \big) 
\nonumber\\
&& \qquad \qquad +\,
(m_a^2 -m_b^2) \,\big(3\, (s - u)^3 - 2 M_{ab}^2 \,(s - u)\, (6\, m_{ab}^2 +r\, (s - u) - 4 \,t)
\nonumber\\
&& \qquad \qquad - \,
    8\, r\,M_{ab}^2\,M_{ab}^2 \,(t - m_{ab}^2 ) \big)  \Big)\,\Big[\bar I^{(2,0)}_{QH,LR}(s,t)- \bar I^{(0,2)}_{QH,LR}(s,t) \Big] \,,
\end{eqnarray}
 and 
\begin{eqnarray}
&&M_{ab}^2\,\bar J^{(1), 4}_{QH,LR} (s,t) = \frac{m_Q^2}{144\,(4\, \pi)^2}\,\Big(77\,(s-u)^2/8+ 25\,M_{ab}^2\,(t- m_{ab}^2 )  \Big)
\nonumber\\
&& \qquad +\,
\frac{1}{384}\,\Big(
11\,(s-u)^2 + 32\,M_{ab}^2\,(t -m_{ab}^2 )
\Big)\,\bar I_Q 
\nonumber\\
&& \qquad +\,\frac{ M_{ab}^2}{16}\,\Big( (s-u)^3/ M_{ab}^2 - 5\,r\,M_{ab}^2 \, (t -m_{ab}^2 ) + (s-u)\,\big(t -m_{ab}^2 - r\, (s-u) \big)\Big)\,\bar I_{QH}(s)
\nonumber\\
&& \qquad 
+\, \frac{M_{ab}^2}{64}
\,(m_a^2-m_b^2)\,(s-u) \,\Big[ \bar I_{QL}(M_a^2) -\bar I_{QR}(M_b^2)\Big]
\nonumber\\
&& \qquad +\, \frac{M_{ab}^2}{768}
\,\Big(-14\,(s-u)^3/M_{ab}^2 + (s-u)\,\big(9\,r\,(s-u) - 14\,t -4\, m_{ab}^2\big) 
\nonumber\\
&& \qquad \qquad
+\, 72\,M_{ab}^2\,r\,(t -m_{ab}^2) \Big) \,\Big[ \bar I_{QL}(M_a^2) +\bar I_{QR}(M_b^2)\Big]
\nonumber\\
&& \qquad +\,
 \frac{M_{ab}^2}{384}
\,\Big( -12\,(\delta_L+\delta_R)\,M_{ab}^2\,m_{ab}^2  + 88\,M_{ab}^2\,m_Q^2\,m_{ab}^2 
\nonumber\\
&& \qquad \qquad
- \, 20\,r^2\,M_{ab}^2\,M_{ab}^2\,m_{ab}^2 - 14\,r\,M_{ab}^2\,m_{ab}^2\,(s-u) + 6\,\delta_H\,(s-u)^2 
\nonumber\\
&& \qquad \qquad
-\, 22\,m_Q^2\,(s-u)^2 + 5\,r^2\,M_{ab}^2\,(s-u)^2 - 5\,m_{ab}^2\,(s-u)^2 
\nonumber\\
&& \qquad \qquad
- \, r\,(s-u)^3 - 4\,(s-u)^4/M_{ab}^2 + t\,M_{ab}^2\,
 \big(12\,(\delta_L + \delta_R ) - 88\,m_Q^2 + 20\,r^2\,M_{ab}^2
 \nonumber\\
&& \qquad \qquad
 + \, 20\,r\,(s-u) - (s-u)^2/M_{ab}^2 \big)\Big) \,\Big[ \bar I_{L,QH}(s) +\bar I_{QH,R}(s)\Big]
\nonumber\\
&& \qquad +\,\frac{M_{ab}^2}{1024}\,\Big(
-24\,(m_a^2-m_b^2)^2\,M_{ab}^2 + 24\,(m_a^2-m_b^2)\,(M_a^2-M_b^2)\,M_{ab}^2 
\nonumber\\
&& \qquad \qquad
-\, 16\,(s-u)^4/M_{ab}^2 + 
 (s-u)^2\,\big(64\,m_{ab}^2 - 8\,r\,(s-u) - 33\,t\big)
 \nonumber\\
&& \qquad \qquad
 + \, 4\,M_{ab}^2\,\big(8\,r\,m_{ab}^2\,(s-u) + 12\,m_{ab}^2\,t - r\,(s-u)\,t - 6\,t^2 \big)
\Big) \,\Big[ \bar I^{(1)}_{L,QH}(s) +\bar I^{(1)}_{QH,R}(s)\Big]
\nonumber\\
&& \qquad -\,
 \frac{1}{192}\,
 \big( (s-u)^2 -4\,M_{ab}^2\,m_{ab}^2  \big)\, \big((s-u)^2 - 4\,M_{ab}^2\,(m_{ab}^2 - t) \big)\,
 \Big[ \bar I^{(2)}_{L,QH}(s) +\bar I^{(2)}_{QH,R}(s)\Big]
\nonumber\\
&& \qquad -\,\frac{M_{ab}^2}{1024}\,
\big(\delta_L - \delta_R - (M_a^2-M_b^2)\big)\,\big(9\,(s-u)^2 - 16\,M_{ab}^2\,(m_{ab}^2 - t)\big)\,
\Big[ \bar I_{L,QH}(s) -\bar I_{QH,R}(s)\Big]
\nonumber\\
&& \qquad +\,\frac{M_{ab}^2}{1024}\,\Big( (m_a^2-m_b^2)\,(s-u)\,\big(28\,M_{ab}^2\,r + 47\,(s-u)\big)
\nonumber\\
&& \qquad \qquad +\, 8\,(M_a^2-M_b^2)\,\big( (s-u)^2 + M_{ab}^2\,(2\,m_{ab}^2 + t) \big)
\Big)\,
\Big[ \bar I^{(1)}_{L,QH}(s) -\bar I^{(1)}_{QH,R}(s)\Big]
\nonumber\\
&& \qquad +\,\frac{M_{ab}^2}{48}\,
(m_a^2-m_b^2)\ \big(-4\,M_{ab}^2\,m_{ab}^2 + (s-u)^2 + 4\,M_{ab}^2\,t \big)\,
\Big[ \bar I^{(2)}_{L,QH}(s) -\bar I^{(2)}_{QH,R}(s)\Big]
\nonumber\\
&& \qquad +\,\frac{M_{ab}^2}{768}\,\Big( 4\,(\delta_L +\delta_R )\,\big((s-u)^2 
-2\,M_{ab}^2\,(t -m_{ab}^2)\big)
-16\,\delta_H\,\big((s-u)^2 
+M_{ab}^2\,(t -m_{ab}^2)\big)
\nonumber\\
&& \qquad \qquad -  \,8\,(s-u)^4/M_{ab}^2 +14\,r\,(s-u)^3 +
20\, (s-u)\,r\,M_{ab}^2\,(t-m_{ab}^2)
\nonumber\\
&& \qquad \qquad -\,2\, (s-u)^2\,\big( 11\,t-20\,m_{ab}^2-14\,m_Q^2 +13\,r^2\,M_{ab}^2/4\big)
\nonumber\\
&& \qquad \qquad -\,8\,M_{ab}^2\, (t- m_{ab}^2)\,\big(t- m_{ab}^2 -20\,m_Q^2+4\,r^2\,M_{ab}^2 \big)
\Big)\,\bar I_{Q,LR}(s)
\nonumber\\
&& \qquad +\,\frac{M_{ab}^2}{96}\,
t\,\big((s-u)^2 + M_{ab}^2\,(t -m_{ab}^2) \big)\,
\Big[ \bar I^{(1)}_{Q,\bar LR}(s) + \bar I^{(1)}_{Q,L\bar R}(s)\Big]
\nonumber\\
&& \qquad +\,\frac{M_{ab}^2}{96}\,
\Big(6\,M_{ab}^2\,(m_a^2-m_b^2)^2  +t\,\big( (s-u)^2 -2\,M_{ab}^2\,(t+ 2\,m_{ab}^2) \big) \Big)\,
\Big[ \bar I^{(2)}_{Q,\bar LR}(s) + \bar I^{(2)}_{Q,L\bar R}(s)\Big]
\nonumber\\
&& \qquad +\,\frac{M_{ab}^2}{192}\,\Big( 
(m_a^2-m_b^2)\,\big( (s-u)^2 - 2\,M_{ab}^2\,(5\,m_{ab}^2 + 3\,r\,(s-u) - 5\,t) \big) 
\nonumber\\
&& \qquad \qquad
- \, 3 \,(M_a^2-M_b^2)\, \big( (s-u)^2 + 2\, M_{ab}^2\, t \big)
\Big)\,
\Big[ \bar I^{(1)}_{Q,\bar LR}(s) - \bar I^{(1)}_{Q,L\bar R}(s)\Big]
\nonumber\\
&& \qquad +\, \frac{M_{ab}^2}{384}\,\Big( 
4\,M_{ab}^2\,M_{ab}^2\,r\,m_{ab}^2\,\big( - 12\,m_Q^2 + M_{ab}^2\,r^2 + m_{ab}^2 \big)
\nonumber\\
&& \qquad \qquad
+ \,    4\,M_{ab}^2\,\big(-14\,m_Q^2 + 3\,M_{ab}^2\,r^2 - m_{ab}^2 \big)\,m_{ab}^2\,(s-u)
\nonumber\\
&& \qquad \qquad    
    - \,     M_{ab}^2\,r\,\big( - 12\,m_Q^2 + M_{ab}^2\,r^2 + 13\,m_{ab}^2 \big)\,(s-u)^2 
\nonumber\\
&& \qquad \qquad    
    + \,     \big(14\,m_Q^2 - 3\,M_{ab}^2\,r^2 + 12\,m_{ab}^2 \big)\,(s-u)^3 + 
    6\,r\,(s-u)^4 - 2\,(s-u)^5/M_{ab}^2 
\nonumber\\
&& \qquad \qquad    
    - \,
 2\,\delta_H\,\big( 4\,(s-u)^3 - M_{ab}^2\,(s-u)\,\big( 4\,m_{ab}^2 + r\,(s-u) - 4\,t \big) + 
    4\,r\,M_{ab}^2\,M_{ab}^2\,(m_{ab}^2 - t) \big)
\nonumber\\
&& \qquad \qquad    
    + \, 3\,(\delta_L + \delta_R)\,\big(-M_{ab}^2\,r\,(s-u)^2 + (s-u)^3 + 4\,r\,M_{ab}^2\,M_{ab}^2\,(m_{ab}^2 - t) \big) 
\nonumber\\
&& \qquad \qquad 
 + \,t\,\big(-8\,r\,M_{ab}^2\,M_{ab}^2\, \big(-6\,m_Q^2 +  r^2\,M_{ab}^2/2 + m_{ab}^2  \big)
\nonumber\\
&& \qquad \qquad 
 + \, M_{ab}^2\,\big( 68\,m_Q^2 + 8\,m_{ab}^2- 27\,r^2\,M_{ab}^2/2  \big)\, (s-u) + 10\,r\,M_{ab}^2\,(s-u)^2  -  15\,(s-u)^3/2 \big)
\nonumber\\
&& \qquad \qquad    
    + \,4\,M_{ab}^2\,t^2\,\big(M_{ab}^2\,r - (s-u)\big)
\Big)\,\bar I_{QH,LR}(s,t)
\nonumber\\
&& \qquad +\,\frac{M_{ab}^2}{128}\,\Big( 
-(s-u)^5/M_{ab}^2 + (s-u)^3\,\big(-2\,\delta_H + \delta_L + \delta_R + 7\,m_{ab}^2 + 2\,r\,(s-u) 
-    4\,t \big) 
\nonumber\\
&& \qquad \qquad +\,4\,M_{ab}^2\,M_{ab}^2\,r\,\big(2\,m_{ab}^2\,m_{ab}^2 - 3\,m_{ab}^2\,t + t^2 \big) 
 + 
 M_{ab}^2\,(s-u)\,\big( - 4\,\delta_H\,t 
\nonumber\\
&& \qquad \qquad
- \,\delta_R \,\big( (m_a^2-m_b^2) + 4\,m_{ab}^2 - 2\, t \big) + \delta_L\,\big( (m_a^2-m_b^2) - 4\,m_{ab}^2 + 2\,t \big) 
\nonumber\\
&& \qquad \qquad   +\, 2\,(m_a^2-m_b^2)^2 - 3\,(m_a^2-m_b^2)\,(M_a^2-M_b^2) + 8\,\delta_H\,m_{ab}^2 - 12\,m_{ab}^2\,m_{ab}^2 
 \nonumber\\
&& \qquad \qquad
 - \,    10\,r\,m_{ab}^2\,(s-u) 
    +  14\,m_{ab}^2\,t + 6\,r\,(s-u)\,t - 4\,t^2  \big)
\Big)\,
\Big[ \,\bar I^{(1,0)}_{QH,LR}(s,t) + \bar I^{(0,1)}_{QH,LR}(s,t)\Big]
\nonumber\\
&& \qquad -\,\frac{M_{ab}^2}{128}\,\Big( 
 2\,(m_a^2-m_b^2)\,M_{ab}^2\,(s-u)\, \big(  \delta_L + \delta_R -2\,\delta_H + 3\,m_{ab}^2 + 2\,r\,(s-u) - 3\, t \big) 
\nonumber\\
&& \qquad \qquad    
- \,(m_a^2-m_b^2)\,(s-u)^3 + 
    (\delta_R - \delta_L)\,M_{ab}^2\,(s-u)\,t + 4\,(m_a^2-m_b^2)\,M_{ab}^2\,M_{ab}^2\,r\,(t -m_{ab}^2) 
\nonumber\\
&& \qquad \qquad
+\,(M_a^2-M_b^2)\,(s-u)\,((s-u)^2 + 3\,M_{ab}^2\,t)
\Big)\,
\Big[ \,\bar I^{(1,0)}_{QH,LR}(s,t) - \bar I^{(0,1)}_{QH,LR}(s,t)\Big]
\nonumber\\
&& \qquad -\,\frac{M_{ab}^2}{192}\,\Big( 
12\,(m_a^2-m_b^2)\,(M_a^2-M_b^2)\,M_{ab}^2\,(s-u) 
+ 
 3\,(m_a^2-m_b^2)^2\,M_{ab}^2\,\big(4\,r\,M_{ab}^2 + (s-u) \big) 
 \nonumber\\
&& \qquad \qquad 
 + \,\big((s-u)-r\,M_{ab}^2  \big)\,\big( (s-u)^2  -4\,M_{ab}^2\,m_{ab}^2 
 \big)^2/M_{ab}^2 
 - \,t^2\,
 M_{ab}^2\,\big( (s-u)+4\,r\,M_{ab}^2  \big)
 \nonumber\\
&& \qquad \qquad
 + \,t\, \,\big( 9\,(s-u)-4\,r\,M_{ab}^2 \big)\,\big(  (s-u)^2- 4\,M_{ab}^2\,m_{ab}^2\big) /2
\Big)\,
 \,\bar I^{(1,1)}_{QH,LR}(s,t) 
\nonumber\\
&& \qquad +\,\frac{M_{ab}^2}{768}\,\Big( 
 (s-u)^3\,\big(16\,m_{ab}^2 + 2\,r\,(s-u) - 9\,t \big) + 32\,r\,M_{ab}^2\,M_{ab}^2\,m_{ab}^2\,(m_{ab}^2 - t) 
 \nonumber\\
&& \qquad \qquad - \,2\,(s-u)^5/M_{ab}^2  - 
 2\,M_{ab}^2\,(s-u)\,\big(3\,(m_a^2-m_b^2)^2 
 + 16\,m_{ab}^2\,m_{ab}^2 + 8\,r\,m_{ab}^2\,(s-u) 
 \nonumber\\
&& \qquad \qquad 
 - 18\,m_{ab}^2\,t - 
    4\, r\, (s-u)\, t + 11 \,t^2 \big)
\Big)\,
\Big[ \,\bar I^{(2,0)}_{QH,LR}(s,t) + \bar I^{(0,2)}_{QH,LR}(s,t)\Big]
\nonumber\\
&& \qquad -\,\frac{M_{ab}^2}{192}\,\Big(
(M_a^2-M_b^2)\,(s-u)\,\big(-4\,M_{ab}^2\,m_{ab}^2 + (s-u)^2 + 4\,M_{ab}^2\,t\big) 
\nonumber\\
&& \qquad \qquad
+ \,
 (m_a^2-m_b^2)\,\big(-(s-u)^3 + 8\,M_{ab}^2\,M_{ab}^2\,r\,(t -m_{ab}^2)
\nonumber\\
&& \qquad \qquad 
 + \,2\,M_{ab}^2\,(s-u)\,(2\,m_{ab}^2 + r\,(s-u) + t) \big)
\Big)\,
\Big[ \,\bar I^{(2,0)}_{QH,LR}(s,t) - \bar I^{(0,2)}_{QH,LR}(s,t)\Big]
\,.
\end{eqnarray}

\bibliography{literature}

\begin{thebibliography}{69}%
\makeatletter
\providecommand \@ifxundefined [1]{%
 \@ifx{#1\undefined}
}%
\providecommand \@ifnum [1]{%
 \ifnum #1\expandafter \@firstoftwo
 \else \expandafter \@secondoftwo
 \fi
}%
\providecommand \@ifx [1]{%
 \ifx #1\expandafter \@firstoftwo
 \else \expandafter \@secondoftwo
 \fi
}%
\providecommand \natexlab [1]{#1}%
\providecommand \enquote  [1]{``#1''}%
\providecommand \bibnamefont  [1]{#1}%
\providecommand \bibfnamefont [1]{#1}%
\providecommand \citenamefont [1]{#1}%
\providecommand \href@noop [0]{\@secondoftwo}%
\providecommand \href [0]{\begingroup \@sanitize@url \@href}%
\providecommand \@href[1]{\@@startlink{#1}\@@href}%
\providecommand \@@href[1]{\endgroup#1\@@endlink}%
\providecommand \@sanitize@url [0]{\catcode `\\12\catcode `\$12\catcode
  `\&12\catcode `\#12\catcode `\^12\catcode `\_12\catcode `\%12\relax}%
\providecommand \@@startlink[1]{}%
\providecommand \@@endlink[0]{}%
\providecommand \url  [0]{\begingroup\@sanitize@url \@url }%
\providecommand \@url [1]{\endgroup\@href {#1}{\urlprefix }}%
\providecommand \urlprefix  [0]{URL }%
\providecommand \Eprint [0]{\href }%
\providecommand \doibase [0]{http://dx.doi.org/}%
\providecommand \selectlanguage [0]{\@gobble}%
\providecommand \bibinfo  [0]{\@secondoftwo}%
\providecommand \bibfield  [0]{\@secondoftwo}%
\providecommand \translation [1]{[#1]}%
\providecommand \BibitemOpen [0]{}%
\providecommand \bibitemStop [0]{}%
\providecommand \bibitemNoStop [0]{.\EOS\space}%
\providecommand \EOS [0]{\spacefactor3000\relax}%
\providecommand \BibitemShut  [1]{\csname bibitem#1\endcsname}%
\let\auto@bib@innerbib\@empty
\bibitem [{\citenamefont {Casalbuoni}\ \emph {et~al.}(1997)\citenamefont
  {Casalbuoni}, \citenamefont {Deandrea}, \citenamefont {Di~Bartolomeo},
  \citenamefont {Gatto}, \citenamefont {Feruglio} \emph
  {et~al.}}]{Casalbuoni:1996pg}%
  \BibitemOpen
  \bibfield  {author} {\bibinfo {author} {\bibfnamefont {R.}~\bibnamefont
  {Casalbuoni}}, \bibinfo {author} {\bibfnamefont {A.}~\bibnamefont
  {Deandrea}}, \bibinfo {author} {\bibfnamefont {N.}~\bibnamefont
  {Di~Bartolomeo}}, \bibinfo {author} {\bibfnamefont {R.}~\bibnamefont
  {Gatto}}, \bibinfo {author} {\bibfnamefont {F.}~\bibnamefont {Feruglio}},
  \emph {et~al.},\ }\href {\doibase 10.1016/S0370-1573(96)00027-0} {\bibfield
  {journal} {\bibinfo  {journal} {Phys.Rept.}\ }\textbf {\bibinfo {volume}
  {281}},\ \bibinfo {pages} {145} (\bibinfo {year} {1997})},\ \Eprint
  {http://arxiv.org/abs/hep-ph/9605342} {arXiv:hep-ph/9605342 [hep-ph]}
  \BibitemShut {NoStop}%
\bibitem [{\citenamefont {Kolomeitsev}\ and\ \citenamefont
  {Lutz}(2004)}]{Kolomeitsev:2003ac}%
  \BibitemOpen
  \bibfield  {author} {\bibinfo {author} {\bibfnamefont {E.~E.}\ \bibnamefont
  {Kolomeitsev}}\ and\ \bibinfo {author} {\bibfnamefont {M.~F.~M.}\
  \bibnamefont {Lutz}},\ }\href {\doibase 10.1016/j.physletb.2003.10.118}
  {\bibfield  {journal} {\bibinfo  {journal} {Phys. Lett. B}\ }\textbf
  {\bibinfo {volume} {582}},\ \bibinfo {pages} {39} (\bibinfo {year} {2004})},\
  \Eprint {http://arxiv.org/abs/hep-ph/0307133} {arXiv:hep-ph/0307133}
  \BibitemShut {NoStop}%
\bibitem [{\citenamefont {Lutz}\ \emph {et~al.}(2016)\citenamefont {Lutz} \emph
  {et~al.}}]{Lutz:2015ejy}%
  \BibitemOpen
  \bibfield  {author} {\bibinfo {author} {\bibfnamefont {M.~F.~M.}\
  \bibnamefont {Lutz}} \emph {et~al.},\ }\href {\doibase
  10.1016/j.nuclphysa.2016.01.070} {\bibfield  {journal} {\bibinfo  {journal}
  {Nucl. Phys.}\ }\textbf {\bibinfo {volume} {A948}},\ \bibinfo {pages} {93}
  (\bibinfo {year} {2016})},\ \Eprint {http://arxiv.org/abs/1511.09353}
  {arXiv:1511.09353 [hep-ph]} \BibitemShut {NoStop}%
\bibitem [{\citenamefont {Chen}\ \emph {et~al.}(2017)\citenamefont {Chen},
  \citenamefont {Chen}, \citenamefont {Liu}, \citenamefont {Liu},\ and\
  \citenamefont {Zhu}}]{Chen:2016spr}%
  \BibitemOpen
  \bibfield  {author} {\bibinfo {author} {\bibfnamefont {H.-X.}\ \bibnamefont
  {Chen}}, \bibinfo {author} {\bibfnamefont {W.}~\bibnamefont {Chen}}, \bibinfo
  {author} {\bibfnamefont {X.}~\bibnamefont {Liu}}, \bibinfo {author}
  {\bibfnamefont {Y.-R.}\ \bibnamefont {Liu}}, \ and\ \bibinfo {author}
  {\bibfnamefont {S.-L.}\ \bibnamefont {Zhu}},\ }\href {\doibase
  10.1088/1361-6633/aa6420} {\bibfield  {journal} {\bibinfo  {journal} {Rept.
  Prog. Phys.}\ }\textbf {\bibinfo {volume} {80}},\ \bibinfo {pages} {076201}
  (\bibinfo {year} {2017})},\ \Eprint {http://arxiv.org/abs/1609.08928}
  {arXiv:1609.08928 [hep-ph]} \BibitemShut {NoStop}%
\bibitem [{\citenamefont {Aoki}\ \emph {et~al.}(2009)\citenamefont {Aoki} \emph
  {et~al.}}]{Aoki:2008sm}%
  \BibitemOpen
  \bibfield  {author} {\bibinfo {author} {\bibfnamefont {S.}~\bibnamefont
  {Aoki}} \emph {et~al.} (\bibinfo {collaboration} {PACS-CS}),\ }\href
  {\doibase 10.1103/PhysRevD.79.034503} {\bibfield  {journal} {\bibinfo
  {journal} {Phys.Rev.}\ }\textbf {\bibinfo {volume} {D79}},\ \bibinfo {pages}
  {034503} (\bibinfo {year} {2009})},\ \Eprint {http://arxiv.org/abs/0807.1661}
  {arXiv:0807.1661 [hep-lat]} \BibitemShut {NoStop}%
\bibitem [{\citenamefont {Mohler}\ and\ \citenamefont
  {Woloshyn}(2011)}]{Mohler:2011ke}%
  \BibitemOpen
  \bibfield  {author} {\bibinfo {author} {\bibfnamefont {D.}~\bibnamefont
  {Mohler}}\ and\ \bibinfo {author} {\bibfnamefont {R.}~\bibnamefont
  {Woloshyn}},\ }\href {\doibase 10.1103/PhysRevD.84.054505} {\bibfield
  {journal} {\bibinfo  {journal} {Phys.Rev.}\ }\textbf {\bibinfo {volume}
  {D84}},\ \bibinfo {pages} {054505} (\bibinfo {year} {2011})},\ \Eprint
  {http://arxiv.org/abs/1103.5506} {arXiv:1103.5506 [hep-lat]} \BibitemShut
  {NoStop}%
\bibitem [{\citenamefont {Na}\ \emph {et~al.}(2012)\citenamefont {Na},
  \citenamefont {Davies}, \citenamefont {Follana}, \citenamefont {Lepage},\
  and\ \citenamefont {Shigemitsu}}]{Na:2012iu}%
  \BibitemOpen
  \bibfield  {author} {\bibinfo {author} {\bibfnamefont {H.}~\bibnamefont
  {Na}}, \bibinfo {author} {\bibfnamefont {C.~T.}\ \bibnamefont {Davies}},
  \bibinfo {author} {\bibfnamefont {E.}~\bibnamefont {Follana}}, \bibinfo
  {author} {\bibfnamefont {G.~P.}\ \bibnamefont {Lepage}}, \ and\ \bibinfo
  {author} {\bibfnamefont {J.}~\bibnamefont {Shigemitsu}},\ }\href {\doibase
  10.1103/PhysRevD.86.054510} {\bibfield  {journal} {\bibinfo  {journal}
  {Phys.Rev.}\ }\textbf {\bibinfo {volume} {D86}},\ \bibinfo {pages} {054510}
  (\bibinfo {year} {2012})},\ \Eprint {http://arxiv.org/abs/1206.4936}
  {arXiv:1206.4936 [hep-lat]} \BibitemShut {NoStop}%
\bibitem [{\citenamefont {Kalinowski}\ and\ \citenamefont
  {Wagner}(2015)}]{Kalinowski:2015bwa}%
  \BibitemOpen
  \bibfield  {author} {\bibinfo {author} {\bibfnamefont {M.}~\bibnamefont
  {Kalinowski}}\ and\ \bibinfo {author} {\bibfnamefont {M.}~\bibnamefont
  {Wagner}},\ }\href {\doibase 10.1103/PhysRevD.92.094508} {\bibfield
  {journal} {\bibinfo  {journal} {Phys. Rev.}\ }\textbf {\bibinfo {volume}
  {D92}},\ \bibinfo {pages} {094508} (\bibinfo {year} {2015})},\ \Eprint
  {http://arxiv.org/abs/1509.02396} {arXiv:1509.02396 [hep-lat]} \BibitemShut
  {NoStop}%
\bibitem [{\citenamefont {Cichy}\ \emph {et~al.}(2016)\citenamefont {Cichy},
  \citenamefont {Kalinowski},\ and\ \citenamefont {Wagner}}]{Cichy:2016bci}%
  \BibitemOpen
  \bibfield  {author} {\bibinfo {author} {\bibfnamefont {K.}~\bibnamefont
  {Cichy}}, \bibinfo {author} {\bibfnamefont {M.}~\bibnamefont {Kalinowski}}, \
  and\ \bibinfo {author} {\bibfnamefont {M.}~\bibnamefont {Wagner}},\ }\href
  {\doibase 10.1103/PhysRevD.94.094503} {\bibfield  {journal} {\bibinfo
  {journal} {Phys. Rev.}\ }\textbf {\bibinfo {volume} {D94}},\ \bibinfo {pages}
  {094503} (\bibinfo {year} {2016})},\ \Eprint
  {http://arxiv.org/abs/1603.06467} {arXiv:1603.06467 [hep-lat]} \BibitemShut
  {NoStop}%
\bibitem [{\citenamefont {Cheung}\ \emph {et~al.}(2016)\citenamefont {Cheung},
  \citenamefont {O'Hara}, \citenamefont {Moir}, \citenamefont {Peardon},
  \citenamefont {Ryan}, \citenamefont {Thomas},\ and\ \citenamefont
  {Tims}}]{Cheung:2016bym}%
  \BibitemOpen
  \bibfield  {author} {\bibinfo {author} {\bibfnamefont {G.~K.~C.}\
  \bibnamefont {Cheung}}, \bibinfo {author} {\bibfnamefont {C.}~\bibnamefont
  {O'Hara}}, \bibinfo {author} {\bibfnamefont {G.}~\bibnamefont {Moir}},
  \bibinfo {author} {\bibfnamefont {M.}~\bibnamefont {Peardon}}, \bibinfo
  {author} {\bibfnamefont {S.~M.}\ \bibnamefont {Ryan}}, \bibinfo {author}
  {\bibfnamefont {C.~E.}\ \bibnamefont {Thomas}}, \ and\ \bibinfo {author}
  {\bibfnamefont {D.}~\bibnamefont {Tims}} (\bibinfo {collaboration} {Hadron
  Spectrum}),\ }\href {\doibase 10.1007/JHEP12(2016)089} {\bibfield  {journal}
  {\bibinfo  {journal} {JHEP}\ }\textbf {\bibinfo {volume} {12}},\ \bibinfo
  {pages} {089} (\bibinfo {year} {2016})},\ \Eprint
  {http://arxiv.org/abs/1610.01073} {arXiv:1610.01073 [hep-lat]} \BibitemShut
  {NoStop}%
\bibitem [{\citenamefont {Moir}\ \emph {et~al.}(2016)\citenamefont {Moir},
  \citenamefont {Peardon}, \citenamefont {Ryan}, \citenamefont {Thomas},\ and\
  \citenamefont {Wilson}}]{Moir:2016srx}%
  \BibitemOpen
  \bibfield  {author} {\bibinfo {author} {\bibfnamefont {G.}~\bibnamefont
  {Moir}}, \bibinfo {author} {\bibfnamefont {M.}~\bibnamefont {Peardon}},
  \bibinfo {author} {\bibfnamefont {S.~M.}\ \bibnamefont {Ryan}}, \bibinfo
  {author} {\bibfnamefont {C.~E.}\ \bibnamefont {Thomas}}, \ and\ \bibinfo
  {author} {\bibfnamefont {D.~J.}\ \bibnamefont {Wilson}},\ }\href {\doibase
  10.1007/JHEP10(2016)011} {\bibfield  {journal} {\bibinfo  {journal} {JHEP}\
  }\textbf {\bibinfo {volume} {10}},\ \bibinfo {pages} {011} (\bibinfo {year}
  {2016})},\ \Eprint {http://arxiv.org/abs/1607.07093} {arXiv:1607.07093
  [hep-lat]} \BibitemShut {NoStop}%
\bibitem [{\citenamefont {Cheung}\ \emph {et~al.}(2021)\citenamefont {Cheung},
  \citenamefont {Thomas}, \citenamefont {Wilson}, \citenamefont {Moir},
  \citenamefont {Peardon},\ and\ \citenamefont {Ryan}}]{Cheung:2020mql}%
  \BibitemOpen
  \bibfield  {author} {\bibinfo {author} {\bibfnamefont {G.~K.~C.}\
  \bibnamefont {Cheung}}, \bibinfo {author} {\bibfnamefont {C.~E.}\
  \bibnamefont {Thomas}}, \bibinfo {author} {\bibfnamefont {D.~J.}\
  \bibnamefont {Wilson}}, \bibinfo {author} {\bibfnamefont {G.}~\bibnamefont
  {Moir}}, \bibinfo {author} {\bibfnamefont {M.}~\bibnamefont {Peardon}}, \
  and\ \bibinfo {author} {\bibfnamefont {S.~M.}\ \bibnamefont {Ryan}} (\bibinfo
  {collaboration} {Hadron Spectrum}),\ }\href {\doibase
  10.1007/JHEP02(2021)100} {\bibfield  {journal} {\bibinfo  {journal} {JHEP}\
  }\textbf {\bibinfo {volume} {02}},\ \bibinfo {pages} {100} (\bibinfo {year}
  {2021})},\ \Eprint {http://arxiv.org/abs/2008.06432} {arXiv:2008.06432
  [hep-lat]} \BibitemShut {NoStop}%
\bibitem [{\citenamefont {Gayer}\ \emph {et~al.}(2021)\citenamefont {Gayer},
  \citenamefont {Lang}, \citenamefont {Ryan}, \citenamefont {Tims},
  \citenamefont {Thomas},\ and\ \citenamefont {Wilson}}]{Gayer:2021xzv}%
  \BibitemOpen
  \bibfield  {author} {\bibinfo {author} {\bibfnamefont {L.}~\bibnamefont
  {Gayer}}, \bibinfo {author} {\bibfnamefont {N.}~\bibnamefont {Lang}},
  \bibinfo {author} {\bibfnamefont {S.~M.}\ \bibnamefont {Ryan}}, \bibinfo
  {author} {\bibfnamefont {D.}~\bibnamefont {Tims}}, \bibinfo {author}
  {\bibfnamefont {C.~E.}\ \bibnamefont {Thomas}}, \ and\ \bibinfo {author}
  {\bibfnamefont {D.~J.}\ \bibnamefont {Wilson}} (\bibinfo {collaboration}
  {Hadron Spectrum}),\ }\href {\doibase 10.1007/JHEP07(2021)123} {\bibfield
  {journal} {\bibinfo  {journal} {JHEP}\ }\textbf {\bibinfo {volume} {07}},\
  \bibinfo {pages} {123} (\bibinfo {year} {2021})},\ \Eprint
  {http://arxiv.org/abs/2102.04973} {arXiv:2102.04973 [hep-lat]} \BibitemShut
  {NoStop}%
\bibitem [{\citenamefont {Fettes}\ and\ \citenamefont
  {Meissner}(2000)}]{Fettes:2000xg}%
  \BibitemOpen
  \bibfield  {author} {\bibinfo {author} {\bibfnamefont {N.}~\bibnamefont
  {Fettes}}\ and\ \bibinfo {author} {\bibfnamefont {U.-G.}\ \bibnamefont
  {Meissner}},\ }\href {\doibase 10.1016/S0375-9474(00)00199-8} {\bibfield
  {journal} {\bibinfo  {journal} {Nucl. Phys. A}\ }\textbf {\bibinfo {volume}
  {676}},\ \bibinfo {pages} {311} (\bibinfo {year} {2000})},\ \Eprint
  {http://arxiv.org/abs/hep-ph/0002162} {arXiv:hep-ph/0002162} \BibitemShut
  {NoStop}%
\bibitem [{\citenamefont {Mai}\ \emph {et~al.}(2009)\citenamefont {Mai},
  \citenamefont {Bruns}, \citenamefont {Kubis},\ and\ \citenamefont
  {Meissner}}]{Mai:2009ce}%
  \BibitemOpen
  \bibfield  {author} {\bibinfo {author} {\bibfnamefont {M.}~\bibnamefont
  {Mai}}, \bibinfo {author} {\bibfnamefont {P.~C.}\ \bibnamefont {Bruns}},
  \bibinfo {author} {\bibfnamefont {B.}~\bibnamefont {Kubis}}, \ and\ \bibinfo
  {author} {\bibfnamefont {U.-G.}\ \bibnamefont {Meissner}},\ }\href {\doibase
  10.1103/PhysRevD.80.094006} {\bibfield  {journal} {\bibinfo  {journal} {Phys.
  Rev. D}\ }\textbf {\bibinfo {volume} {80}},\ \bibinfo {pages} {094006}
  (\bibinfo {year} {2009})},\ \Eprint {http://arxiv.org/abs/0905.2810}
  {arXiv:0905.2810 [hep-ph]} \BibitemShut {NoStop}%
\bibitem [{\citenamefont {Huang}\ \emph {et~al.}(2017)\citenamefont {Huang},
  \citenamefont {Zhang}, \citenamefont {Li},\ and\ \citenamefont
  {Kaiser}}]{Huang:2017bmx}%
  \BibitemOpen
  \bibfield  {author} {\bibinfo {author} {\bibfnamefont {B.-L.}\ \bibnamefont
  {Huang}}, \bibinfo {author} {\bibfnamefont {J.-S.}\ \bibnamefont {Zhang}},
  \bibinfo {author} {\bibfnamefont {Y.-D.}\ \bibnamefont {Li}}, \ and\ \bibinfo
  {author} {\bibfnamefont {N.}~\bibnamefont {Kaiser}},\ }\href {\doibase
  10.1103/PhysRevD.96.016021} {\bibfield  {journal} {\bibinfo  {journal} {Phys.
  Rev. D}\ }\textbf {\bibinfo {volume} {96}},\ \bibinfo {pages} {016021}
  (\bibinfo {year} {2017})},\ \Eprint {http://arxiv.org/abs/1701.06018}
  {arXiv:1701.06018 [nucl-th]} \BibitemShut {NoStop}%
\bibitem [{\citenamefont {Andersen}\ \emph {et~al.}(2019)\citenamefont
  {Andersen}, \citenamefont {Bulava}, \citenamefont {H\"orz},\ and\
  \citenamefont {Morningstar}}]{Andersen:2019ktw}%
  \BibitemOpen
  \bibfield  {author} {\bibinfo {author} {\bibfnamefont {C.~W.}\ \bibnamefont
  {Andersen}}, \bibinfo {author} {\bibfnamefont {J.}~\bibnamefont {Bulava}},
  \bibinfo {author} {\bibfnamefont {B.}~\bibnamefont {H\"orz}}, \ and\ \bibinfo
  {author} {\bibfnamefont {C.}~\bibnamefont {Morningstar}},\ }\href {\doibase
  10.22323/1.363.0039} {\bibfield  {journal} {\bibinfo  {journal} {PoS}\
  }\textbf {\bibinfo {volume} {LATTICE2019}},\ \bibinfo {pages} {039} (\bibinfo
  {year} {2019})},\ \Eprint {http://arxiv.org/abs/1911.10021} {arXiv:1911.10021
  [hep-lat]} \BibitemShut {NoStop}%
\bibitem [{\citenamefont {Lang}\ and\ \citenamefont
  {Verduci}(2013)}]{Lang:2012db}%
  \BibitemOpen
  \bibfield  {author} {\bibinfo {author} {\bibfnamefont {C.~B.}\ \bibnamefont
  {Lang}}\ and\ \bibinfo {author} {\bibfnamefont {V.}~\bibnamefont {Verduci}},\
  }\href {\doibase 10.1103/PhysRevD.87.054502} {\bibfield  {journal} {\bibinfo
  {journal} {Phys. Rev. D}\ }\textbf {\bibinfo {volume} {87}},\ \bibinfo
  {pages} {054502} (\bibinfo {year} {2013})},\ \Eprint
  {http://arxiv.org/abs/1212.5055} {arXiv:1212.5055 [hep-lat]} \BibitemShut
  {NoStop}%
\bibitem [{\citenamefont {Detmold}\ and\ \citenamefont
  {Nicholson}(2016)}]{Detmold:2015qwf}%
  \BibitemOpen
  \bibfield  {author} {\bibinfo {author} {\bibfnamefont {W.}~\bibnamefont
  {Detmold}}\ and\ \bibinfo {author} {\bibfnamefont {A.}~\bibnamefont
  {Nicholson}},\ }\href {\doibase 10.1103/PhysRevD.93.114511} {\bibfield
  {journal} {\bibinfo  {journal} {Phys. Rev.}\ }\textbf {\bibinfo {volume}
  {D93}},\ \bibinfo {pages} {114511} (\bibinfo {year} {2016})},\ \Eprint
  {http://arxiv.org/abs/1511.02275} {arXiv:1511.02275 [hep-lat]} \BibitemShut
  {NoStop}%
\bibitem [{\citenamefont {Andersen}\ \emph {et~al.}(2018)\citenamefont
  {Andersen}, \citenamefont {Bulava}, \citenamefont {H\"orz},\ and\
  \citenamefont {Morningstar}}]{Andersen:2017una}%
  \BibitemOpen
  \bibfield  {author} {\bibinfo {author} {\bibfnamefont {C.~W.}\ \bibnamefont
  {Andersen}}, \bibinfo {author} {\bibfnamefont {J.}~\bibnamefont {Bulava}},
  \bibinfo {author} {\bibfnamefont {B.}~\bibnamefont {H\"orz}}, \ and\ \bibinfo
  {author} {\bibfnamefont {C.}~\bibnamefont {Morningstar}},\ }\href {\doibase
  10.1103/PhysRevD.97.014506} {\bibfield  {journal} {\bibinfo  {journal} {Phys.
  Rev. D}\ }\textbf {\bibinfo {volume} {97}},\ \bibinfo {pages} {014506}
  (\bibinfo {year} {2018})},\ \Eprint {http://arxiv.org/abs/1710.01557}
  {arXiv:1710.01557 [hep-lat]} \BibitemShut {NoStop}%
\bibitem [{\citenamefont {Silvi}\ \emph {et~al.}(2021)\citenamefont {Silvi}
  \emph {et~al.}}]{Silvi:2021uya}%
  \BibitemOpen
  \bibfield  {author} {\bibinfo {author} {\bibfnamefont {G.}~\bibnamefont
  {Silvi}} \emph {et~al.},\ }\href {\doibase 10.1103/PhysRevD.103.094508}
  {\bibfield  {journal} {\bibinfo  {journal} {Phys. Rev. D}\ }\textbf {\bibinfo
  {volume} {103}},\ \bibinfo {pages} {094508} (\bibinfo {year} {2021})},\
  \Eprint {http://arxiv.org/abs/2101.00689} {arXiv:2101.00689 [hep-lat]}
  \BibitemShut {NoStop}%
\bibitem [{\citenamefont {Bulava}\ \emph
  {et~al.}(2023{\natexlab{a}})\citenamefont {Bulava}, \citenamefont {Hanlon},
  \citenamefont {H\"orz}, \citenamefont {Morningstar}, \citenamefont
  {Nicholson}, \citenamefont {Romero-L\'opez}, \citenamefont {Skinner},
  \citenamefont {Vranas},\ and\ \citenamefont {Walker-Loud}}]{Bulava:2022vpq}%
  \BibitemOpen
  \bibfield  {author} {\bibinfo {author} {\bibfnamefont {J.}~\bibnamefont
  {Bulava}}, \bibinfo {author} {\bibfnamefont {A.~D.}\ \bibnamefont {Hanlon}},
  \bibinfo {author} {\bibfnamefont {B.}~\bibnamefont {H\"orz}}, \bibinfo
  {author} {\bibfnamefont {C.}~\bibnamefont {Morningstar}}, \bibinfo {author}
  {\bibfnamefont {A.}~\bibnamefont {Nicholson}}, \bibinfo {author}
  {\bibfnamefont {F.}~\bibnamefont {Romero-L\'opez}}, \bibinfo {author}
  {\bibfnamefont {S.}~\bibnamefont {Skinner}}, \bibinfo {author} {\bibfnamefont
  {P.}~\bibnamefont {Vranas}}, \ and\ \bibinfo {author} {\bibfnamefont
  {A.}~\bibnamefont {Walker-Loud}},\ }\href {\doibase
  10.1016/j.nuclphysb.2023.116105} {\bibfield  {journal} {\bibinfo  {journal}
  {Nucl. Phys. B}\ }\textbf {\bibinfo {volume} {987}},\ \bibinfo {pages}
  {116105} (\bibinfo {year} {2023}{\natexlab{a}})},\ \Eprint
  {http://arxiv.org/abs/2208.03867} {arXiv:2208.03867 [hep-lat]} \BibitemShut
  {NoStop}%
\bibitem [{\citenamefont {Bulava}\ \emph
  {et~al.}(2023{\natexlab{b}})\citenamefont {Bulava} \emph
  {et~al.}}]{Bulava:2023gfx}%
  \BibitemOpen
  \bibfield  {author} {\bibinfo {author} {\bibfnamefont {J.}~\bibnamefont
  {Bulava}} \emph {et~al.},\ }\href@noop {} {\  (\bibinfo {year}
  {2023}{\natexlab{b}})},\ \Eprint {http://arxiv.org/abs/2307.13471}
  {arXiv:2307.13471 [hep-lat]} \BibitemShut {NoStop}%
\bibitem [{\citenamefont {Bulava}\ \emph
  {et~al.}(2023{\natexlab{c}})\citenamefont {Bulava} \emph
  {et~al.}}]{Bulava:2023rmn}%
  \BibitemOpen
  \bibfield  {author} {\bibinfo {author} {\bibfnamefont {J.}~\bibnamefont
  {Bulava}} \emph {et~al.},\ }\href@noop {} {\  (\bibinfo {year}
  {2023}{\natexlab{c}})},\ \Eprint {http://arxiv.org/abs/2307.10413}
  {arXiv:2307.10413 [hep-lat]} \BibitemShut {NoStop}%
\bibitem [{\citenamefont {Alexandrou}\ \emph {et~al.}(2023)\citenamefont
  {Alexandrou}, \citenamefont {Bacchio}, \citenamefont {Koutsou}, \citenamefont
  {Leontiou}, \citenamefont {Paul}, \citenamefont {Petschlies},\ and\
  \citenamefont {Pittler}}]{Alexandrou:2023elk}%
  \BibitemOpen
  \bibfield  {author} {\bibinfo {author} {\bibfnamefont {C.}~\bibnamefont
  {Alexandrou}}, \bibinfo {author} {\bibfnamefont {S.}~\bibnamefont {Bacchio}},
  \bibinfo {author} {\bibfnamefont {G.}~\bibnamefont {Koutsou}}, \bibinfo
  {author} {\bibfnamefont {T.}~\bibnamefont {Leontiou}}, \bibinfo {author}
  {\bibfnamefont {S.}~\bibnamefont {Paul}}, \bibinfo {author} {\bibfnamefont
  {M.}~\bibnamefont {Petschlies}}, \ and\ \bibinfo {author} {\bibfnamefont
  {F.}~\bibnamefont {Pittler}},\ }\href@noop {} {\  (\bibinfo {year} {2023})},\
  \Eprint {http://arxiv.org/abs/2307.12846} {arXiv:2307.12846 [hep-lat]}
  \BibitemShut {NoStop}%
\bibitem [{\citenamefont {Guo}\ \emph {et~al.}(2018{\natexlab{a}})\citenamefont
  {Guo}, \citenamefont {Heo},\ and\ \citenamefont {Lutz}}]{Guo:2018kno}%
  \BibitemOpen
  \bibfield  {author} {\bibinfo {author} {\bibfnamefont {X.-Y.}\ \bibnamefont
  {Guo}}, \bibinfo {author} {\bibfnamefont {Y.}~\bibnamefont {Heo}}, \ and\
  \bibinfo {author} {\bibfnamefont {M.~F.~M.}\ \bibnamefont {Lutz}},\ }\href
  {\doibase 10.1103/PhysRevD.98.014510} {\bibfield  {journal} {\bibinfo
  {journal} {Phys. Rev.}\ }\textbf {\bibinfo {volume} {D98}},\ \bibinfo {pages}
  {014510} (\bibinfo {year} {2018}{\natexlab{a}})},\ \Eprint
  {http://arxiv.org/abs/1801.10122} {arXiv:1801.10122 [hep-lat]} \BibitemShut
  {NoStop}%
\bibitem [{\citenamefont {Guo}\ \emph {et~al.}(2021)\citenamefont {Guo},
  \citenamefont {Heo},\ and\ \citenamefont {Lutz}}]{Guo:2021kdo}%
  \BibitemOpen
  \bibfield  {author} {\bibinfo {author} {\bibfnamefont {X.-Y.}\ \bibnamefont
  {Guo}}, \bibinfo {author} {\bibfnamefont {Y.}~\bibnamefont {Heo}}, \ and\
  \bibinfo {author} {\bibfnamefont {M.~F.~M.}\ \bibnamefont {Lutz}},\ }in\
  \href@noop {} {\emph {\bibinfo {booktitle} {{38th International Symposium on
  Lattice Field Theory}}}}\ (\bibinfo {year} {2021})\ \Eprint
  {http://arxiv.org/abs/2107.12284} {arXiv:2107.12284 [hep-lat]} \BibitemShut
  {NoStop}%
\bibitem [{\citenamefont {Hofmann}\ and\ \citenamefont
  {Lutz}(2004)}]{Hofmann:2003je}%
  \BibitemOpen
  \bibfield  {author} {\bibinfo {author} {\bibfnamefont {J.}~\bibnamefont
  {Hofmann}}\ and\ \bibinfo {author} {\bibfnamefont {M.~F.~M.}\ \bibnamefont
  {Lutz}},\ }\href {\doibase 10.1016/j.nuclphysa.2003.12.013} {\bibfield
  {journal} {\bibinfo  {journal} {Nucl. Phys.}\ }\textbf {\bibinfo {volume}
  {A733}},\ \bibinfo {pages} {142} (\bibinfo {year} {2004})},\ \Eprint
  {http://arxiv.org/abs/hep-ph/0308263} {arXiv:hep-ph/0308263 [hep-ph]}
  \BibitemShut {NoStop}%
\bibitem [{\citenamefont {Lutz}\ and\ \citenamefont
  {Soyeur}(2008)}]{Lutz:2007sk}%
  \BibitemOpen
  \bibfield  {author} {\bibinfo {author} {\bibfnamefont {M.~F.~M.}\
  \bibnamefont {Lutz}}\ and\ \bibinfo {author} {\bibfnamefont {M.}~\bibnamefont
  {Soyeur}},\ }\href {\doibase 10.1016/j.nuclphysa.2008.09.003} {\bibfield
  {journal} {\bibinfo  {journal} {Nucl.Phys.}\ }\textbf {\bibinfo {volume}
  {A813}},\ \bibinfo {pages} {14} (\bibinfo {year} {2008})},\ \Eprint
  {http://arxiv.org/abs/0710.1545} {arXiv:0710.1545 [hep-ph]} \BibitemShut
  {NoStop}%
\bibitem [{\citenamefont {Liu}\ \emph {et~al.}(2013)\citenamefont {Liu},
  \citenamefont {Orginos}, \citenamefont {Guo}, \citenamefont {Hanhart},\ and\
  \citenamefont {Mei{\ss}ner}}]{Liu:2012zya}%
  \BibitemOpen
  \bibfield  {author} {\bibinfo {author} {\bibfnamefont {L.}~\bibnamefont
  {Liu}}, \bibinfo {author} {\bibfnamefont {K.}~\bibnamefont {Orginos}},
  \bibinfo {author} {\bibfnamefont {F.-K.}\ \bibnamefont {Guo}}, \bibinfo
  {author} {\bibfnamefont {C.}~\bibnamefont {Hanhart}}, \ and\ \bibinfo
  {author} {\bibfnamefont {U.-G.}\ \bibnamefont {Mei{\ss}ner}},\ }\href
  {\doibase 10.1103/PhysRevD.87.014508} {\bibfield  {journal} {\bibinfo
  {journal} {Phys.Rev.}\ }\textbf {\bibinfo {volume} {D87}},\ \bibinfo {pages}
  {014508} (\bibinfo {year} {2013})},\ \Eprint {http://arxiv.org/abs/1208.4535}
  {arXiv:1208.4535 [hep-lat]} \BibitemShut {NoStop}%
\bibitem [{\citenamefont {Altenbuchinger}\ \emph {et~al.}(2014)\citenamefont
  {Altenbuchinger}, \citenamefont {Geng},\ and\ \citenamefont
  {Weise}}]{Altenbuchinger:2013vwa}%
  \BibitemOpen
  \bibfield  {author} {\bibinfo {author} {\bibfnamefont {M.}~\bibnamefont
  {Altenbuchinger}}, \bibinfo {author} {\bibfnamefont {L.~S.}\ \bibnamefont
  {Geng}}, \ and\ \bibinfo {author} {\bibfnamefont {W.}~\bibnamefont {Weise}},\
  }\href {\doibase 10.1103/PhysRevD.89.014026} {\bibfield  {journal} {\bibinfo
  {journal} {Phys. Rev.}\ }\textbf {\bibinfo {volume} {D89}},\ \bibinfo {pages}
  {014026} (\bibinfo {year} {2014})},\ \Eprint {http://arxiv.org/abs/1309.4743}
  {arXiv:1309.4743 [hep-ph]} \BibitemShut {NoStop}%
\bibitem [{\citenamefont {Cleven}\ \emph {et~al.}(2014)\citenamefont {Cleven},
  \citenamefont {Grie{\ss}hammer}, \citenamefont {Guo}, \citenamefont
  {Hanhart},\ and\ \citenamefont {Mei{\ss}ner}}]{Cleven:2014oka}%
  \BibitemOpen
  \bibfield  {author} {\bibinfo {author} {\bibfnamefont {M.}~\bibnamefont
  {Cleven}}, \bibinfo {author} {\bibfnamefont {H.~W.}\ \bibnamefont
  {Grie{\ss}hammer}}, \bibinfo {author} {\bibfnamefont {F.-K.}\ \bibnamefont
  {Guo}}, \bibinfo {author} {\bibfnamefont {C.}~\bibnamefont {Hanhart}}, \ and\
  \bibinfo {author} {\bibfnamefont {U.-G.}\ \bibnamefont {Mei{\ss}ner}},\
  }\href {\doibase 10.1140/epja/i2014-14149-y} {\bibfield  {journal} {\bibinfo
  {journal} {Eur. Phys. J.}\ }\textbf {\bibinfo {volume} {A50}},\ \bibinfo
  {pages} {149} (\bibinfo {year} {2014})},\ \Eprint
  {http://arxiv.org/abs/1405.2242} {arXiv:1405.2242 [hep-ph]} \BibitemShut
  {NoStop}%
\bibitem [{\citenamefont {Du}\ \emph {et~al.}(2016)\citenamefont {Du},
  \citenamefont {Guo}, \citenamefont {Mei{\ss}ner},\ and\ \citenamefont
  {Yao}}]{Du:2016tgp}%
  \BibitemOpen
  \bibfield  {author} {\bibinfo {author} {\bibfnamefont {M.-L.}\ \bibnamefont
  {Du}}, \bibinfo {author} {\bibfnamefont {F.-K.}\ \bibnamefont {Guo}},
  \bibinfo {author} {\bibfnamefont {U.-G.}\ \bibnamefont {Mei{\ss}ner}}, \ and\
  \bibinfo {author} {\bibfnamefont {D.-L.}\ \bibnamefont {Yao}},\ }\href
  {\doibase 10.1103/PhysRevD.94.094037} {\bibfield  {journal} {\bibinfo
  {journal} {Phys. Rev.}\ }\textbf {\bibinfo {volume} {D94}},\ \bibinfo {pages}
  {094037} (\bibinfo {year} {2016})},\ \Eprint
  {http://arxiv.org/abs/1610.02963} {arXiv:1610.02963 [hep-ph]} \BibitemShut
  {NoStop}%
\bibitem [{\citenamefont {Huang}\ \emph {et~al.}(2022)\citenamefont {Huang},
  \citenamefont {Lin}, \citenamefont {Chen},\ and\ \citenamefont
  {Zhu}}]{Huang:2022cag}%
  \BibitemOpen
  \bibfield  {author} {\bibinfo {author} {\bibfnamefont {B.-L.}\ \bibnamefont
  {Huang}}, \bibinfo {author} {\bibfnamefont {Z.-Y.}\ \bibnamefont {Lin}},
  \bibinfo {author} {\bibfnamefont {K.}~\bibnamefont {Chen}}, \ and\ \bibinfo
  {author} {\bibfnamefont {S.-L.}\ \bibnamefont {Zhu}},\ }\href@noop {} {\
  (\bibinfo {year} {2022})},\ \Eprint {http://arxiv.org/abs/2205.02619}
  {arXiv:2205.02619 [hep-ph]} \BibitemShut {NoStop}%
\bibitem [{\citenamefont {Lutz}\ \emph {et~al.}(2015)\citenamefont {Lutz},
  \citenamefont {Kolomeitsev},\ and\ \citenamefont {Korpa}}]{Lutz:2015lca}%
  \BibitemOpen
  \bibfield  {author} {\bibinfo {author} {\bibfnamefont {M.~F.~M.}\
  \bibnamefont {Lutz}}, \bibinfo {author} {\bibfnamefont {E.~E.}\ \bibnamefont
  {Kolomeitsev}}, \ and\ \bibinfo {author} {\bibfnamefont {C.~L.}\ \bibnamefont
  {Korpa}},\ }\href {\doibase 10.1103/PhysRevD.92.016003} {\bibfield  {journal}
  {\bibinfo  {journal} {Phys. Rev. D}\ }\textbf {\bibinfo {volume} {92}},\
  \bibinfo {pages} {016003} (\bibinfo {year} {2015})},\ \Eprint
  {http://arxiv.org/abs/1506.02375} {arXiv:1506.02375 [hep-ph]} \BibitemShut
  {NoStop}%
\bibitem [{\citenamefont {Lutz}\ and\ \citenamefont
  {Korpa}(2018)}]{Lutz:2018kaz}%
  \BibitemOpen
  \bibfield  {author} {\bibinfo {author} {\bibfnamefont {M.~F.~M.}\
  \bibnamefont {Lutz}}\ and\ \bibinfo {author} {\bibfnamefont {C.~L.}\
  \bibnamefont {Korpa}},\ }\href {\doibase 10.1103/PhysRevD.98.076003}
  {\bibfield  {journal} {\bibinfo  {journal} {Phys. Rev. D}\ }\textbf {\bibinfo
  {volume} {98}},\ \bibinfo {pages} {076003} (\bibinfo {year} {2018})},\
  \Eprint {http://arxiv.org/abs/1808.08695} {arXiv:1808.08695 [hep-ph]}
  \BibitemShut {NoStop}%
\bibitem [{\citenamefont {Lutz}\ \emph {et~al.}(2022)\citenamefont {Lutz},
  \citenamefont {Guo}, \citenamefont {Heo},\ and\ \citenamefont
  {Korpa}}]{Lutz:2022enz}%
  \BibitemOpen
  \bibfield  {author} {\bibinfo {author} {\bibfnamefont {M.~F.~M.}\
  \bibnamefont {Lutz}}, \bibinfo {author} {\bibfnamefont {X.-Y.}\ \bibnamefont
  {Guo}}, \bibinfo {author} {\bibfnamefont {Y.}~\bibnamefont {Heo}}, \ and\
  \bibinfo {author} {\bibfnamefont {C.~L.}\ \bibnamefont {Korpa}},\ }\href
  {\doibase 10.1103/PhysRevD.106.114038} {\bibfield  {journal} {\bibinfo
  {journal} {Phys. Rev. D}\ }\textbf {\bibinfo {volume} {106}},\ \bibinfo
  {pages} {114038} (\bibinfo {year} {2022})},\ \Eprint
  {http://arxiv.org/abs/2209.10601} {arXiv:2209.10601 [hep-ph]} \BibitemShut
  {NoStop}%
\bibitem [{\citenamefont {Korpa}\ \emph {et~al.}(2023)\citenamefont {Korpa},
  \citenamefont {Lutz}, \citenamefont {Guo},\ and\ \citenamefont
  {Heo}}]{Korpa:2022voo}%
  \BibitemOpen
  \bibfield  {author} {\bibinfo {author} {\bibfnamefont {C.~L.}\ \bibnamefont
  {Korpa}}, \bibinfo {author} {\bibfnamefont {M.~F.~M.}\ \bibnamefont {Lutz}},
  \bibinfo {author} {\bibfnamefont {X.-Y.}\ \bibnamefont {Guo}}, \ and\
  \bibinfo {author} {\bibfnamefont {Y.}~\bibnamefont {Heo}},\ }\href {\doibase
  10.1103/PhysRevD.107.L031505} {\bibfield  {journal} {\bibinfo  {journal}
  {Phys. Rev. D}\ }\textbf {\bibinfo {volume} {107}},\ \bibinfo {pages}
  {L031505} (\bibinfo {year} {2023})},\ \Eprint
  {http://arxiv.org/abs/2211.03508} {arXiv:2211.03508 [hep-ph]} \BibitemShut
  {NoStop}%
\bibitem [{\citenamefont {Gasparyan}\ and\ \citenamefont
  {Lutz}(2010)}]{Gasparyan:2010xz}%
  \BibitemOpen
  \bibfield  {author} {\bibinfo {author} {\bibfnamefont {A.}~\bibnamefont
  {Gasparyan}}\ and\ \bibinfo {author} {\bibfnamefont {M.~F.~M.}\ \bibnamefont
  {Lutz}},\ }\href {\doibase 10.1016/j.nuclphysa.2010.08.006} {\bibfield
  {journal} {\bibinfo  {journal} {Nucl.Phys.}\ }\textbf {\bibinfo {volume}
  {A848}},\ \bibinfo {pages} {126} (\bibinfo {year} {2010})},\ \Eprint
  {http://arxiv.org/abs/1003.3426} {arXiv:1003.3426 [hep-ph]} \BibitemShut
  {NoStop}%
\bibitem [{\citenamefont {Danilkin}\ \emph
  {et~al.}(2011{\natexlab{a}})\citenamefont {Danilkin}, \citenamefont
  {Gasparyan},\ and\ \citenamefont {Lutz}}]{Danilkin:2010xd}%
  \BibitemOpen
  \bibfield  {author} {\bibinfo {author} {\bibfnamefont {I.}~\bibnamefont
  {Danilkin}}, \bibinfo {author} {\bibfnamefont {A.}~\bibnamefont {Gasparyan}},
  \ and\ \bibinfo {author} {\bibfnamefont {M.~F.~M.}\ \bibnamefont {Lutz}},\
  }\href {\doibase 10.1016/j.physletb.2011.01.036} {\bibfield  {journal}
  {\bibinfo  {journal} {Phys.Lett.}\ }\textbf {\bibinfo {volume} {B697}},\
  \bibinfo {pages} {147} (\bibinfo {year} {2011}{\natexlab{a}})},\ \Eprint
  {http://arxiv.org/abs/1009.5928} {arXiv:1009.5928 [hep-ph]} \BibitemShut
  {NoStop}%
\bibitem [{\citenamefont {Danilkin}\ \emph
  {et~al.}(2011{\natexlab{b}})\citenamefont {Danilkin}, \citenamefont {Gil},\
  and\ \citenamefont {Lutz}}]{Danilkin:2011fz}%
  \BibitemOpen
  \bibfield  {author} {\bibinfo {author} {\bibfnamefont {I.~V.}\ \bibnamefont
  {Danilkin}}, \bibinfo {author} {\bibfnamefont {L.~I.~R.}\ \bibnamefont
  {Gil}}, \ and\ \bibinfo {author} {\bibfnamefont {M.~F.~M.}\ \bibnamefont
  {Lutz}},\ }\href {\doibase 10.1016/j.physletb.2011.08.001} {\bibfield
  {journal} {\bibinfo  {journal} {Phys. Lett. B}\ }\textbf {\bibinfo {volume}
  {703}},\ \bibinfo {pages} {504} (\bibinfo {year} {2011}{\natexlab{b}})},\
  \Eprint {http://arxiv.org/abs/1106.2230} {arXiv:1106.2230 [hep-ph]}
  \BibitemShut {NoStop}%
\bibitem [{\citenamefont {Gasparyan}\ \emph {et~al.}(2012)\citenamefont
  {Gasparyan}, \citenamefont {Lutz},\ and\ \citenamefont
  {Epelbaum}}]{Gasparyan:2012km}%
  \BibitemOpen
  \bibfield  {author} {\bibinfo {author} {\bibfnamefont {A.~M.}\ \bibnamefont
  {Gasparyan}}, \bibinfo {author} {\bibfnamefont {M.~F.~M.}\ \bibnamefont
  {Lutz}}, \ and\ \bibinfo {author} {\bibfnamefont {E.}~\bibnamefont
  {Epelbaum}},\ }\href {\doibase 10.1140/epja/i2013-13115-7} {\bibfield
  {journal} {\bibinfo  {journal} {Eur. Phys. J. A}\ }\textbf {\bibinfo {volume}
  {49}},\ \bibinfo {pages} {115} (\bibinfo {year} {2012})},\ \Eprint
  {http://arxiv.org/abs/1212.3057} {arXiv:1212.3057 [nucl-th]} \BibitemShut
  {NoStop}%
\bibitem [{\citenamefont {Yao}\ \emph {et~al.}(2015)\citenamefont {Yao},
  \citenamefont {Du}, \citenamefont {Guo},\ and\ \citenamefont
  {Mei{\ss}ner}}]{Yao:2015qia}%
  \BibitemOpen
  \bibfield  {author} {\bibinfo {author} {\bibfnamefont {D.-L.}\ \bibnamefont
  {Yao}}, \bibinfo {author} {\bibfnamefont {M.-L.}\ \bibnamefont {Du}},
  \bibinfo {author} {\bibfnamefont {F.-K.}\ \bibnamefont {Guo}}, \ and\
  \bibinfo {author} {\bibfnamefont {U.-G.}\ \bibnamefont {Mei{\ss}ner}},\
  }\href {\doibase 10.1007/JHEP11(2015)058} {\bibfield  {journal} {\bibinfo
  {journal} {JHEP}\ }\textbf {\bibinfo {volume} {11}},\ \bibinfo {pages} {058}
  (\bibinfo {year} {2015})},\ \Eprint {http://arxiv.org/abs/1502.05981}
  {arXiv:1502.05981 [hep-ph]} \BibitemShut {NoStop}%
\bibitem [{\citenamefont {Du}\ \emph {et~al.}(2017)\citenamefont {Du},
  \citenamefont {Guo}, \citenamefont {Mei{\ss}ner},\ and\ \citenamefont
  {Yao}}]{Du:2017ttu}%
  \BibitemOpen
  \bibfield  {author} {\bibinfo {author} {\bibfnamefont {M.-L.}\ \bibnamefont
  {Du}}, \bibinfo {author} {\bibfnamefont {F.-K.}\ \bibnamefont {Guo}},
  \bibinfo {author} {\bibfnamefont {U.-G.}\ \bibnamefont {Mei{\ss}ner}}, \ and\
  \bibinfo {author} {\bibfnamefont {D.-L.}\ \bibnamefont {Yao}},\ }\href
  {\doibase 10.1140/epjc/s10052-017-5287-6} {\bibfield  {journal} {\bibinfo
  {journal} {Eur. Phys. J.}\ }\textbf {\bibinfo {volume} {C77}},\ \bibinfo
  {pages} {728} (\bibinfo {year} {2017})},\ \Eprint
  {http://arxiv.org/abs/1703.10836} {arXiv:1703.10836 [hep-ph]} \BibitemShut
  {NoStop}%
\bibitem [{\citenamefont {Fuchs}\ \emph {et~al.}(2003)\citenamefont {Fuchs},
  \citenamefont {Gegelia}, \citenamefont {Japaridze},\ and\ \citenamefont
  {Scherer}}]{Fuchs:2003qc}%
  \BibitemOpen
  \bibfield  {author} {\bibinfo {author} {\bibfnamefont {T.}~\bibnamefont
  {Fuchs}}, \bibinfo {author} {\bibfnamefont {J.}~\bibnamefont {Gegelia}},
  \bibinfo {author} {\bibfnamefont {G.}~\bibnamefont {Japaridze}}, \ and\
  \bibinfo {author} {\bibfnamefont {S.}~\bibnamefont {Scherer}},\ }\href
  {\doibase 10.1103/PhysRevD.68.056005} {\bibfield  {journal} {\bibinfo
  {journal} {Phys.Rev.}\ }\textbf {\bibinfo {volume} {D68}},\ \bibinfo {pages}
  {056005} (\bibinfo {year} {2003})},\ \Eprint
  {http://arxiv.org/abs/hep-ph/0302117} {arXiv:hep-ph/0302117 [hep-ph]}
  \BibitemShut {NoStop}%
\bibitem [{\citenamefont {Djukanovic}\ \emph {et~al.}(2006)\citenamefont
  {Djukanovic}, \citenamefont {Gegelia},\ and\ \citenamefont
  {Scherer}}]{Djukanovic:2006xc}%
  \BibitemOpen
  \bibfield  {author} {\bibinfo {author} {\bibfnamefont {D.}~\bibnamefont
  {Djukanovic}}, \bibinfo {author} {\bibfnamefont {J.}~\bibnamefont {Gegelia}},
  \ and\ \bibinfo {author} {\bibfnamefont {S.}~\bibnamefont {Scherer}},\ }\href
  {\doibase 10.1140/epja/i2006-10096-6} {\bibfield  {journal} {\bibinfo
  {journal} {Eur.Phys.J.}\ }\textbf {\bibinfo {volume} {A29}},\ \bibinfo
  {pages} {337} (\bibinfo {year} {2006})},\ \Eprint
  {http://arxiv.org/abs/hep-ph/0604164} {arXiv:hep-ph/0604164 [hep-ph]}
  \BibitemShut {NoStop}%
\bibitem [{\citenamefont {Schindler}\ \emph {et~al.}(2008)\citenamefont
  {Schindler}, \citenamefont {Djukanovic}, \citenamefont {Gegelia},\ and\
  \citenamefont {Scherer}}]{Schindler:2007dr}%
  \BibitemOpen
  \bibfield  {author} {\bibinfo {author} {\bibfnamefont {M.~R.}\ \bibnamefont
  {Schindler}}, \bibinfo {author} {\bibfnamefont {D.}~\bibnamefont
  {Djukanovic}}, \bibinfo {author} {\bibfnamefont {J.}~\bibnamefont {Gegelia}},
  \ and\ \bibinfo {author} {\bibfnamefont {S.}~\bibnamefont {Scherer}},\ }\href
  {\doibase 10.1016/j.nuclphysa.2008.01.023} {\bibfield  {journal} {\bibinfo
  {journal} {Nucl.Phys.}\ }\textbf {\bibinfo {volume} {A803}},\ \bibinfo
  {pages} {68} (\bibinfo {year} {2008})},\ \Eprint
  {http://arxiv.org/abs/0707.4296} {arXiv:0707.4296 [hep-ph]} \BibitemShut
  {NoStop}%
\bibitem [{\citenamefont {Yan}\ \emph {et~al.}(1992)\citenamefont {Yan},
  \citenamefont {Cheng}, \citenamefont {Cheung}, \citenamefont {Lin},
  \citenamefont {Lin},\ and\ \citenamefont {Yu}}]{Yan:1992gz}%
  \BibitemOpen
  \bibfield  {author} {\bibinfo {author} {\bibfnamefont {T.-M.}\ \bibnamefont
  {Yan}}, \bibinfo {author} {\bibfnamefont {H.-Y.}\ \bibnamefont {Cheng}},
  \bibinfo {author} {\bibfnamefont {C.-Y.}\ \bibnamefont {Cheung}}, \bibinfo
  {author} {\bibfnamefont {G.-L.}\ \bibnamefont {Lin}}, \bibinfo {author}
  {\bibfnamefont {Y.~C.}\ \bibnamefont {Lin}}, \ and\ \bibinfo {author}
  {\bibfnamefont {H.-L.}\ \bibnamefont {Yu}},\ }\href {\doibase
  10.1103/PhysRevD.46.1148, 10.1103/PhysRevD.55.5851} {\bibfield  {journal}
  {\bibinfo  {journal} {Phys. Rev.}\ }\textbf {\bibinfo {volume} {D46}},\
  \bibinfo {pages} {1148} (\bibinfo {year} {1992})},\ \bibinfo {note}
  {[Erratum: Phys. Rev.D55,5851(1997)]}\BibitemShut {NoStop}%
\bibitem [{\citenamefont {Guo}\ \emph {et~al.}(2008)\citenamefont {Guo},
  \citenamefont {Hanhart}, \citenamefont {Krewald},\ and\ \citenamefont
  {Mei{\ss}ner}}]{Guo:2008gp}%
  \BibitemOpen
  \bibfield  {author} {\bibinfo {author} {\bibfnamefont {F.-K.}\ \bibnamefont
  {Guo}}, \bibinfo {author} {\bibfnamefont {C.}~\bibnamefont {Hanhart}},
  \bibinfo {author} {\bibfnamefont {S.}~\bibnamefont {Krewald}}, \ and\
  \bibinfo {author} {\bibfnamefont {U.-G.}\ \bibnamefont {Mei{\ss}ner}},\
  }\href {\doibase 10.1016/j.physletb.2008.07.060} {\bibfield  {journal}
  {\bibinfo  {journal} {Phys. Lett.}\ }\textbf {\bibinfo {volume} {B666}},\
  \bibinfo {pages} {251} (\bibinfo {year} {2008})},\ \Eprint
  {http://arxiv.org/abs/0806.3374} {arXiv:0806.3374 [hep-ph]} \BibitemShut
  {NoStop}%
\bibitem [{\citenamefont {Geng}\ \emph {et~al.}(2010)\citenamefont {Geng},
  \citenamefont {Kaiser}, \citenamefont {Martin-Camalich},\ and\ \citenamefont
  {Weise}}]{Geng:2010vw}%
  \BibitemOpen
  \bibfield  {author} {\bibinfo {author} {\bibfnamefont {L.~S.}\ \bibnamefont
  {Geng}}, \bibinfo {author} {\bibfnamefont {N.}~\bibnamefont {Kaiser}},
  \bibinfo {author} {\bibfnamefont {J.}~\bibnamefont {Martin-Camalich}}, \ and\
  \bibinfo {author} {\bibfnamefont {W.}~\bibnamefont {Weise}},\ }\href
  {\doibase 10.1103/PhysRevD.82.054022} {\bibfield  {journal} {\bibinfo
  {journal} {Phys. Rev. D}\ }\textbf {\bibinfo {volume} {82}},\ \bibinfo
  {pages} {054022} (\bibinfo {year} {2010})},\ \Eprint
  {http://arxiv.org/abs/1008.0383} {arXiv:1008.0383 [hep-ph]} \BibitemShut
  {NoStop}%
\bibitem [{\citenamefont {Jiang}\ \emph {et~al.}(2019)\citenamefont {Jiang},
  \citenamefont {Liu},\ and\ \citenamefont {Yang}}]{Jiang:2019hgs}%
  \BibitemOpen
  \bibfield  {author} {\bibinfo {author} {\bibfnamefont {S.-Z.}\ \bibnamefont
  {Jiang}}, \bibinfo {author} {\bibfnamefont {Y.-R.}\ \bibnamefont {Liu}}, \
  and\ \bibinfo {author} {\bibfnamefont {Q.-H.}\ \bibnamefont {Yang}},\ }\href
  {\doibase 10.1103/PhysRevD.99.074018} {\bibfield  {journal} {\bibinfo
  {journal} {Phys. Rev. D}\ }\textbf {\bibinfo {volume} {99}},\ \bibinfo
  {pages} {074018} (\bibinfo {year} {2019})},\ \Eprint
  {http://arxiv.org/abs/1901.09479} {arXiv:1901.09479 [hep-ph]} \BibitemShut
  {NoStop}%
\bibitem [{\citenamefont {Guo}\ \emph {et~al.}(2018{\natexlab{b}})\citenamefont
  {Guo}, \citenamefont {Heo},\ and\ \citenamefont {Lutz}}]{Guo:2018poy}%
  \BibitemOpen
  \bibfield  {author} {\bibinfo {author} {\bibfnamefont {X.-Y.}\ \bibnamefont
  {Guo}}, \bibinfo {author} {\bibfnamefont {Y.}~\bibnamefont {Heo}}, \ and\
  \bibinfo {author} {\bibfnamefont {M.~F.~M.}\ \bibnamefont {Lutz}},\ }\href
  {\doibase 10.22323/1.334.0085} {\bibfield  {journal} {\bibinfo  {journal}
  {PoS}\ }\textbf {\bibinfo {volume} {LATTICE2018}},\ \bibinfo {pages} {085}
  (\bibinfo {year} {2018}{\natexlab{b}})},\ \Eprint
  {http://arxiv.org/abs/1811.00478} {arXiv:1811.00478 [hep-lat]} \BibitemShut
  {NoStop}%
\bibitem [{\citenamefont {Gasser}\ and\ \citenamefont
  {Leutwyler}(1985)}]{Gasser:1984gg}%
  \BibitemOpen
  \bibfield  {author} {\bibinfo {author} {\bibfnamefont {J.}~\bibnamefont
  {Gasser}}\ and\ \bibinfo {author} {\bibfnamefont {H.}~\bibnamefont
  {Leutwyler}},\ }\href {\doibase 10.1016/0550-3213(85)90492-4} {\bibfield
  {journal} {\bibinfo  {journal} {Nucl. Phys.}\ }\textbf {\bibinfo {volume}
  {B250}},\ \bibinfo {pages} {465} (\bibinfo {year} {1985})}\BibitemShut
  {NoStop}%
\bibitem [{\citenamefont {Lutz}\ \emph {et~al.}(2018)\citenamefont {Lutz},
  \citenamefont {Heo},\ and\ \citenamefont {Guo}}]{Lutz:2018cqo}%
  \BibitemOpen
  \bibfield  {author} {\bibinfo {author} {\bibfnamefont {M.~F.~M.}\
  \bibnamefont {Lutz}}, \bibinfo {author} {\bibfnamefont {Y.}~\bibnamefont
  {Heo}}, \ and\ \bibinfo {author} {\bibfnamefont {X.-Y.}\ \bibnamefont
  {Guo}},\ }\href {\doibase 10.1016/j.nuclphysa.2018.05.007} {\bibfield
  {journal} {\bibinfo  {journal} {Nucl. Phys. A}\ }\textbf {\bibinfo {volume}
  {977}},\ \bibinfo {pages} {146} (\bibinfo {year} {2018})},\ \Eprint
  {http://arxiv.org/abs/1801.06417} {arXiv:1801.06417 [hep-lat]} \BibitemShut
  {NoStop}%
\bibitem [{\citenamefont {Bavontaweepanya}\ \emph {et~al.}(2018)\citenamefont
  {Bavontaweepanya}, \citenamefont {Guo},\ and\ \citenamefont
  {Lutz}}]{Bavontaweepanya:2018yds}%
  \BibitemOpen
  \bibfield  {author} {\bibinfo {author} {\bibfnamefont {R.}~\bibnamefont
  {Bavontaweepanya}}, \bibinfo {author} {\bibfnamefont {X.-Y.}\ \bibnamefont
  {Guo}}, \ and\ \bibinfo {author} {\bibfnamefont {M.~F.~M.}\ \bibnamefont
  {Lutz}},\ }\href {\doibase 10.1103/PhysRevD.98.056005} {\bibfield  {journal}
  {\bibinfo  {journal} {Phys. Rev. D}\ }\textbf {\bibinfo {volume} {98}},\
  \bibinfo {pages} {056005} (\bibinfo {year} {2018})},\ \Eprint
  {http://arxiv.org/abs/1801.10522} {arXiv:1801.10522 [hep-ph]} \BibitemShut
  {NoStop}%
\bibitem [{\citenamefont {Lutz}\ \emph {et~al.}(2020)\citenamefont {Lutz},
  \citenamefont {Sauerwein},\ and\ \citenamefont {Timmermans}}]{Lutz:2020dfi}%
  \BibitemOpen
  \bibfield  {author} {\bibinfo {author} {\bibfnamefont {M.~F.~M.}\
  \bibnamefont {Lutz}}, \bibinfo {author} {\bibfnamefont {U.}~\bibnamefont
  {Sauerwein}}, \ and\ \bibinfo {author} {\bibfnamefont {R.~G.~E.}\
  \bibnamefont {Timmermans}},\ }\href {\doibase 10.1140/epjc/s10052-020-8417-5}
  {\bibfield  {journal} {\bibinfo  {journal} {Eur. Phys. J. C}\ }\textbf
  {\bibinfo {volume} {80}},\ \bibinfo {pages} {844} (\bibinfo {year} {2020})},\
  \Eprint {http://arxiv.org/abs/2003.10158} {arXiv:2003.10158 [hep-lat]}
  \BibitemShut {NoStop}%
\bibitem [{\citenamefont {Sauerwein}\ \emph {et~al.}(2022)\citenamefont
  {Sauerwein}, \citenamefont {Lutz},\ and\ \citenamefont
  {Timmermans}}]{Sauerwein:2021jxb}%
  \BibitemOpen
  \bibfield  {author} {\bibinfo {author} {\bibfnamefont {U.}~\bibnamefont
  {Sauerwein}}, \bibinfo {author} {\bibfnamefont {M.~F.~M.}\ \bibnamefont
  {Lutz}}, \ and\ \bibinfo {author} {\bibfnamefont {R.~G.~E.}\ \bibnamefont
  {Timmermans}},\ }\href {\doibase 10.1103/PhysRevD.105.054005} {\bibfield
  {journal} {\bibinfo  {journal} {Phys. Rev. D}\ }\textbf {\bibinfo {volume}
  {105}},\ \bibinfo {pages} {054005} (\bibinfo {year} {2022})},\ \Eprint
  {http://arxiv.org/abs/2105.06755} {arXiv:2105.06755 [hep-ph]} \BibitemShut
  {NoStop}%
\bibitem [{\citenamefont {Passarino}\ and\ \citenamefont
  {Veltman}(1979)}]{Passarino:1978jh}%
  \BibitemOpen
  \bibfield  {author} {\bibinfo {author} {\bibfnamefont {G.}~\bibnamefont
  {Passarino}}\ and\ \bibinfo {author} {\bibfnamefont {M.~J.~G.}\ \bibnamefont
  {Veltman}},\ }\href {\doibase 10.1016/0550-3213(79)90234-7} {\bibfield
  {journal} {\bibinfo  {journal} {Nucl. Phys.}\ }\textbf {\bibinfo {volume}
  {B160}},\ \bibinfo {pages} {151} (\bibinfo {year} {1979})}\BibitemShut
  {NoStop}%
\bibitem [{\citenamefont {Semke}\ and\ \citenamefont
  {Lutz}(2006)}]{Semke:2005sn}%
  \BibitemOpen
  \bibfield  {author} {\bibinfo {author} {\bibfnamefont {A.}~\bibnamefont
  {Semke}}\ and\ \bibinfo {author} {\bibfnamefont {M.~F.~M.}\ \bibnamefont
  {Lutz}},\ }\href {\doibase 10.1016/j.nuclphysa.2006.07.043} {\bibfield
  {journal} {\bibinfo  {journal} {Nucl. Phys. A}\ }\textbf {\bibinfo {volume}
  {778}},\ \bibinfo {pages} {153} (\bibinfo {year} {2006})},\ \Eprint
  {http://arxiv.org/abs/nucl-th/0511061} {arXiv:nucl-th/0511061} \BibitemShut
  {NoStop}%
\bibitem [{\citenamefont {Chetyrkin}\ and\ \citenamefont
  {Tkachov}(1981)}]{Chetyrkin:1981qh}%
  \BibitemOpen
  \bibfield  {author} {\bibinfo {author} {\bibfnamefont {K.~G.}\ \bibnamefont
  {Chetyrkin}}\ and\ \bibinfo {author} {\bibfnamefont {F.~V.}\ \bibnamefont
  {Tkachov}},\ }\href {\doibase 10.1016/0550-3213(81)90199-1} {\bibfield
  {journal} {\bibinfo  {journal} {Nucl. Phys. B}\ }\textbf {\bibinfo {volume}
  {192}},\ \bibinfo {pages} {159} (\bibinfo {year} {1981})}\BibitemShut
  {NoStop}%
\bibitem [{\citenamefont {Tkachov}(1981)}]{Tkachov:1981wb}%
  \BibitemOpen
  \bibfield  {author} {\bibinfo {author} {\bibfnamefont {F.~V.}\ \bibnamefont
  {Tkachov}},\ }\href {\doibase 10.1016/0370-2693(81)90288-4} {\bibfield
  {journal} {\bibinfo  {journal} {Phys. Lett. B}\ }\textbf {\bibinfo {volume}
  {100}},\ \bibinfo {pages} {65} (\bibinfo {year} {1981})}\BibitemShut
  {NoStop}%
\bibitem [{\citenamefont {Tarasov}(1996)}]{Tarasov:1996br}%
  \BibitemOpen
  \bibfield  {author} {\bibinfo {author} {\bibfnamefont {O.~V.}\ \bibnamefont
  {Tarasov}},\ }\href {\doibase 10.1103/PhysRevD.54.6479} {\bibfield  {journal}
  {\bibinfo  {journal} {Phys. Rev. D}\ }\textbf {\bibinfo {volume} {54}},\
  \bibinfo {pages} {6479} (\bibinfo {year} {1996})},\ \Eprint
  {http://arxiv.org/abs/hep-th/9606018} {arXiv:hep-th/9606018} \BibitemShut
  {NoStop}%
\bibitem [{\citenamefont {Duplancic}\ and\ \citenamefont
  {Nizic}(2004)}]{Duplancic:2003tv}%
  \BibitemOpen
  \bibfield  {author} {\bibinfo {author} {\bibfnamefont {G.}~\bibnamefont
  {Duplancic}}\ and\ \bibinfo {author} {\bibfnamefont {B.}~\bibnamefont
  {Nizic}},\ }\href {\doibase 10.1140/epjc/s2004-01723-7} {\bibfield  {journal}
  {\bibinfo  {journal} {Eur. Phys. J. C}\ }\textbf {\bibinfo {volume} {35}},\
  \bibinfo {pages} {105} (\bibinfo {year} {2004})},\ \Eprint
  {http://arxiv.org/abs/hep-ph/0303184} {arXiv:hep-ph/0303184} \BibitemShut
  {NoStop}%
\bibitem [{\citenamefont {Denner}\ and\ \citenamefont
  {Dittmaier}(2006)}]{Denner:2005nn}%
  \BibitemOpen
  \bibfield  {author} {\bibinfo {author} {\bibfnamefont {A.}~\bibnamefont
  {Denner}}\ and\ \bibinfo {author} {\bibfnamefont {S.}~\bibnamefont
  {Dittmaier}},\ }\href {\doibase 10.1016/j.nuclphysb.2005.11.007} {\bibfield
  {journal} {\bibinfo  {journal} {Nucl. Phys. B}\ }\textbf {\bibinfo {volume}
  {734}},\ \bibinfo {pages} {62} (\bibinfo {year} {2006})},\ \Eprint
  {http://arxiv.org/abs/hep-ph/0509141} {arXiv:hep-ph/0509141} \BibitemShut
  {NoStop}%
\bibitem [{\citenamefont {Battistel}\ and\ \citenamefont
  {Dallabona}(2006)}]{Battistel:2006zq}%
  \BibitemOpen
  \bibfield  {author} {\bibinfo {author} {\bibfnamefont {O.~A.}\ \bibnamefont
  {Battistel}}\ and\ \bibinfo {author} {\bibfnamefont {G.}~\bibnamefont
  {Dallabona}},\ }\href {\doibase 10.1140/epjc/s2005-02437-0} {\bibfield
  {journal} {\bibinfo  {journal} {Eur. Phys. J. C}\ }\textbf {\bibinfo {volume}
  {45}},\ \bibinfo {pages} {721} (\bibinfo {year} {2006})}\BibitemShut
  {NoStop}%
\bibitem [{\citenamefont {Guillet}\ \emph {et~al.}(2019)\citenamefont
  {Guillet}, \citenamefont {Pilon}, \citenamefont {Shimizu},\ and\
  \citenamefont {Zidi}}]{Guillet:2018cdm}%
  \BibitemOpen
  \bibfield  {author} {\bibinfo {author} {\bibfnamefont {J.~P.}\ \bibnamefont
  {Guillet}}, \bibinfo {author} {\bibfnamefont {E.}~\bibnamefont {Pilon}},
  \bibinfo {author} {\bibfnamefont {Y.}~\bibnamefont {Shimizu}}, \ and\
  \bibinfo {author} {\bibfnamefont {M.~S.}\ \bibnamefont {Zidi}},\ }\href
  {\doibase 10.1093/ptep/ptz114} {\bibfield  {journal} {\bibinfo  {journal}
  {PTEP}\ }\textbf {\bibinfo {volume} {2019}},\ \bibinfo {pages} {113B05}
  (\bibinfo {year} {2019})},\ \Eprint {http://arxiv.org/abs/1811.03550}
  {arXiv:1811.03550 [hep-ph]} \BibitemShut {NoStop}%
\bibitem [{\citenamefont {Li}(2023)}]{Li:2022cbx}%
  \BibitemOpen
  \bibfield  {author} {\bibinfo {author} {\bibfnamefont {T.}~\bibnamefont
  {Li}},\ }\href {\doibase 10.1007/JHEP07(2023)051} {\bibfield  {journal}
  {\bibinfo  {journal} {JHEP}\ }\textbf {\bibinfo {volume} {07}},\ \bibinfo
  {pages} {051} (\bibinfo {year} {2023})},\ \Eprint
  {http://arxiv.org/abs/2209.11428} {arXiv:2209.11428 [hep-ph]} \BibitemShut
  {NoStop}%
\bibitem [{\citenamefont {Lutz}(2000)}]{Lutz:1999yr}%
  \BibitemOpen
  \bibfield  {author} {\bibinfo {author} {\bibfnamefont {M.}~\bibnamefont
  {Lutz}},\ }\href {\doibase 10.1016/S0375-9474(00)00206-2} {\bibfield
  {journal} {\bibinfo  {journal} {Nucl. Phys. A}\ }\textbf {\bibinfo {volume}
  {677}},\ \bibinfo {pages} {241} (\bibinfo {year} {2000})},\ \Eprint
  {http://arxiv.org/abs/nucl-th/9906028} {arXiv:nucl-th/9906028} \BibitemShut
  {NoStop}%
\bibitem [{\citenamefont {Lutz}\ \emph {et~al.}(2014)\citenamefont {Lutz},
  \citenamefont {Bavontaweepanya}, \citenamefont {Kobdaj},\ and\ \citenamefont
  {Schwarz}}]{Lutz:2014oxa}%
  \BibitemOpen
  \bibfield  {author} {\bibinfo {author} {\bibfnamefont {M.~F.~M.}\
  \bibnamefont {Lutz}}, \bibinfo {author} {\bibfnamefont {R.}~\bibnamefont
  {Bavontaweepanya}}, \bibinfo {author} {\bibfnamefont {C.}~\bibnamefont
  {Kobdaj}}, \ and\ \bibinfo {author} {\bibfnamefont {K.}~\bibnamefont
  {Schwarz}},\ }\href {\doibase 10.1103/PhysRevD.90.054505} {\bibfield
  {journal} {\bibinfo  {journal} {Phys. Rev.}\ }\textbf {\bibinfo {volume}
  {D90}},\ \bibinfo {pages} {054505} (\bibinfo {year} {2014})},\ \Eprint
  {http://arxiv.org/abs/1401.7805} {arXiv:1401.7805 [hep-lat]} \BibitemShut
  {NoStop}%
\end{thebibliography}%
\bibliographystyle{apsrev4-1}
\end{document}